\ifdefined\JOURNALMODE
  \documentclass[acmsmall,screen]{acmart}
\else
  \documentclass[acmsmall,screen,nonacm]{acmart}
\fi

\usepackage{CJKutf8}

\usepackage{graphicx}
\usepackage{wrapfig}
\hypersetup{hidelinks}
\usepackage{amsmath}
\usepackage{needspace}
\usepackage{xurl}       % Better URL breaking - replaces url package
\usepackage{textcomp}   % Fixes \textregistered font warning
\usepackage{microtype}  % Better text justification
\usepackage{ragged2e}   % Better ragged typesetting
\usepackage{listings}   % Support code listings
\usepackage{array}
\usepackage{tabularx}
\newcolumntype{L}{>{\raggedright\arraybackslash}X}
\newcolumntype{P}[1]{>{\raggedright\arraybackslash}p{#1}}

\usepackage{placeins}

\usepackage{csquotes}
\IfFileExists{latexml.sty}{\usepackage{latexml}}{\newif\iflatexml}
\usepackage{caption}

\ifdefined\JOURNALMODE
  \hypersetup{colorlinks=true,linkcolor=black,citecolor=black,
              urlcolor=black}
\else
  \hypersetup{colorlinks=true,linkcolor=blue,citecolor=blue,
              urlcolor=blue}
\fi

\ifdefined\JOURNALMODE
\else
\fi

\hypersetup{
  bookmarksdepth=4,
  pdftitle={End-to-End Data Movement: Paradigm Reexamination and Principles for Efficiency},
  pdfauthor={Chin~Fang, Timothy~Stitt, Michael~J.~McManus, and
            Toshio~Moriya},
  pdfsubject={End-to-End Data Movement, Co-design},
  pdfkeywords={End-to-End Data Movement, Data Movement Appliances,
               Burst Buffer, System-level Optimization, Cloud, 
               Cross-Domain Data Flow, Research and Education Networks},
}

\iflatexml
  \newcommand{\zhname}[1]{#1}%
\else
  \newcommand{\zhname}[1]{\begin{CJK*}{UTF8}{bsmi}#1\end{CJK*}}%
\fi

\providecommand{\Description}[1]{}
\begin{document}
\raggedbottom
% arXiv HTML (LaTeXML) only: acmart.cls.ltxml doesn't set acmart's
% numeric mode, so natbib defaults to author-year and the bare
% \bibitem{key} labels render as raw cite keys. Force numeric here so
% the HTML matches the PDF ([n], grouped cites combined). pdflatex is
% unaffected (\iflatexml is false there).
\iflatexml\setcitestyle{numbers,square,comma}\fi
\newcommand{\tblheader}[1]{\textbf{\textnormal{#1}}}
\title{End-to-End Data Movement: Paradigm Reexamination and
  Principles for Efficiency}

% ===== Authors (ACM + arXiv safe) =====
\author{\texorpdfstring{\hyperref[bio:fangchin]{Chin~Fang}}{Chin Fang}}
\orcid{0009-0001-2619-6430}
\affiliation{%
  \institution{Zettar Inc.}
  \country{USA}
}
\email{fangchin@zettar.com}

\author{\texorpdfstring{\hyperref[bio:stitt]{Timothy~Stitt}}{Timothy Stitt}}
\orcid{0009-0005-9696-0683}
\affiliation{%
  \institution{F. Hoffmann-La Roche Ltd.}
  \country{Switzerland}
}

\author{\texorpdfstring{\hyperref[bio:mcmanus]{Michael~J.~McManus}}{Michael J. McManus}}
\orcid{0000-0003-3476-6242}
\affiliation{%
  \institution{Intel Corporation (Retired)} \country{USA} }

\author{\texorpdfstring{\hyperref[bio:moriya]{Toshio~Moriya}}{Toshio Moriya}}
\orcid{0000-0001-7226-5487}
\affiliation{%
  \institution{Institute of Materials Structure Science, High Energy
    Accelerator Research Organization (KEK)}
  \country{Japan}
}

\ifdefined\JOURNALMODE
  \authornote{Corresponding author. This manuscript is original, and
    is not under review in any other venue. All authors have approved
    this work and declare no conflicts of interest.}
\fi

\ifdefined\JOURNALMODE
  % ACM submission: no preprint page
\else
  \clearpage
  \thispagestyle{empty}
  \null\vfill
  \long\def\preprintbody{%
        \textbf{To read the Abstract}, please see
        \url{https://arxiv.org/abs/2512.15028}.\\[1em]

      This report is actively maintained and revised as hardware,
      software, and operational practices evolve.\\[1em]
        
      To avoid any external delays, it is kept only on preprint
      servers. This way, we can continuously update, extend, and
      refine the report. Iterative, real-world refinement is
      precisely what makes engineering literature valuable. We invite
      rigorous scrutiny of the presented findings. Our only standard:
      the operational truth.\\[1em]

      This work introduces four principal contributions: a co-design principle
      for efficient data movement, holistically designed data movement
      appliances, a burst-buffer and data-staging method for sustained
      wide-area transfer (with a quantitative capacity-sizing bound),
      and the \emph{Drainage Basin Pattern} (Fig.~\ref{fig:dbp})
      conceptual model. The model guides appliance selection across
      the Core/Mini+/Mini spectrum, matching network bandwidth, burst
      buffer capacity, and compute resources to workflow
      requirements. It also provides a framework for reasoning about
      bottlenecks, resource allocation, and risk in
      data-transfer-intensive projects.\\[1em]

      The prevalent underutilization of provisioned network bandwidth
      is defined using the concept of \enquote{fidelity gap}. A burst
      buffer subsystem, together with data staging, is introduced at
      every tier to decouple data movement from erratic production
      storage. All tiers run the same Zettar zx unified software data
      mover~\cite{ZettarProducts}. Its design has always integrated
      storage, computing, networking, and security.\\[1em]

      The Gen1 appliance ($\le$ 100~Gbps) was completed in
      2023. Though Gen2 (intrinsically scale-out and HA; $>$ 100~Gbps)
      is under development, its architecture was already proven in
      September 2018~\cite{ESnet2018Record}. See also
      Fig.~\ref{fig:slac-testbed}.\\[1em]

      \textbf{License:} CC BY 4.0\\
      \small\url{https://creativecommons.org/licenses/by/4.0/}\\[1em]

      For the latest version, please see:\\
      \url{https://arxiv.org/abs/2512.15028}\\[1em]
      
      \textbf{Authors’ Contact Information:}\\
      \hyperref[bio:fangchin]{Chin Fang} (\zhname{方智}), Ph.D.,
      Zettar Inc., USA, fangchin@zettar.com\\
      \hyperref[bio:stitt]{Timothy Stitt}, Ph.D., F. Hoffmann-La Roche
      Ltd., Switzerland\\
      \hyperref[bio:mcmanus]{Michael J. McManus}, Ph.D., Intel Corporation (Retired), USA\\
      \hyperref[bio:moriya]{Toshio Moriya} (\zhname{守屋 俊夫}),
      Ph.D., Institute of Materials Structure Science, KEK, Japan

}% end \preprintbody
  \begin{center}
    {\Large\bfseries Evolving Systems Engineering Report}\\[0.75em]
    {\normalsize \today}\iflatexml\else\\[1.5em]

    \begin{minipage}{0.8\textwidth} \raggedright
      \preprintbody
    \end{minipage}

    \vfill\fi
  \end{center}
  \iflatexml{\raggedright\preprintbody\par}\fi
  \vfill

  \clearpage
  % Start main article on page 1
  \setcounter{page}{1}

  % ELIMINATE RUNNING HEADERS FOR arXiv PREPRINT
  \makeatletter
  \def\@evenhead{}%
  \def\@oddhead{}%
  \def\@evenfoot{\hfil\thepage\hfil}%   % page numbers bottom center
  \def\@oddfoot{\hfil\thepage\hfil}%    % page numbers bottom center
  \makeatother
\fi

\iflatexml\else
\begin{abstract}
High-performance data transfer is often viewed through raw bandwidth,
with 100+ Gbps international links seen as the primary enabler. Yet
this network-centric view confuses provisioned speed with sustainable
throughput. Suboptimal rates occur even on 10 Gbps links, and faster
networks only magnify the issue. We examine six paradigms - network
latency, TCP congestion control, CPU performance, virtualization, and
others - that critically impact data movement workflows. These reflect
common engineering assumptions shaping system design, procurement, and
operations. To bridge the gap between raw bandwidth and
application-level throughput, we introduce the "Drainage Basin
Pattern" - a conceptual model for reasoning about end-to-end
constraints across heterogeneous hardware and software at varying
target rates. Our findings are validated via production-scale
deployments, from 10 Gbps links to U.S. DOE ESnet technical
evaluations and transcontinental trials over 100 Gbps operational
links. Results show that bottlenecks typically lie outside the network
core, and that holistic hardware-software co-design delivers
consistent, predictable performance for demanding bulk and streaming
transfers. A burst buffer subsystem, together with data staging, is
introduced at every tier to decouple data movement from erratic
production storage and sustain wide-area transfer, with a quantitative
bound for sizing the buffer capacity it requires. The primary goal is
to transform such transfers from unpredictable struggles into routine,
line-rate operations accessible to any regular user. Finally, we
correct two industry misconceptions: using aggregated traffic rate as
a measure of application efficiency, and conflating operational
complexity with technical expertise.
\end{abstract}
\fi

\ifdefined\JOURNALMODE
  \maketitle
\fi


\ifdefined\JOURNALMODE

\fi

% ---- TOC: every level, down to \paragraph ====
\renewcommand{\contentsname}{Table of Contents}
\setcounter{secnumdepth}{4}
\setcounter{tocdepth}{4}
\begingroup
\footnotesize           % \small: 9 - \footnotesize 8 - \scriptsize 7
\tableofcontents
\endgroup 
\clearpage

% Include the converted content with two-column spanning figures/tables

\section{Introduction}\label{sec:intro}

For modern data-intensive enterprises and scientific collaborations,
efficient movement of data from edge locations to core data centers
and cloud resources is an unresolved challenge. To address it,
high-bandwidth, long-distance networks have been a common
response. For example, ESnet6 provides transcontinental and
international connectivity essential for global scientific
collaborations, with speeds from 400~Gbps to 1.2~Tbps
\cite{ESnetAbout}. Nevertheless, a predominant focus on high raw
network bandwidth often obscures a more critical truth: sustainable
end-to-end data movement is constrained by the full environment,
including storage, host architectures, software design, and security
measures, along the data path rather than by the network alone. This
holds true from 10~Gbps to 100~Gbps and faster.

\begin{figure*}[!b]
  \centering
  
  \includegraphics[width=0.98\textwidth]{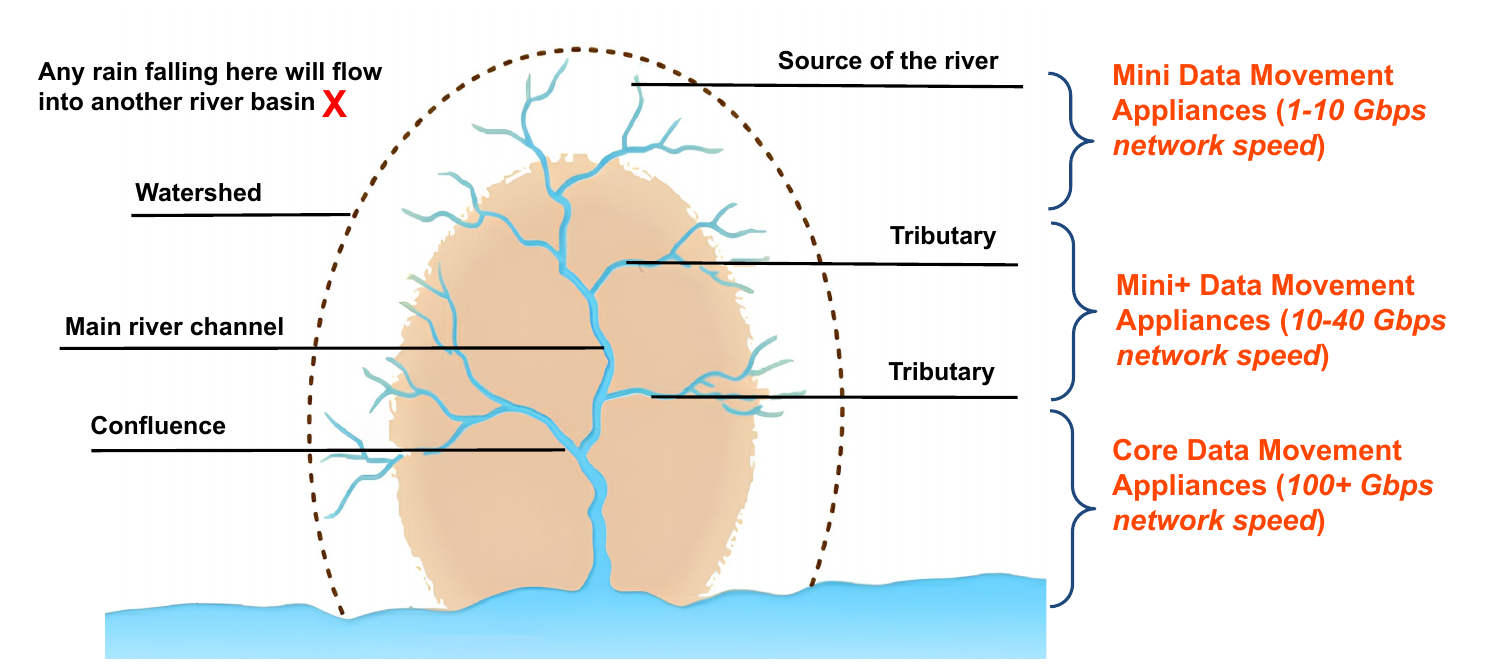}
  \caption{The experience of moving data for most practitioners is
    typically limited to the source of the river. This corresponds to
    end-user activities, such as transferring photos and videos from
    mobile phones or moving spreadsheets from one folder to another. A
    petabyte? It is so often approached as though it were merely a
    larger folder. This prevalent misconception motivated the
    introduction of the Drainage Basin Pattern. It shows the reality.
  }
  \Description{A conceptual diagram showing the Drainage Basin
    Pattern for data movement. Illustrates how data flows from edge
    sources through networks to core data centers, similar to water
    flowing from streams into rivers to ocean.}
  \label{fig:dbp}
\end{figure*}

This report analyzes production-scale deployments over
transcontinental links and edge-to-core transfers, describes
first-hand insight from designing real systems since
2016~\cite{Fang2016SLAC, Fang2016CHEP}, and warns what the wrong tools
cost~\cite{Fang2021Rsync}.

We define the discrepancy between theoretical link capacity and actual
application-level throughput as a \enquote{fidelity gap}.  This gap
manifests as lower than desired data rates, even on 10~Gbps links and
modest commodity hardware.  Henceforth, we define a \enquote{fast}
network as 10~Gbps up to (but excluding) 100~Gbps, and a
\enquote{high-speed} network as 100~Gbps and faster. The gap's effect
amplifies as the network speed increases. This is not theoretical.
for an actual 2025 example involving a Chinese AI company, please
see~\cite{WSJ2025AI,Toms2025AI}. Over a properly provisioned 100~Gbps
link and co-designed data movement appliance, the same data can be
moved in under a week (Table~\ref{tab:tab5}).

The \enquote{Drainage Basin Pattern} conceptual model
(Fig.~\ref{fig:dbp}) addresses the gap by shifting the focus from
isolated network optimization to systemic co-design. The efficacy of
this approach is evidenced by achieving near-line-rate performance on
a \enquote{high-speed} 100~Gbps transcontinental link ($\sim$84~Gbps
in a pharmaceutical production environment, with full TLS
encryption). Figures~\ref{fig:bulk-sweep-plain},
\ref{fig:streaming-sweep}, and \ref{fig:hpe-reproduction} further
validate this across bulk transfers, streaming workloads, and
independent reproduction---all with full TLS encryption. It also
enables efficient 1--10~Gbps transfers at the edge.  Detailed
methodology and measurement results appear in
Sections~\ref{sec:paradigms} and~\ref{sec:reproducibility}, where we
analyze how host configurations influence end-to-end throughput across
multiple deployment scenarios.

Drawing on over a decade of operational experience, our work
demonstrates that once a link is in place, whether fast or high-speed,
host-side configurations become the primary determinants of
performance. A key architectural component is the use of high-speed
burst buffers (BB)~\cite{Bhimji2016Burst} as staging areas, enabling
reliable, near-line-rate movement when the end-to-end path is properly
co-designed. These principles have been validated across years of
production trials and independent evaluations, such as Linac Coherent
Light Source (LCLS) petabyte transfers with ESnet and the Overall
Winner finish at the Inaugural Supercomputing Asia 2019 Data Mover
Challenge. This body of evidence shows that the bottlenecks have
shifted from the WAN links to the internal systems architecture of the
endpoints~\cite{Fang2016SLAC}, where these factors, not the network,
set the true performance limits of data movement regardless of the
network's nominal speed.

A central insight: co-design yields not only high peak performance but
also crucial operational benefits. As we detail in
Section~\ref{sec:appliances}, the ESnet evaluation found that a
properly co-designed system can sustain consistent throughput across
file sizes from 1~MiB to 1~TiB with a single configuration. This
stands in contrast to software-centric approaches that often require
file-size-range-based optimization. Such operational simplicity is not
merely a convenience; it is a direct consequence of architectural
co-design and a prerequisite for deployment in production environments
where datasets and workflows vary widely.

After establishing the necessary background and architectural
foundations (Section~\ref{sec:foundations}), the report addresses two
more recent misconceptions (Section~\ref{sec:misconceptions}) and then
evaluates the six long-standing paradigms (Section~\ref{sec:paradigms}
and tabulated below). They continue to shape expectations, system
design, and investment decisions in production end-to-end data
movement.

Through production deployments and operational trials, we show why
they often fail to predict achievable performance across the full
edge-to-core spectrum.  Reliable and predictable data movement emerges
from holistic, co-designed, and tightly integrated hardware-software
systems, instead of isolated optimization of individual
components. The end-to-end behavior of these systems can be reasoned
about using the Drainage Basin Pattern. We examine how the proposed
principles generalize to emerging network speeds beyond 100~Gbps.

\medskip
\noindent
\begin{minipage}{\linewidth}
  \centering
  \small
  \begin{tabularx}{\linewidth}{|c|X|c|}
    \hline
    \textbf{\#} & \textbf{The common belief} & \textbf{Section} \\
    \hline
    1 & Network latency is the primary performance constraint
      & \ref{sec:latency} \\
    \hline
    2 & Packet loss dominates, so the TCP Congestion Control
      Algorithm (CCA) is what matters
      & \ref{sec:loss-cca} \\
    \hline
    3 & Dedicated lines are essential for high-speed
      data transfer testing
      & \ref{sec:dedicated-lines} \\
    \hline
    4 & More link bandwidth directly yields higher application
      transfer rates
      & \ref{sec:bandwidth} \\
    \hline
    5 & Premium CPUs are essential at every endpoint
      & \ref{sec:cpus} \\
    \hline
    6 & Virtualization and cloud are universally useful for
      high-throughput workflows
      & \ref{sec:cloud} \\
    \hline
  \end{tabularx}
\end{minipage}
\medskip
\noindent

Throughout this report, all referenced systems are assumed to run Red
Hat Enterprise Linux (RHEL) 9.6 \cite{RedHatRHEL} or free rebuilds
such as Rocky Linux 9.6 \cite{RockyLinux}.

\section{Foundations}\label{sec:foundations}

Research programs such as the Global Research Platform
(GRP)~\cite{GRPAbout} have advanced high-speed networking and dynamic
path provisioning. A recent workshop~\cite{GRP2025SCA} pushed
programmable 400~Gbps-class connectivity even further. But a
network-first focus relegates the data mover and storage to
afterthoughts. We invert that order: this work treats data movement as
a first-class systems problem~\cite{Fang2016SLAC,Fang2019Samsung}, by
examining how storage behavior, host architecture, and data mover
design interact with high-speed networks in production. The network is
necessary. Nevertheless, it is not always where the bottleneck lives.

\subsection{The Drainage Basin Pattern (DBP)}\label{sec:dbp}

\textbf{The DBP is not only descriptive; it is prescriptive}---it
defines the bounded terrain within which end-to-end data movement can
actually be engineered. A drainage basin collects flows from many
small headwaters, gathers the water through tributaries that meet at
confluences, and carries the final flow along a main channel toward
the outlet. Production data movement has the same shape: edge sources
are the headwaters; aggregation links are the tributaries; aggregation
points are the confluences; the high-speed backbone is the main
channel; and a core data center, cloud, or destination store is an
outlet (Fig.~\ref{fig:dbp}).

\noindent
\begin{minipage}{\linewidth}
  \centering
  \small
  \begin{tabularx}{\linewidth}{|p{3.6cm}|L|}
    \hline
    \textbf{Basin element} & \textbf{Data-movement counterpart} \\
    \hline
    Headwaters & Edge data sources: instruments, sensors, devices \\
    \hline
    Tributaries & Aggregation links carrying converging flows \\
    \hline
    Confluences & Aggregation points where flows merge \\
    \hline
    Main channel & High-speed backbone, e.g., R\&E networks \\
    \hline
    Outlet & Core data center, cloud, or destination store \\
    \hline
    Watershed & Bounded membership: the security perimeter \\
    \hline
  \end{tabularx}
\end{minipage}

A single basin drains to one outlet---that convergence is what makes
it a basin. Separate basins can be joined; an approach that has a long
history is the use of canals. For example, the Grand Canal of China (a
UNESCO World Heritage site, built in the 7th century) was unified into
a single national system in the Sui dynasty, linking the country's
five river basins to move grain and knit its north and south
together~\cite{GrandCanalUNESCO}. A data canal is no different in
principle. The U.S.\ DOE, for instance, operates three major
supercomputing facilities---the Argonne and Oak Ridge Leadership
Computing Facilities and NERSC~\cite{ALCFAbout, OLCFAbout,
  NERSCAbout}---each the outlet of its own basin, and DOE's own
high-speed backbone, ESnet, is the canal that connects them.  Movement
among them is smooth because a single authority engineers and governs
that canal and every endpoint it joins is already an enrolled
member. A watershed bounds \emph{membership}, not connectivity---it
keeps outsiders out, not insiders apart; canals join the insiders by
design, not by flood.

What marks a basin is its \emph{watershed}: the ridgeline that bounds
which flows belong to it. This boundary is the pattern's prescriptive
content, and it is not an abstraction---it maps directly to the
network security boundary that every enterprise already
operates. Membership is bounded and enumerable: a flow is either
inside the watershed or it is not. In the architecture presented here,
that boundary is drawn by mutual, reciprocal enrollment of
endpoints---two sites exchange data only after each has admitted the
other, so consent itself traces the basin's perimeter.

\begin{figure}[tb]
  \centering
  \includegraphics[width=0.80\linewidth]{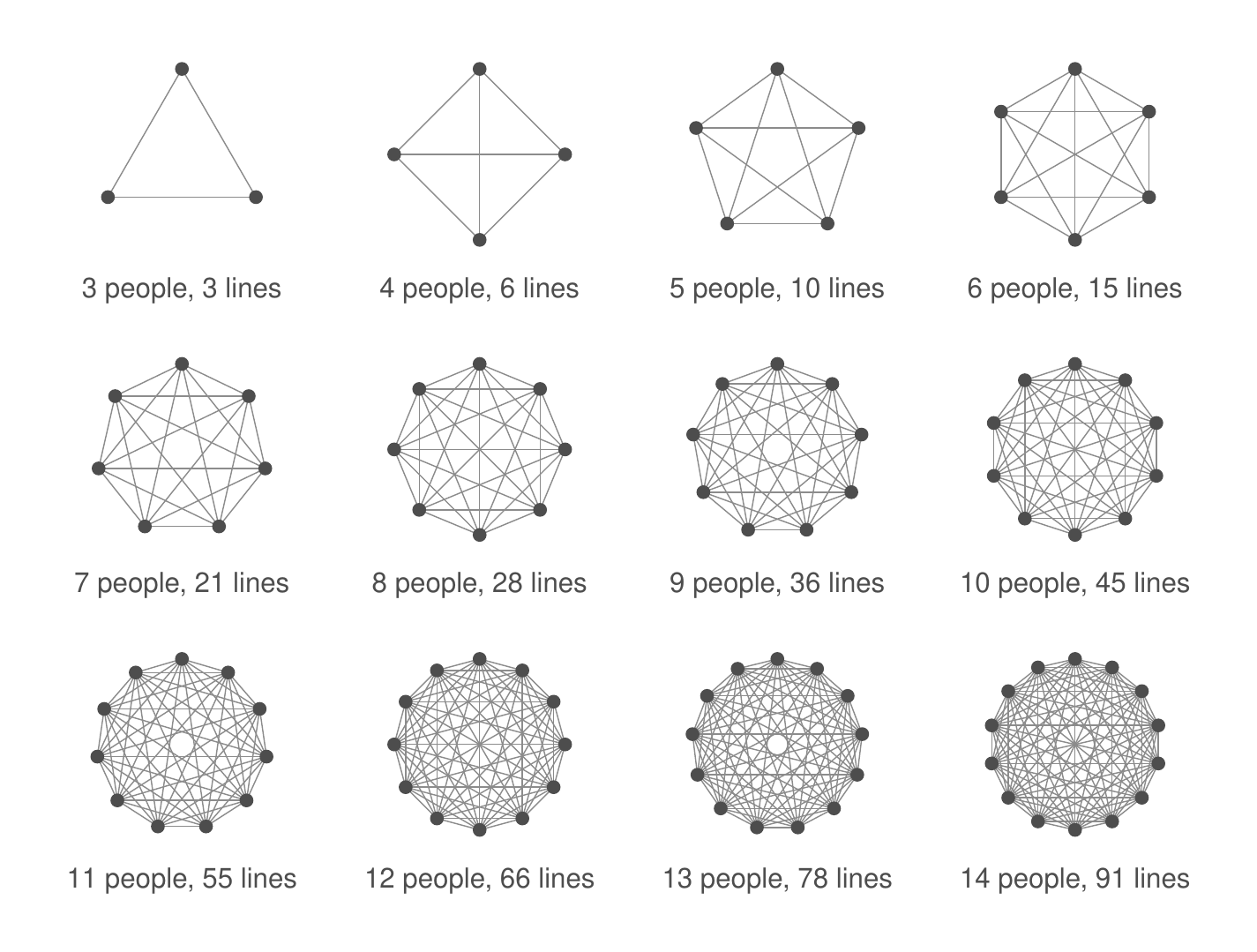}
  \caption{Pairwise links in an all-to-all mesh grow as
    $\binom{n}{2}=n(n-1)/2$: from 3 among 3 participants to 91 among 14.
    A bounded watershed never incurs this $O(n^{2})$ cost, because the
    unbounded mesh cannot form.}
  \Description{Twelve panels, each a complete graph of $n$ nodes evenly
    spaced on a circle with every pair connected, for $n$ from 3 to 14.
    Each is labelled with the number of people and the number of
    connecting lines, equal to $n(n-1)/2$: 3 people 3 lines, 4 people 6
    lines, up to 14 people 91 lines. The count rises steeply as $n$
    grows.}
  \label{fig:cn2}
\end{figure}

Because the watershed is bounded, the counts that matter are finite.
The headwater flows, the tributary flows, and the confluences are
finite in number, so the $O(n^{2})$ coordination cost of an unbounded
mesh~(Fig.~\ref{fig:cn2}, Eq.~\ref{eq:cn2}) never arises---not because
it is mitigated, but because it cannot form. At each confluence the
number of participating users is likewise finite, which renders a
common objection---``how will this scale to tens of thousands of
users?''---moot by construction.  Bounded membership is also the only
setting in which predictable data rates can be engineered at all:
without a perimeter, there is nothing finite to design against.

\begin{equation}\label{eq:cn2}
  \text{Number of lines} = \binom{n}{2} = \frac{n(n-1)}{2}.
\end{equation}

Furthermore, a watershed is more than a boundary; it is a specific
ecosystem, shaped by its own terrain, climate, and inhabitants. That
maps to organization-specific customizations, such as security
policy, storage backends, identity integration, and compliance
obligations.  An open, auto-joining federation does not dig a canal
between basins; it floods them together and dissolves the
divides. With unbounded membership there is no enumerable perimeter,
no per-organization ecosystem, and no finite structure to engineer
against. However well such a flood serves open sharing, it is
ill-fitting for demanding, bounded, production data movement in the
real world.

Movement across watershed boundaries is a distinct and genuinely
harder problem. As an example, a large biopharmaceutical enterprise
routinely exchanges data with several external Contract Research
Organizations (CROs) that sit outside its perimeter, in separate
watersheds of their own. Unlike the facilities within a single
institution, these parties share neither a watershed nor a canal
already joining them, so there is no bounded membership to engineer
against; practitioners consistently report such cross-boundary
exchange as painful and complex. The prescriptive response is not to
flood the two perimeters into one open federation but to dig a canal
between them, i.e., a single, bounded, mutually consented channel that
each authority explicitly agrees to and governs. Both watersheds stay
intact; only the agreed flow crosses. Transfer task
forwarding~(Fig.~\ref{fig:transfer-forwarding} and
Table~\ref{tab:forwarding}) is the canal.

The bounded model is not merely a preference; practice already
converges on it. Federated learning, for instance, honors the
watershed exactly: computation stays on-premises within each boundary,
and only results cross it. The productive response to a boundary is to
respect it, not to erase it.

\subsection{Burst Buffers and Data Staging}\label{sec:burst-buffers}

\textbf{BBs decouple the data mover from slow, unpredictable
  production storage.} Originally, supercomputer designers pioneered
the BB to absorb data for high-performance computing (HPC) nodes at
extreme velocities; we reshape this concept for sustained, wide-area
data movement, drawing on the far older principle of flow decoupling
and rate mismatch management.

In 256~BCE, Li~Bing (\zhname{李冰})~\cite{BaiduLiBing} built the
Dujiangyan (\zhname{都江堰}, a World Heritage Site)~\cite{UNESCODujiangyan}
on the Min River, taming it without a dam by dividing and smoothing the
flow to decouple irrigation from flood; it still runs today.

Li adopted the oldest Chinese water management principle. Four
thousand years ago, Gun (\zhname{鯀})~\cite{Shiji} dammed the flood
and failed. But his son Yu (\zhname{禹})~\cite{Shiji} guided the water
to the sea and prevailed: blocking is inferior to guiding
(\zhname{堵不如疏})~\cite{BaiduDayu}.

Dams can hold back a flood. But if they are breached, whether by human
hand or by structural failure, the consequences are catastrophic. Such
disasters are not confined to any one nation or era. In more modern
times, the 1938 Yellow River flood~\cite{Huayuankou1938} and the 1941
Dnieper dam destruction~\cite{Dnieper1941} fit the same pattern: a dam
or dike destroyed to flood the enemy's path for short-term tactical
gain, at lasting human and environmental cost. A dam or levee can also
fail via structural issues, causing flooding disasters, for instance
the 1928 St.\ Francis Dam collapse~\cite{StFrancisDam1928}, the 1959
Malpasset Dam failure~\cite{MalpassetDam1959}, the 1975 Banqiao
cascade~\cite{Banqiao1975}, and the 2005 New Orleans levee
failure~\cite{ILITKatrina2006}. Yet a dam's deepest harm isn't even
the breach; it's the slow debt even a sound one runs up, hoarding the
sediment that builds deltas, walling off fish migration, silting
toward its eventual demise. The artificial levee is the dam's partner
in crime. Together, they disconnect the river from its landscape,
causing the delta to wither.

Dujiangyan only channels, passes the silt, and has run 2,250 years
owing the river nothing. The lesson is also architectural: many
channels, no single wall, i.e., scale-out, not scale-up. Our
appliances make the same choice: demanding data movement spread across
coordinated, co-designed nodes, not one monolithic server or
store. See also~\cite{ESnet2018Record}, Fig.~\ref{fig:slac-testbed},
and Fig.~\ref{fig:bigtwin-hardware}. Using the guiding principle,
Section~\ref{sec:forwarding} shows how to carry the flow across a
basin's security perimeter, its watershed boundary.

A burst buffer does the same for data: a fast, low-latency layer that
data is staged through, mitigating the impact of production storage's
low throughput and high latency on the flow.

A BB is built on Non-Volatile Memory express (NVMe) Solid State Drives
(SSDs), with a dedicated software layer on top, e.g., a local file
system or a scale-out file store. It serves as a staging area between
production storage and the network.  A vendor-agnostic data movement
appliance design employs a unified software data mover (Zettar zx
\cite{ZettarProducts}) and BBs to create a predictable performance
envelope.  This enables sustained near-line-rate data movement between
BBs across both fast ($< 100~\mathrm{Gbps}$) and high-speed ($\ge
100~\mathrm{Gbps}$) WAN links. See Table~\ref{tab:tab1} for a summary
of zx's features and capabilities.

\begin{table}[hbp]
  \centering
  \caption{Key functional capabilities and core features of the 
    Zettar zx data movement software.}
  \label{tab:tab1}
  \small
  \begin{tabularx}{\textwidth}{|>{\hsize=0.8\hsize}X
                               |>{\hsize=1.2\hsize}X |}
    \hline
    \textbf{Capability} & \textbf{Notes} \\
    \hline
    Bulk and streaming transfers & Utilizes TCP for transport 
    \\
    \hline
    Predictable and sustained performance & Regardless of encryption,
    compression, latency and file sizes \\
    \hline
    Cluster architecture & Linear scaling \\
    \hline
    Single application coverage & File, object, locally and over 
    distance, on-prem and cloud \\
    \hline
    Symmetric concurrent send/receive & \texttt{rsync} and 
    \texttt{bbcp} alike \\
    \hline
    Embeddable on specialized hardware & Support for high-speed biomedical
    instruments\\
    \hline
    Quality of Service (QoS) & Built-in support for traffic 
    prioritization \\
    \hline
    Integration with typical IAM & Active Directory (AD) and 
    OpenID Connect (OIDC) support \\
    \hline
  \end{tabularx}
\end{table}

The BB serves both as a fast storage tier and a deliberate decoupling
mechanism.  It isolates the data mover from production storage
services, which are often optimized for capacity or ease of use,
rather than consistent high throughput or low latency. This decoupling
allows each layer to operate within its effective performance regime
while staging coordinates data movement across mismatched
behaviors. In other words, the BB acts as a low-jitter interface that
buffers the stochastic throughput and latency of the non-deterministic
source (production storage) to ensure a deterministic, high-bandwidth
supply to the high-speed sink.

In this report, data staging is defined as the movement of data between
production storage and BBs. It is a critical coordinating
process. This operation must be simple, predictable, and capable of
supporting multiple users concurrently. High efficiency is mandatory,
as any delay in staging fundamentally negates the performance benefits
of burst buffering. The need for staging arises because production
storage systems are frequently constrained by throughput, latency, and
protocol limitations, making them incapable of feeding high-speed WAN
links directly.

At DBP aggregation points, where multiple flows converge and expose
the mismatch between distributed behavior and storage services,
staging is critical. Without it, such convergence shifts the
bottleneck from the network to storage, underutilizing the attainable
connectivity.  A natural objection: these rates are BB to BB, while
production data resides in production storage, and staging into the BB
takes time. The objection assumes the two operations are serial. They
are not; they can happen concurrently.

To make the concurrency precise, consider a dataset of size $D$
moved through three pipeline stages: reading from source production
storage at sustained rate $r_s$, crossing the WAN at line rate
$r_n$, and writing to destination production storage at sustained
rate $r_d$. Coupled (serial) execution pays for every stage in
turn~\cite{Fang2021Rsync}, whereas staging overlaps them behind the BB:
\begin{equation}\label{eq:stage-time}
  T_{\text{coupled}} = D\left(\frac{1}{r_s} + \frac{1}{r_n}
    + \frac{1}{r_d}\right),
  \qquad
  T_{\text{staged}} = \frac{D}{\min(r_s, r_n, r_d)} + t_{0},
\end{equation}
where $t_{0}$ is the one-time cost of the initial filling and the
final draining of the pipeline.  It is negligible for large datasets.
The resulting staging speedup is
\begin{equation}\label{eq:stage-speedup}
  S \;\equiv\; \frac{T_{\text{coupled}}}{T_{\text{staged}}}
  \;\xrightarrow{\,D \to \infty\,}\;
  \min(r_s, r_n, r_d)\left(\frac{1}{r_s} + \frac{1}{r_n}
    + \frac{1}{r_d}\right),
  \qquad 1 \le S \le 3,
\end{equation}
with equality $S = 3$ if and only if $r_s = r_n = r_d$, i.e.,
\textit{balanced}. Staging thus returns up to a threefold reduction in
wall-clock time, and it delivers the \emph{full} factor precisely when
the stages are balanced, which is exactly what the holistic co-design
of endpoints and network is for. An unbalanced pipeline instead
is dominated by its slowest stage, the $\min(r_s, r_n, r_d)$ ceiling
that the end-to-end rate meets in any case.

Staging also targets a local BB rather than the wide-area path, so
production storage is read at its own pace. Let $r_{\text{stage}}$ be
the sustained staging rate of production storage, and $r_{\text{link}}
= \min(r_{\text{bb}}, r_n)$ the attainable fast-path rate, the smaller
of the BB-to-BB rate $r_{\text{bb}}$ and the network bandwidth
$r_n$. When $r_{\text{stage}} < r_{\text{link}}$, no steady-state
arrangement recovers the difference: the buffer drains faster than
production storage can refill it, the sustained end-to-end rate
settles at $r_{\text{stage}}$, and the link is left underutilized. A
way to compensate is to increase BB capacity, trading space for time,
but only over a \emph{finite} transfer.

To hold the full rate $r_{\text{link}}$ across a dataset of size $D$
in spite of slower production storage, the BB must absorb the
accumulated deficit:
\begin{equation}\label{eq:stage-capacity}
  B \;\ge\; D\left(1 - \frac{r_{\text{stage}}}{r_{\text{link}}}\right).
\end{equation}
The required capacity grows linearly with both the dataset size and
the fractional shortfall $1 - r_{\text{stage}}/r_{\text{link}}$: a
tier delivering half the link rate demands a buffer as large as half
the entire dataset. Nevertheless, the cost of high-performance NVMe
SSDs often makes this mathematically valid sizing not realizable. A
remedy is to raise $r_{\text{stage}}$ itself: tune zx for staging, or
increase the production-storage performance, often a very challenging
endeavor~\cite{Egersdoerfer2025STELLAR}.

The first, however, draws on zx itself, which was developed as the
foundation for the holistic co-design principle central to this
work. Its architecture integrates all necessary functions, including
over-distance transfer, internal data staging (between production
storage and BBs), support for diverse protocols (file and object), and
handling of both bulk and streaming transfers. This single,
concurrent, and scale-out data mover manages the complete data
placement workflow, from source storage through transit to destination
storage. Burst Buffers and Data Staging are thus foundational elements
of our Co-design Engineering Principle. A six-minute recorded overview
of this workflow is available~\cite{Fang2026DataStaging}.

\subsubsection{The Staging Workflow}\label{sec:staging-workflow}

\begin{figure}[tb]
  \centering
  \includegraphics[width=\textwidth]{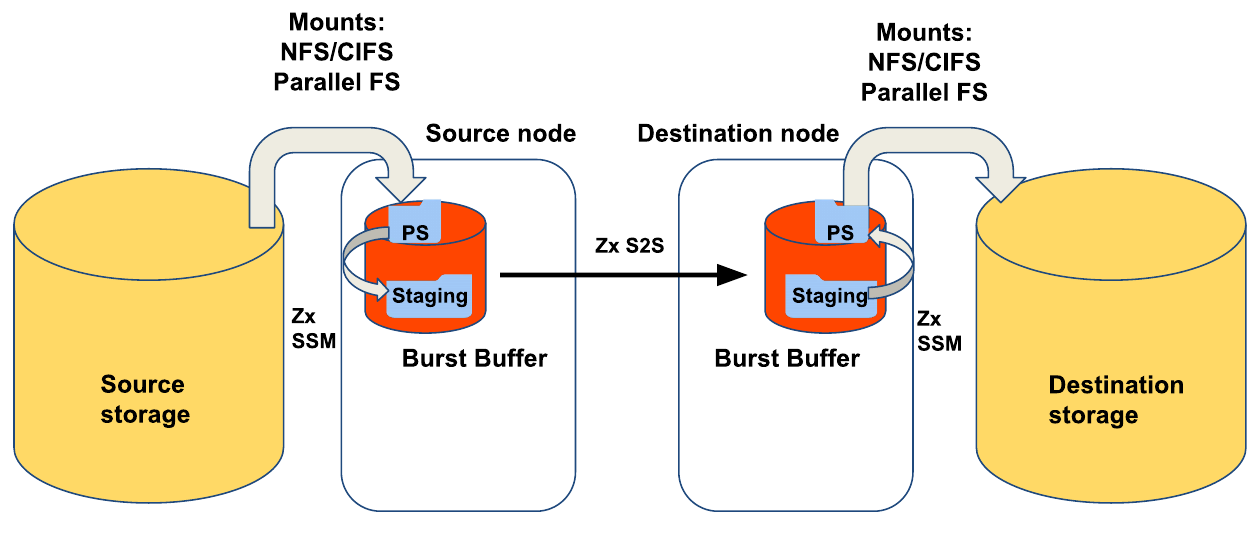}
  \caption{The data staging app runs completely on a user’s
    workstation. It leverages Zettar zx’s ability to do local transfer
    using its single-site-mode (SSM), like rsync and cp, but much
    faster and scale-out-capable. zx’s task forwarding is employed to
    transport staged data from source to destination.}
  \Description{A high-level diagram of the data staging workflow. On the
    left, source storage connects to a source node over NFS/CIFS or a
    parallel file system. Inside the source node a burst buffer holds two
    folders, PS and Staging; a local Zettar zx task (the SSM step) copies
    data from PS to Staging. A Zettar zx site-to-site transfer (S2S) then
    carries the data across the network from the source node's Staging
    folder to the destination node's Staging folder. On the destination
    node a second local zx task copies from Staging back to PS, and the
    node writes to destination storage over NFS/CIFS or a parallel file
    system.}
  \label{fig:staging-workflow}
\end{figure}

Because the three steps are separate
(Fig.~\ref{fig:staging-workflow}), they run concurrently on different
pieces of the data — one crossing the network while the next is staged
in and the previous written out. That overlap is what lets staging
outpace a direct transfer, where the slow storage would gate the whole
flow.

\subsubsection{A Worked Example}\label{sec:staging-example}

To make Eqs.~(\ref{eq:stage-time})--(\ref{eq:stage-capacity}) concrete,
consider a single user staging a $D = 100$~TiB dataset
($\approx 8.8\times10^{5}$~Gbit) between two appliances. Production
storage is deliberately \emph{asymmetric}: source read $r_s = 25$~Gbps
and destination write $r_d = 20$~Gbps (writing commits a change of state
and is the structurally slower half). The $100$~Gbps link carries roughly
$15\%$ other traffic, leaving $r_n = 85$~Gbps, and a conservative burst
buffer gives $r_{\text{bb}} = 70$~Gbps. Because $r_{\text{bb}} < r_n$, the
crossing stage runs at $r_{\text{link}} = \min(r_{\text{bb}}, r_n) =
70$~Gbps---the rate that enters the middle term of
Eqs.~(\ref{eq:stage-time})--(\ref{eq:stage-speedup}), coinciding with
$r_n$ only when the buffer does not gate it. The user's share is
$B = 30$~TiB, well below the dataset.

The slowest stage is the $20$~Gbps destination write, so the staged
pipeline sustains $\min(r_s, r_{\text{link}}, r_d) = 20$~Gbps:
\begin{equation*}
  T_{\text{staged}} \approx \frac{D}{20~\text{Gbps}} \approx 12.2~\text{h},
  \qquad
  T_{\text{coupled}} = D\!\left(\tfrac{1}{25} + \tfrac{1}{70}
    + \tfrac{1}{20}\right) \approx 25.5~\text{h},
\end{equation*}
a speedup of
\begin{equation*}
  S = 20\!\left(\tfrac{1}{25} + \tfrac{1}{70} + \tfrac{1}{20}\right)
  \approx 2.1,
\end{equation*}
short of the threefold ceiling because the two production stages sit well
below the crossing. A coupled tool cannot recover this difference: the
standard utilities serialize read, transfer, and
write~\cite{Fang2021Rsync}, so the $\approx 2.1\times$ is the overlap that
staging alone provides.

Sustaining the full crossing rate across the entire dataset would
instead require, from Eq.~(\ref{eq:stage-capacity}),
\begin{equation*}
  B \ge D\!\left(1 - \frac{20}{70}\right) \approx 71.4~\text{TiB},
\end{equation*}
far beyond the $30$~TiB on hand---a concrete instance of the
mathematically valid but unrealizable sizing noted above. It is also a
non-goal here: the transfer is bounded by the $20$~Gbps write, not by the
crossing, so the buffer need only keep the three stages overlapped---a
sliding working set far smaller than $30$~TiB. Indeed the
source--destination mismatch accumulates in the destination buffer at only
$25 - 20 = 5$~Gbps, which would take about $14.7$~h to fill $30$~TiB,
longer than the whole $12.2$~h transfer, so the buffer never fills. Under
realistic, unbalanced conditions a single user still gains a clean
$\approx 2.1\times$ from overlap, and a modest share suffices; the
bottleneck is production storage, not the network or the buffer.

\subsection{Bulk and Streaming Transfers}\label{sec:bulk-streaming}

\textbf{Streaming and bulk transfers differ fundamentally, and both
reward removing a central coordinator from the data path.} The
data movement workloads supported by zx fall into two primary types:

\begin{itemize}
\item
  \textbf{Bulk Transfer}: the movement of a static, pre-existing
  dataset where the complete data is at rest in the source storage
  before the transfer initiates. This is typical for migrating
  archives, databases, or completed experimental runs.
\item
  \textbf{Streaming Transfer}: the movement of a dynamic, actively
  growing dataset where data is transferred concurrently as it is
  being generated and written to the source storage. This type is
  essential for real-time workloads, such as data acquisition from
  scientific instruments like LCLS-II \cite{Thayer2019LCLS} and
  large-scale bioscience imaging operations (e.g., Cryo-EM, virtual
  staining).  The high rates of such sources demand concurrent,
  line-rate data movement for efficient production workflows.
\end{itemize}

This distinction is critical because streaming transfers demand both
high storage throughput and low storage latency, stressing storage
subsystems, memory hierarchies, and scheduling mechanisms in ways that
differ fundamentally from bulk transfer. As a result, streaming
workloads require architectures designed for sustained concurrency and
flow.

Traditional High-Performance Computing (HPC) environments commonly
integrate data movement into centralized batch-scheduled workflows, in
which transfers are initiated, throttled, and terminated by external
resource managers. While effective for coarse-grained compute
allocation, this model introduces non-deterministic start times,
queuing delay, and run-to-run jitter into the data path. These
effects are not incidental: by treating data movement as a schedulable
job rather than a continuously executing system service, the resource
manager disrupts the temporal continuity required for sustained,
near-line-rate streaming and bulk transfers of long duration.

The architecture presented in this work eliminates this source of
variability. Data movement is executed as a peer-to-peer service,
coordinated implicitly through asynchronous buffer state rather than
explicit global scheduling. In this model, transfer cadence emerges
from the zx-managed interaction among co-designed hardware resources,
the host operating system, and local burst buffers, rather than from
external queuing policies. As a result, throughput stability is
bounded by physical system limits, enabling predictable performance
across heterogeneous deployment environments---even fully air-gapped.

This design resolves the critical mismatch that occurs when high
network speed meets slow production storage---a problem that
lies beyond the expertise of most operational IT staff. It does so
without requiring them to master the application either.

\subsection{Integrated Appliances vs Software-Centric
Approaches}\label{sec:appliances}

While valuable, the network-centric approach discussed previously does
not reliably achieve target data rates in production. Empirical evidence, including our own
experience, also indicates that providing software alone---even backed
by proven performance in validation trials
\cite{ESnet2017Petabyte,ESnet2018Record,ICM2019Poland,NSCC2019DMC}---can
create significant operational challenges. Hardware selection, system
integration, tuning, and maintenance require deep, multi-disciplinary
expertise (parent page of~\cite{ESnetHardware}). In many cases,
hardware is also significantly overprovisioned, with enterprise
servers deployed for workloads that could be served efficiently by
compact and far less expensive systems. This combination presents a
substantial total cost of ownership that is not affordable to many
organizations. Furthermore, the resulting variability makes
performance unpredictable and hard to reproduce across different
organizations and environments.

In addition, the distinction between integrated appliances and
software-centric approaches manifests not only in peak performance but
also in operational complexity and configuration overhead.

\noindent
\begin{minipage}{\linewidth}
  \centering
  \small
  \begin{tabularx}{\linewidth}{|p{3.2cm}|L|L|}
    \hline
    & \textbf{Zettar Co-Designed Appliance}
    & \textbf{Software-Centric Solution} \\
    \hline
    \textbf{Tuning}
    & Robust performance with minimal tuning; a single configuration
    across a wide range of file sizes
    & Per-dataset optimization; good results require expert
    tuning \\
    \hline
    \textbf{Workload diversity}
    & Consistently high performance across diverse workloads
    & Fragile and labor-intensive in production with heterogeneous
    data \\
    \hline
    \textbf{Performance stability}
    & Consistent performance under realistic operating conditions
    & Good benchmark results only under carefully tuned parameters
    and selected datasets \\
    \hline
    \textbf{Operator burden}
    & Minimal configuration effort
    & Places a significant burden on operators \\
    \hline
  \end{tabularx}
\end{minipage}

\subsubsection{Evidence: The ESnet Tuning Study}\label{sec:esnet-tuning}

This difference is illustrated by an evaluation conducted by
ESnet, reported in 2020~\cite{Kissel2020ESnet}. The authors compared
the tuning requirements of a co-designed data mover with those of a
representative commercial software-centric solution
(\texttt{globus-connect-server}). Section 10 of the report, Lessons
Learned, is particularly instructive:

\begin{quote}
``With a well-tuned infrastructure, it is feasible for a software data
mover to use a single setting for a wide range of file sizes and
various size distributions (i.e., file size histograms).''
\end{quote}

\begin{quote}
``Although tuning takes the adjustments of only a few parameters and
is straightforward, the results from properly and insufficiently
tuned can be very significant (e.g., $\ge 200\%$).''
\end{quote}

Performance stability under realistic operating conditions is
therefore as important as peak throughput. In large
infrastructures---where datasets, workflows, and file-size
distributions vary widely---fragile, tuning-dependent performance
directly translates into operational inefficiency and reduced
productivity.

The report's Fig.~1 (page~9) documents this contrast directly.

Such a single, broadly effective configuration---referred to as global
tuning---does not imply rigidity. A well-architected system can support
hierarchical tuning, in which global settings provide sensible
defaults for most workloads, while per-task parameters---accessible
through an intuitive interface such as a built-in WebUI or REST
APIs---allow advanced users to override defaults for specialized
transfers. These per-task settings take precedence, providing
flexibility without sacrificing ease of use.

\subsubsection{One Principle, Many Form-Factors}\label{sec:form-factors}

This report explains that the end-to-end data movement problem,
despite its variations across workflows, can be addressed through a
\emph{single} co-design engineering principle. This work builds on
extensive prior advances in high-speed networking and host tuning, and
focuses on making such expertise deployable and repeatable across
production environments. We present a unified appliance architecture
that embodies this principle across the entire data movement workflow
spectrum. This co-design principle is applied across a range of
hardware form-factors, supporting data rates from 1~Gbps up to
100~Gbps and beyond:

\noindent
\begin{minipage}{\linewidth}
  \centering
  \small
  \begin{tabularx}{\linewidth}{|p{4.2cm}|L|}
    \hline
    \textbf{Form-Factor} & \textbf{Example deployment} \\
    \hline
    Compact, low-power node
    & Edge mini-appliance (Fig.~\ref{fig:core-bom}) \\
    \hline
    SFF workstation
    & HP (Fig.~\ref{fig:core-bom}) \\
    \hline
    High-speed biomedical instrument
    & Instrument-native movement (Section~\ref{sec:instrument-native}) \\
    \hline
    Specialized Data Processing Unit (DPU)
    & Co-designed data path (Section~\ref{sec:cloud-antidote}) \\
    \hline
    Certified enterprise server
    & Tier-1 vendor, e.g., HPE (Fig.~\ref{fig:core-bom}) \\
    \hline
  \end{tabularx}
\end{minipage} These
appliances encapsulate the necessary hardware/software co-design and
tuning that was previously a manual, expert-driven process.

\begin{figure}[t]
  \centering
  \includegraphics[width=0.98\textwidth]{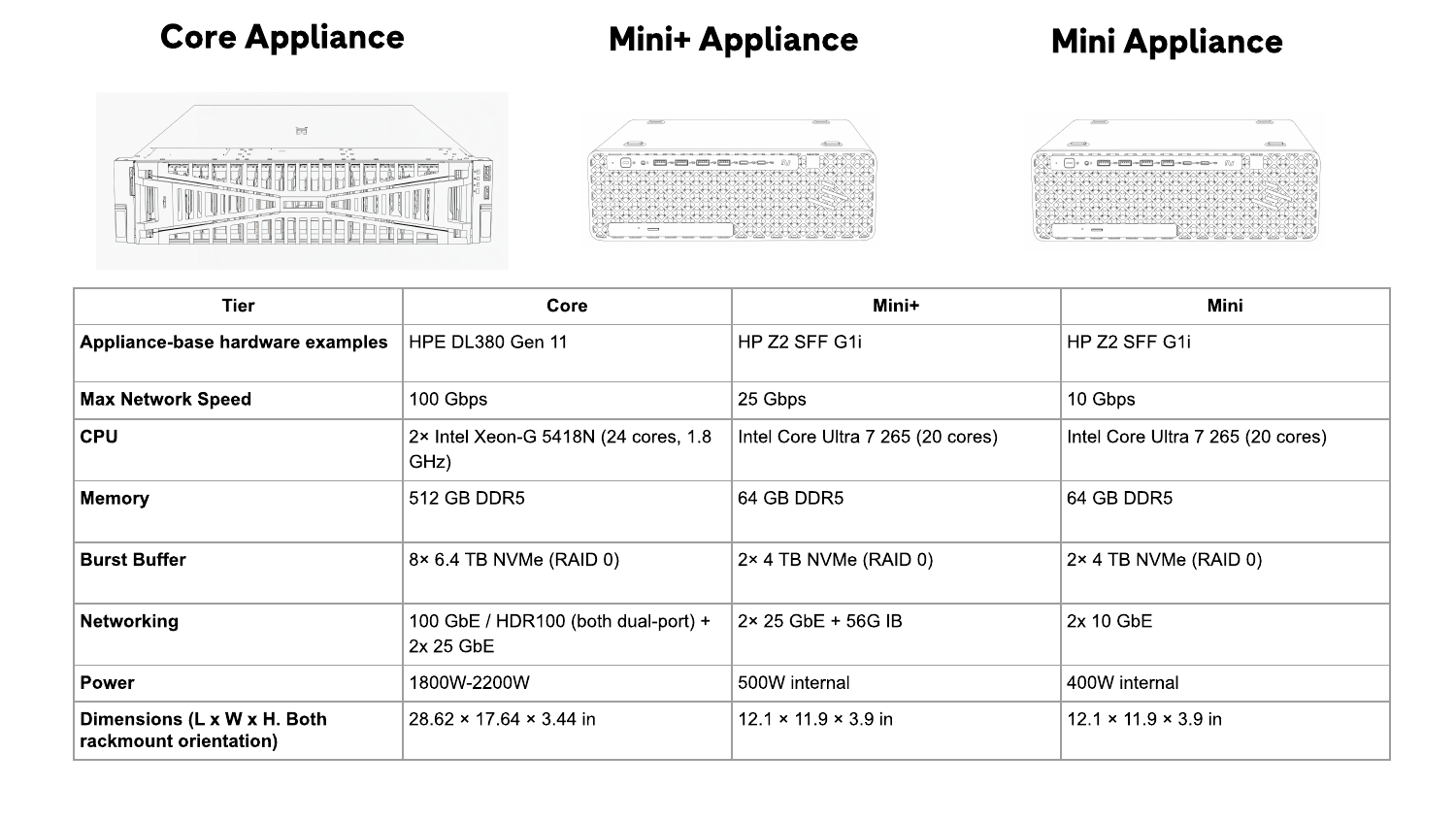}
  \caption{Bill of Materials (BOM) and component details for the Core
    (e.g., HPE DL380 Gen 11), Mini+ (e.g., HP Z2 SFF G1i with~25~Gbps
    network interface card (NIC)), and Mini (e.g., HP Z2 SFF G1i
    with~10~Gbps NIC) appliances, demonstrating the vendor- and
    form-factor-agnostic unified data movement appliance design, which
    achieves consistent performance across implementations.}
  \Description{A comparison table showing Bill of Materials and
    component details for Core (HPE DL380) and Mini (HP Z2 SFF G1i)
    data movement appliances.}
  \label{fig:core-bom}
\end{figure}

The overall approach was developed over the past decade to meet the
ambitious data movement requirements of the premier United States
(U.S.) Department of Energy (DOE) Exascale Computing Preparation
Project, LCLS-II \cite{Thayer2019LCLS}. The core challenges and
performance targets defined by the project directly motivated the
architecture. Furthermore, the ongoing dialogue with LCLS-II and other
tier-1 scientific entities continues to advance and validate the
evolution of this holistic approach to data movement systems. The core
software's performance has been independently validated across many
production trials employing long-distance 100~Gbps links over the past
decade\footnote{Between 2020 and 2025 the work continued without
pause: invited talks at SC21 and SC22 on at-scale AI data
migration~\cite{Fang2021Nvidia,Fang2022Nvidia}, and a 2024 MiTAC and
NVIDIA collaboration in which a single NVIDIA BlueField-3 network
processor running zx sustained up to 500~TB per day across any
distance~\cite{Zettar2024HPCwire}.} (see also
Figs.~\ref{fig:bulk-sweep-plain}-\ref{fig:streaming-sweep} and the
footnote).

The outcomes documented by independent institutions, e.g.,
\cite{ESnet2017Petabyte}, \cite{ESnet2018Record}, \cite{NSCC2019DMC}
are reproducible; nevertheless, our experience shows that operational
reproducibility depends critically on the availability of
pre-validated system configurations rather than ad-hoc assembly at
each site.

\begin{table}[htbp]
  \centering
  \footnotesize
  \begin{tabularx}{\linewidth}{|c|p{3.4cm}|L|c|}
    \hline
    \textbf{Year} & \textbf{Partner / Event}
    & \textbf{Result / Milestone} & \textbf{Ref} \\
    \hline
    2017 & SLAC, AIC, ESnet
    & Petabyte datasets moved at unprecedented speed via ESnet
    & \cite{ESnet2017Petabyte} \\
    \hline
    2018 & SLAC, ESnet (LCLS-II)
    & Record-setting transfer of 1 petabyte of data in
    $\sim$29~h---LCLS-II production trial
    & \cite{ESnet2018Record, Thayer2019LCLS} \\
    \hline
    2019 & Supercomputing Asia Data Mover Challenge
    & Overall Winner among seven international teams
    & \cite{NSCC2019DMC, Lee2023GRP} \\
    \hline
    2019 & ICM (Poland--Singapore)
    & First Poland--Singapore transcontinental data transfer
    production trial
    & \cite{ICM2019Poland} \\
    \hline
    2020 & ESnet DTN evaluation
    & Independent \texttt{zx} evaluation for ESnet DTNs
    & \cite{Kissel2020ESnet} \\
    \hline
    2025 & NCHC / TWAREN (Taiwan)
    & Production data movement over TWAREN (December 2025)
    & \cite{TWARENAbout} \\
    \hline
  \end{tabularx}
  \caption{Independent, production-scale validation of the
    co-design principle across a decade (2017--2025): milestones
    include the 2018 one-petabyte LCLS-II trial and the 2019
    Supercomputing Asia Data Mover Challenge overall win.}
\end{table}

This co-design principle has been implemented in
commercially available appliances, with implementations ranging from
regular server-based units from tier-1 vendors such as Hewlett Packard
Enterprise (HPE) to Small Form Factor (SFF) workstations from HP
(Fig.~\ref{fig:core-bom}). The resulting appliances are simpler,
faster, more energy-efficient, space-saving, and already validated,
while also reducing deployment cost through the use of standard
hardware and software components.

\subsubsection{What Makes a Data Movement Appliance Different}\label{sec:appliance-diff}

It is important to distinguish the data movement appliances discussed
in this report from typical enterprise appliances---a firewall,
switches, or NAS. A data movement appliance is far more than a server
with software installed. \emph{It packages expertise.}

\begin{table}[htbp]
  \centering
  \footnotesize
  \begin{tabularx}{\linewidth}{|p{2.5cm}|X|X|}
    \hline
    & \textbf{Typical Enterprise Appliance} (firewall, switch, NAS)
    & \textbf{Data Movement Appliance} \\
    \hline
    \textbf{Scope}
    & Manages a single stack: security, network, or storage
    & Simultaneously orchestrates storage I/O, compute, high-speed
    networking, and security---while running a parallel, concurrent
    data mover at near-line-rate \\
    \hline
    \textbf{Platform}
    & Often a proprietary operating system; a locked-down black box
    & Commercial Off-The-Shelf (COTS) hardware and standard software
    components; not a black box \\
    \hline
    \textbf{Openness}
    & Closed
    & Open by necessity: controlled, auditable access for
    storage-client installation \\
    \hline
    \textbf{Storage clients}
    & N/A
    & Third-party vendor storage clients run on the appliance
    itself---NAS (NFS or SMB) and parallel file systems (e.g., Lustre, GPFS,
    BeeGFS) \\
    \hline
    \textbf{Software install}
    & Pre-loaded
    & Post-deployment \textbf{storage} client installation (may require
    site-specific licensing); open but secure, limited to restricted
    admin activities \\
    \hline
  \end{tabularx}
\end{table}

The co-design principle handles such differences by enabling the use
of regular COTS hardware and standard software components, yielding a
solution that is both cost-effective and operationally more
manageable---even though the proprietary expertise required to achieve
this integration resides with the appliance designers. This stands in
contrast to two common but misguided assumptions: that one can simply
install software on arbitrary hardware, and expect good results; or
that enterprise IT teams can treat these appliances as just another
compute server to be assembled and tuned locally. Both approaches
overlook the deterministic, cross-layer optimization that
distinguishes a true data movement appliance from a general-purpose
system running data movement software.

Reliable, efficient data movement is achieved most effectively through
complete systems engineering \cite{Kissel2020Setup,
  Kissel2020Env}. The underlying principle is akin to a
high-performance electric vehicle (EV): engineered for speed and
efficiency, yet perfectly capable of simple, reliable operation
for everyday tasks. Just as an EV is not merely a carriage with an
electric motor swapped in, a purposely-built data movement appliance
is not simply a server with powerful CPUs, NVMe SSDs, and
high-bandwidth NICs; the hardware Bill of Materials (BOM) is
meticulously formed to maximize operational and cost efficiency. The
components are orchestrated via software to eliminate the
non-deterministic bottlenecks inherent in general-purpose computing.
In passing, we note that attempts to introduce legacy, non-optimized
software data movers on such a deterministic platform result in
performance regressions, negating the advantages of the co-design
principle.

\subsubsection{From Data Center to Edge}\label{sec:edge}

Despite the recognized challenges of end-to-end data movement, the
prevailing industry and research approach continues to favor complex
tuning methodologies and reliance on prohibitively expensive
hardware. Such factors are particularly acute at the network
edge---the \enquote{headwaters} of the \enquote{Drainage Basin
  Pattern} (Fig.~\ref{fig:dbp}), where resource-constrained
environments such as hospitals, clinics, and remote laboratories
typically utilize 1--10~Gbps links. Deploying reliable,
high-efficiency data transfer in these settings becomes
cost-prohibitive when conventional solutions require specialized IT
staff and expensive hardware.

The architecture proposed in this report directly addresses these
barriers by validating a different principle: that architectural
co-design can inherently reduce both operational complexity and
capital expenditure across the entire data spectrum. This principle is
practically realized in a mini-appliance designed for resource
constrained edge sites. The hardware costs approximately \$5,500
(Fig.~\ref{fig:core-bom} Mini Appliance), including fast BBs. This
demonstrates that high-efficiency 1--10~Gbps data transfer is
achievable simply and affordably, without complex system integration
or the utilization of inappropriate virtualization options
(Section~\ref{sec:cloud}). This cost efficiency and simplicity are a
direct consequence of the holistic approach detailed in the following
sections.

\subsection{Instrument-Native Data Movement}\label{sec:instrument-native}

\textbf{The logical conclusion of the co-design principle is to endow
  high-speed instruments with native data-moving capability.} A modern
instrument---such as a next-generation sequencer, or a Cryo-EM
microscope---already contains substantial computational resources:
CPUs, memory, storage, and fast networking (i.e., $<$~100~Gbps).
These resources are sufficient to run a lightweight data mover
alongside the instrument's primary data acquisition and processing
tasks. After all, many such instruments are actually based on
commodity servers and ODMed by server vendors.

In this model, the instrument acquires the ability to move its own
data. Data flows from the instrument's internal BB directly
to its final destination (a data collection device, core data center,
cloud, or collaborator site) without additional software, or user
intervention. The researcher simply defines the intent (e.g.,
\enquote{send tonight's sequencing run to the AI factory}), and the
instrument executes the transfer autonomously.

This represents the ultimate friction removal: data movement is no 
longer a separate concern. It is a native capability of the 
instrument, as integrated as its data acquisition system. The same 
co-design principle that enabled line-rate performance on dedicated 
appliances applies equally to instrument-native deployment, further 
demonstrating the generality and scalability of the approach.

Early validation of this concept already exists. The KEK Cryo-EM
workflow described in Section~\ref{sec:cloud-evidence} demonstrates
instrument-to-cloud data movement using zx as the transport layer. The
next step is to embed zx directly into the instrument---not as an
add-on, but as a native service.

The implications are significant. When every instrument becomes a 
first-class citizen of the data spectrum:
\begin{itemize}
\item \textbf{Research acceleration:} Data moves as it is created, not 
      hours or days later.
\item \textbf{Operational overhead disappears:} No staging, no 
      manual transfers, no \enquote{sneakernet}.
\item \textbf{Security improves:} The instrument authenticates once 
      and pushes data over encrypted, audited paths.
\item \textbf{Faster science:} Researchers spend time on 
      analysis, not on data logistics.
\end{itemize}

The co-design principle thus extends beyond the data center to the
edge---not just to \$5,500 mini-appliances, but to the instruments
themselves. This is not a future fantasy. The hardware is already
there. So is the software (zx). The remaining work is integration and
standardization.

\section{Common Misconceptions}\label{sec:misconceptions}

Roughly during the past decade, two beliefs emerged. They are more
recent than the long-standing paradigms (Section~\ref{sec:paradigms}) that follow
and no less consequential. We examine them below.

\subsection{AI Agentic: What Works, What Does Not}\label{sec:ai-agentic}

\textbf{AI agents cannot autonomously discover production data
  movement}; the fix is a tool intelligent enough to make the agent's
job trivial.  A recent work should prove this point: an LLM agent
built solely to autotune the 159-plus parameters of Lustre~2.12.5 and
the roughly 1{,}536 of Ceph~\cite{Egersdoerfer2025STELLAR}. The work
lists the challenges storage tuning presents and answers them with a
bigger agent rather than a tool that removes the need to tune. On the
one hand, the rise of AI agentic applications has generated
significant excitement; on the other hand, the canonical
examples---resume parsing, spreadsheet manipulation, API
orchestration---operate in low-stakes, well-structured domains where
failure is cheap and recovery is simple. Moving data at scale and
speed across production networks is the opposite.

Production data movement involves too many unknown or dynamic factors
for an agent to discover or optimize autonomously:

\begin{itemize}
\item \textbf{Storage backend behavior:} Parallel file systems (e.g.,
  Lustre, GPFS, BeeGFS) exhibit non-linear performance profiles
  depending on striping, client counts, and competing workloads. An
  agent cannot \enquote{learn} these without destabilizing production
  transfers.
\item \textbf{Network conditions:} While well-engineered backbones
      (Section~\ref{sec:loss-cca}) have negligible packet loss, edge
      networks do not. Agents that probe for available bandwidth risk
      inducing congestion or misinterpreting transient blips as persistent
      constraints.
\item \textbf{Security policies:} Multi-domain transfers often require
      staged authentication, JWT tokens, or approval workflows. An agent
      that cannot present the right credentials at the right time will
      fail---and automating credential management is itself a hard
      security problem.
\item \textbf{Tacit knowledge:} The most valuable optimizations
  (Section~\ref{sec:appliances}) are not written down. They exist in
  the collective experience of operators who have debugged many
  transfers. Agent training on public data will face an
  expertise-acquisition challenge.
\end{itemize}

Thus, an agent expecting to \enquote{solve} demanding data movement
will likely fail---not because agents are useless, but because the
problem domain is fundamentally underspecified for autonomous
discovery. The solution is not a smarter agent, but a tool with
sufficient built-in intelligence that the agent's job becomes trivial.

Zettar zx has been designed to work well with an AI agent-oriented
approach. It is a unified data mover rich in capability yet simple to
automate. Its REST API and CLI expose every function needed for
large-scale transfers: start, stop, monitor, retry, and integrate with
IAM. An agent does not need to understand storage striping, TCP
tuning, or BB sizing. It simply calls the zx API.

This division of labor is not a limitation. The hard part, such as
co-design, tuning, validation, is encapsulated in the appliance. The
easy part, e.g., orchestration, scheduling, user intent, can be handed
to an agent with confidence that the underlying transfer will complete
at line rate, every time.

Early examples already exist. The KEK Cryo-EM workflow
(Section~\ref{sec:cloud-evidence}) uses zx as a callable
service. Instrument-native movement
(Section~\ref{sec:instrument-native}) defines intent, not
implementation. In both cases, an agent could drive the transfer
regardless.  That is the goal: make data movement utility-like so that
even an agent can do it.

\subsection{Aggregated Traffic Data Rates Are Not the Same as
End-to-End Data Transport Rates}\label{sec:aggregation}

A recurring confusion in the data movement literature is treating
\textbf{network-level traffic aggregation} as equivalent to
\textbf{application-level end-to-end data transport}. These are
different problems, with different constraints, failure modes, and
measures of success, among others.

\begin{table}[htbp]
  \centering
  \small
  \begin{tabularx}{\linewidth}{|p{3.2cm}|L|L|}
    \hline
    & \textbf{Network Traffic Aggregation}
    & \textbf{End-to-End Data Transport} \\
    \hline
    \textbf{What it measures}
    & Aggregate throughput across many flows, many paths,
    or both
    & Throughput of a single logical dataset from source
    storage to destination storage \\
    \hline
    \textbf{Where measured}
    & At network switches, routers, or aggregation points
    & At application layer, storage-to-storage \\
    \hline
    \textbf{What it ignores}
    & Per-flow fairness, storage I/O, encryption overhead,
    small file penalties, metadata costs
    & Nothing---it is the full stack \\
    \hline
    \textbf{Who reports it}
    & Network operators
    & Data movers (should be everyone, but is not) \\
    \hline
  \end{tabularx}
  \caption{Conceptual differences between network aggregation
    and end-to-end transport}
\end{table}

\subsubsection{Historical Precedent (Correct Use)}\label{sec:aggregation-precedent}

When Comcast and Google reported high aggregate rates, they were
explicit about what they were measuring:

\begin{itemize}
    \item \textbf{Comcast (2013)}: ``The Evolution of a Transport
      Network'' (NANOG)~\cite{Comcast2013NANOG}---network traffic
      engineering across a national backbone.
    \item \textbf{Google (2013)}: ``B4: Experience with a
      Globally-Deployed Software Defined WAN'' (ACM
      SIGCOMM)~\cite{GoogleB4}---WAN traffic engineering using SDN.
\end{itemize}

Note that they were \emph{actual production engineering}, not academic
research, carried out more than a decade ago! Both papers are about
moving bits across networks.  They are not about:
\begin{itemize}
    \item Moving files from storage to storage
    \item Preserving data integrity across the transfer
    \item Handling encryption for the data
    \item Handling small files or mixed file sizes
    \item Operating over untrusted paths with encryption
    \item Sustaining line rate for a single logical flow
\end{itemize}

They never claimed to solve these problems. The category error occurs
when others miscite these papers as evidence of data movement
capability. In fact, it is a stretch to treat 100~Gbps, 400~Gbps, or
even a cumulative 1~Tbps aggregated network-layer statistic as some
sort of groundbreaking, novel milestone in 2025 and beyond.

\subsubsection{The Faulty Extrapolation Examples}\label{sec:aggregation-faulty}

\begin{center}
  \small
  \begin{tabularx}{\linewidth}{|P{3.3cm}|P{3.3cm}|L|}
    \hline
    \textbf{Paper} & \textbf{What They Cite}
    & \textbf{Why It Is Wrong} \\
    \hline
    A tier-1 server vendor (2021)~\cite{DellVcinity2021}
    & Reports ``1~PB in just over 23 hours,'' aggregate rates across
    6 server pairs
    & An aggregation of parallel flows, with data deleted
    mid-test---not a sustained, storage-to-storage transfer of a
    single logical dataset\\
    \hline
    A CMS/LHC group (2025)~\cite{XRootD2025}
    & Reports $\sim$260~Gbps as a single-server ceiling at 0~ms
    round-trip time (RTT) toward a 400~Gbps per-site target; data
    served from \texttt{tmpfs}, uniform 4~GB files
    & RAM-backed storage, thus persistent-storage I/O is never exercised;
    a single XRootD instance does not scale past $\sim$16 cores, and
    4 origins (64 cores) reach only $\sim$100~Gbps at production RTTs \\
    \hline
  \end{tabularx}
  \captionof{table}{Examples of faulty extrapolation from network aggregation
    to data transport}
\end{center}

\subsubsection{The Bicyclist Analogy}\label{sec:bicyclist}

A bicyclist who reaches 20 miles per hour and shouts ``Look! I am
so fast!'' is not wrong about 20 mph. But to compare that to a
freight train's cargo capacity is a category error.

\noindent
\begin{minipage}{\linewidth}
  \centering
  \small
  \begin{tabularx}{\linewidth}{|p{2.6cm}|X|}
    \hline
    \textbf{Bicyclist} & 20 mph, one passenger \\
    \hline
    \textbf{Freight train} & 60 mph, thousands of tons of cargo \\
    \hline
    \textbf{The error} & Treating a lightweight, no-cargo
    demonstration as equivalent to heavy freight transport \\
    \hline
  \end{tabularx}
\end{minipage}

Similarly, 260 Gbps on \texttt{tmpfs} with uniform large files is not
comparable to 100 Gbps end-to-end, storage-to-storage, encrypted, over
a wide file size range from 1 KiB to 1 TiB.

\subsubsection{Why This Matters for Reproducibility}\label{sec:aggregation-repro}

If the field accepts aggregate rates as evidence of data mover
performance:

\begin{itemize}
    \item Vendors hide behind parallel flows; no single flow is
          ever optimized
    \item Storage bottlenecks are ignored; the real problem
          remains unsolved
    \item Small files are excluded; most real-world datasets
          contain small files
    \item Encryption is omitted; production transfers often require TLS
\end{itemize}

\emph{Per-instance efficiency composes upward; aggregate throughput
does not decompose downward.} In other words, per-instance efficiency
is measurable and composable, and aggregate-throughput-as-efficiency
is neither. The former introduces quantifiable ROI. The latter
introduces hidden waste.

Stated mathematically, aggregate throughput is the sum of the $N$
per-instance rates $T_i$,
\[
  T_{\mathrm{aggregate}} = \sum_i T_i,
\]
but a given per-instance rate cannot be recovered by dividing
that aggregate evenly:
\[
  T_i \neq \frac{T_{\mathrm{aggregate}}}{N}.
\]

\subsubsection{Zettar's Testing Methodology}\label{sec:testing-methodology}

For transparency and completeness, we publish our testing
methodology. We invite others to do the same.

\paragraph{Four Products, One Unified Mover}\label{sec:four-products}

The Zettar zx software data mover comprises four tightly integrated products:

\noindent
\begin{minipage}{\linewidth}
  \centering
  \small
  \begin{tabularx}{\linewidth}{|p{3cm}|X|}
    \hline
    \textbf{zx-File} & Bulk file and file-to-object transfer
    \cite{ZxFile} \\
    \hline
    \textbf{zx-Object} & S3-compatible object movement
    \cite{ZxObject} \\
    \hline
    \textbf{zx-Single-site} & Scale-out local copy and sync
    \cite{ZxSingleSite} \\
    \hline
    \textbf{zx-Streaming} & Live append streaming
    \cite{ZxStreaming} \\
    \hline
  \end{tabularx}
\end{minipage}

Each product undergoes rigorous, systematic testing.

\paragraph{Testing Dimensions}\label{sec:testing-dimensions}

For zx-File, zx-Object, and zx-Streaming, every test covers:

\noindent
\begin{minipage}{\linewidth}
  \centering
  \small
  \begin{tabularx}{\linewidth}{|c|p{3.2cm}|X|}
    \hline
    \textbf{\#} & \textbf{Dimension} & \textbf{Specification} \\
    \hline
    1 & \textbf{Latency values} & 10\,ms, 50\,ms, and 100\,ms
    (one-way latency; approximately half of RTT as measured by
    \texttt{ping}). \\
    \hline
    2 & \textbf{Hyperscale datasets} & A wide range of file sizes.
    zx-File and zx-Object: 1\,KiB to 1\,TiB; zx-Streaming:
    4\,MiB to 1\,TiB. \\
    \hline
    3 & \textbf{Data Transfer Sweep} & Transferring each hyperscale
    dataset sequentially. Elaborated further in Section~\ref{sec:tcp-cca}. \\
    \hline
    4 & \textbf{Parallel tasks} & Simultaneous transfers with both
    uniform and randomly selected priorities. \\
    \hline
    
  \end{tabularx}
\end{minipage}

\vspace{2ex}

Reproducible performance results are the foundation of engineering
practice.  Both the traditional Chinese martial competition
(\zhname{打擂台})~\cite{BaiduLeitai} and IO500~\cite{IO500About}
exemplify this principle: genuine skill is distinguished from
reputation only through open, agreed demonstration.  A claim that
cannot be reproduced under defined, transparent conditions remains
unproven. We apply the same standard to our own work and welcome
others to do likewise.

\paragraph{An Open Invitation}\label{sec:open-invitation}

Below is the complete testing methodology we use to validate the zx
data mover. We publish it so that any party may independently verify,
challenge, or extend these results. We believe the field advances
fastest when methodology is open and claims are falsifiable.

\begin{itemize}
    \item Complete testing methodology.
    \item Test harness (open source).
    \item Results for a single-instance, end-to-end, storage-to-storage,
          encrypted transfer over a real production 100 Gbps WAN link,
          with file-size sweeps as specified above.
    \item Sustained near-line-rate throughput at 100~Gbps and faster
          (Figs.~\ref{fig:bulk-sweep-plain}--\ref{fig:streaming-sweep}),
          at which a petabyte is a day's work (Table~\ref{tab:tab5}), not
          weeks.
    \item A \emph{single} configuration that holds across most of the
          file-size range and across repeated runs, not a result
          confined to a favorable regime or a hand-tuned dataset.
    \item Evidence that the procedure is operable by a non-specialist:
          a first-attempt result obtained from the documentation alone,
          without custom tuning or an assembly of many tools.
    \item Coverage beyond a single mode: bulk, streaming, and transfer
          forwarding; local (host and cluster) as well as over-distance;
          file and object. (\textit{Object protocols such as S3 do not support
          append-streaming; that is a protocol constraint, not a
          limitation of the mover}.)
\end{itemize}

We meet these criteria in the open. The file-size sweeps are automated
and documented; the reproduction in Fig.~\ref{fig:hpe-reproduction}
came from an independent HPE engineer on his first attempt, using only the
published procedure. A sustained rate is either measured or it is not;
the figures above are the measurement.

\section{Six Common Paradigms}\label{sec:paradigms}

\subsection{The Latency Killer Paradigm}\label{sec:latency}

A prevalent belief in data-intensive computing is that \emph{network
latency is the ultimate data movement killer.} While this thinking is
common among many experienced IT professionals, our empirical testing
and architectural analysis motivates the re-examination of this
paradigm. The perceived severity of latency is fundamentally governed
by TCP\textquotesingle s Bandwidth-Delay Product (BDP), which dictates
the congestion window required for full bandwidth utilization.

This paradigm can present significant challenges for distributed,
data-intensive engineering and scientific endeavors, an impact
particularly relevant in the biopharma and life sciences sector. It is
the top global data generator, with data volume that RBC projected to
grow at a Compound Annual Growth Rate (CAGR) of 36\% through 2025
\cite{RBC2025Health}. Some data volumes are already
concrete~\cite{IDCDataAge2025, Stephens2015Genomical,
  Kiryati2021Imaging}. The challenges of moving data at this scale are
a primary concern for industry leaders, including organizations such
as tier-1 biopharma businesses and enterprise computing providers like
HPE. Inefficient data movement negatively impacts drug discovery,
precision medicine development, and research timelines. As we
demonstrate, these challenges can be significantly mitigated through
architectural and software approaches that reduce sensitivity to
latency.

\begin{figure}[hbp]
  \centering
  \includegraphics[width=0.98\textwidth]{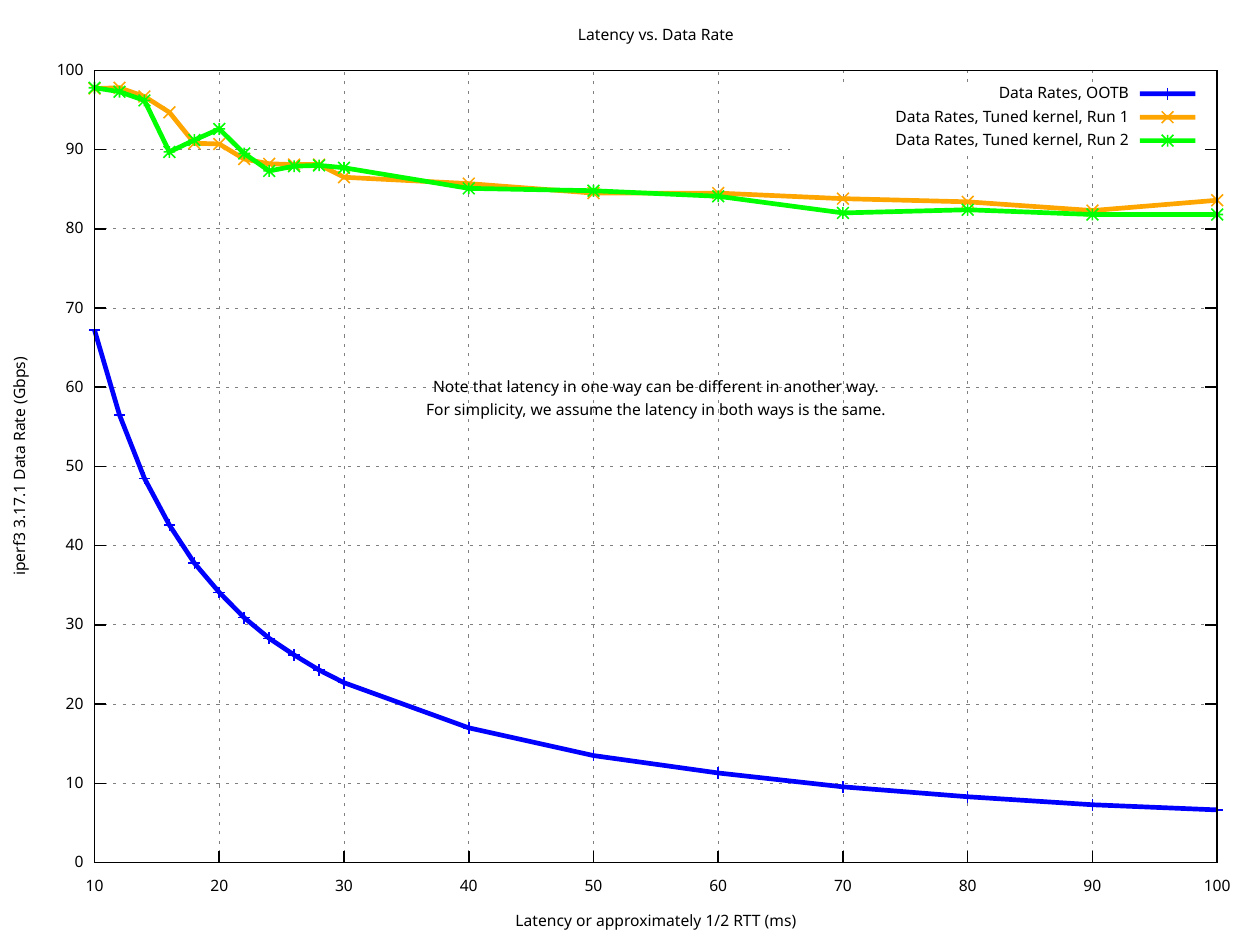}
  \caption{\texttt{iperf3} latency sweep results obtained using two
    HPE DL380 Gen 11 server-based appliances
    (Fig.~\ref{fig:swindon-testbed}) on a latency simulation-capable
    testbed (max network speed 100~Gbps) established in the former
    Intel Swindon Lab, Swindon, U.K., in 2024. While default kernel
    networking settings (OOTB) show severe performance degradation
    under high latency, proper kernel tuning substantially reduces
    this penalty.}  \Description{A line graph showing iperf3 latency
    sweep results comparing default vs. tuned kernel settings on a
    100~Gbps connection with simulated latencies from 10~ms to
    100~ms.}
  \label{fig:iperf3-latency-sweep}
\end{figure}

For years, established resources such as the U.S. DOE
ESnet\textquotesingle s fasterdata knowledge base
\cite{ESnetFasterdata} have documented methods to mitigate
latency\textquotesingle s impact on data rates. While ESnet provides
valuable guidance as a network service provider, a comprehensive
solution requires consideration of the entire environment, not just
the network---a point we expand in
Section~\ref{sec:dedicated-lines}. The reproducible test results
summarized in Fig.~\ref{fig:iperf3-latency-sweep} and the appliance
design in Fig.~\ref{fig:core-bom} together exemplify this holistic
approach. See Table~\ref{tab:tab2} for Ethernet adapter ring buffer
settings and Table~\ref{tab:tab3} for tuned Linux kernel parameters.

The test utilized \texttt{iperf3}, a benchmark tool maintained by the
U.S. DOE ESnet \cite{iperf3Github}. Beginning with version 3.16 (released
November 30, 2023 \cite{iperf3Rel}), the tool was re-architected to be
genuinely multithreaded, making it significantly easier to tune for
high-latency, high-speed links (e.g., $\sim$160~Gbps
memory-to-memory over a 200~Gbps ESnet testbed data path) \cite{iperf3FAQ}.

Section~\ref{sec:dedicated-lines} provides descriptions of the testbed
design. Among many uses of the testbed, it enables the testing of
different latency values automatically
(Fig.~\ref{fig:swindon-testbed}). Such a testbed was first constructed
in the former Intel Swindon lab, in Swindon, United Kingdom (U.K.).
Note that RTT, as reported by tools like \texttt{ping}
\cite{LinuxPing}, is approximately 2 $\times$ one-way latency.

For reproducibility, the employed HPE server-based appliance,
including NIC settings, tuned Linux kernel parameters, and simulated
latency values are described in
Tables~\ref{tab:tab2}-\ref{tab:tab4}. More related materials are
publicly available in a GitHub repository \cite{FangPieee}.

\begin{table}[hbp]
\centering
\caption{rx and tx ring buffer values for Intel® Ethernet Network
  Adapter E810-2CQDA2.}
\label{tab:tab2}
\small
\setlength{\tabcolsep}{6pt}  % Optional: normal column spacing
\begin{tabular}{@{}ll@{}}
\hline
\tblheader{Parameter} & \tblheader{Value} \\
\hline
rx\_value & 8160 \\
tx\_value & 8160 \\
\hline
\end{tabular}
\end{table}

\begin{table*}[!t]
  \centering
  \caption{Tuned Linux kernel parameters. This specific set was
    empirically determined to be optimal for the zx data mover's
    concurrency model, whereas other data movers may require more
    extensive tuning. For example, see ESnet fasterdata, Linux
    Tuning~\cite{ESnetTuning}.}
  \label{tab:tab3}
  \small
  \begin{tabularx}{\textwidth}{|>{\hsize=1.0\hsize}X 
                               |>{\hsize=1.0\hsize}X |}
    \hline
    \textbf{Parameter} & \textbf{Value} \\
    \hline
    net.core.rmem\_max & 2147483647 \\
    \hline
    net.core.wmem\_max & 2147483647 \\
    \hline
    net.ipv4.tcp\_rmem & 4096 67108864 1073741824 \\
    \hline
    net.ipv4.tcp\_wmem & 4096 67108864 1073741824 \\
    \hline
    net.ipv4.tcp\_mtu\_probing & 1 \\
    \hline
    net.core.default\_qdisc & fq\_codel \\
    \hline
    net.ipv4.tcp\_congestion\_control & cubic \\
    \hline
    net.core.netdev\_max\_backlog & 8192 \\
    \hline
  \end{tabularx}
\end{table*}

\begin{table*}[!t]
  \centering
  \caption{Some common latency ranges. 10\,ms, 50\,ms,
    and 100\,ms were used for latency simulation.}
  \label{tab:tab4}
  \small
  \begin{tabularx}{\textwidth}{|>{\hsize=1.1\hsize}X 
                               |>{\hsize=0.9\hsize}X |}
    \hline
    \textbf{Distance category} & \textbf{Latency range (ms)} \\
    \hline
    Metropolitan (same city or nearby) & 1 -- 10 \\
    \hline
    Interstate (within the same country) & 10 -- 50 \\
    \hline
    Cross Continent & 50 -- 100 \\
    \hline
  \end{tabularx}
\end{table*}

Resources such as the ESnet fasterdata knowledge base provide
invaluable guidance. However, practitioners sometimes apply published
tuning parameters as static prescriptions rather than adaptable
guidelines. This approach can lead to suboptimal performance when it
doesn\textquotesingle t account for interactions between specific data
mover software, hardware stacks, and application workloads. Effective
high-performance data movement requires understanding and adaptation
beyond copy and paste. Note that the results in Fig.~\ref{fig:iperf3-latency-sweep}
and the modest parameter set in Tables~\ref{tab:tab2}-\ref{tab:tab3}
demonstrate how an architectural approach also can reduce complexity
and potential for misconfiguration.

\subsection{The Presumed Packet Loss and TCP CCA Paradigm}\label{sec:loss-cca}

This section provides a systematic examination of two interconnected
paradigms of end-to-end data movement: the assumption of universal
packet loss and the impact of TCP CCA selection. We present empirical
evidence that challenges the two prior assumptions in high-speed
environments. A real-world case study, drawn from a Wall Street
Journal report, illustrates the practical consequences of these
paradigms. To clarify the scale of these workflows,
Table~\ref{tab:tab5} quantifies the daily data volume achievable at
common network speeds. Finally, we discuss why conflating different
classes of workflows often leads to suboptimal outcomes.

\begin{table*}[!t]
  \centering
  \caption{Daily data volume achievable at three common network
    speeds.}
  \label{tab:tab5}
  \small
  \begin{tabularx}{\textwidth}{|>{\hsize=1.0\hsize}X 
                               |>{\hsize=1.0\hsize}X 
                               |>{\hsize=1.0\hsize}X |}
    \hline
    \textbf{Data rate (Gbps)} & \textbf{Data (TB)/day} & 
    \textbf{Notes} \\
    \hline
    1 & 10 & 5G speed often $\leq$ 0.5~Gbps \\
    \hline
    10 & 100 & High-performance edge transfer \\
    \hline
    100 & 1000 & i.e. 1 Petabyte \\
    \hline
  \end{tabularx}
\end{table*}

While the end-to-end movement of data often appears simple to
application users, the underlying mechanisms are complex. The
\enquote{Drainage Basin Pattern} illustrates this full spectrum of
data movement workflows, revealing considerations that may not be
immediately apparent.

While most users experience only the \enquote{source of the river,}
where consumer-grade networks may indeed have packet loss, the
\enquote{high-speed backbone}---the deep main river channel---is an
entirely different environment. This recognition is based on extensive
experience since 2015 with 100~Gbps and faster connections, gained
through collaboration with the U.S. DOE ESnet and various tier-1
organizations. Consequently, our observations confirm that in the
well-engineered 100~Gbps+ Research and Education (R\&E) networks
discussed here---principally ESnet and Internet2
\cite{Internet2About}---packet loss is effectively negligible during
normal operation. It is also a key characteristic of peer R\&E
networks such as GÉANT \cite{GeantAbout},
AARNet \cite{AARNetAbout}, and SingAREN \cite{SingARENAbout}.

\subsubsection{Packet Loss}\label{sec:packet-loss}

The nature of data transmission at high speeds differs fundamentally
in both engineering and measurement from casual, end-user level data
transfers. For high-speed networks, \enquote{packets} often vary in
size and are transmitted at rates where bit-level impairments dominate
over per-packet events. For example, for nodes connected to 10~Gbps
and faster networks, it is common to employ \enquote{jumbo frame}
Maximum Transmission Unit (MTU) of 9,014 or 9,018 bytes, rather than a
constant 1,500-byte frame. Consequently, high-speed network device
vendors typically specify performance using Bit Error Rates (BER)
\cite{NISTBER} rather than packet loss. While operators historically
reported BERs more explicitly, modern practitioners typically must
extract such information indirectly via monitoring infrastructures
such as the perfSONAR lookup service \cite{perfSONARStats}.

To place the current argument in context, we cite a 2009 ESnet
reference to show the historical lower bounds of key performance
metrics. Subsequently, we present a practical 2018 case study to
demonstrate that packet loss is not the dominant constraint for
high-volume, high-speed data transfer. ESnet\textquotesingle s
Services and Service Level Descriptions (SLD) (Appendix A in
\cite{ESnetSLA}) specifies guaranteed network performance via its
Loss Thresholds Table. This table typically lists a Frame
Loss Rate (FLR) range of 10\textsuperscript{-7} to
10\textsuperscript{-10} and includes a formula for converting the FLR
to BER. Furthermore, an accompanying ESnet Fasterdata Knowledge Base
(KB) article on Packet Loss provides an approximate translation of BER
to end-to-end packet loss, citing a typical scenario of 1 packet out
of 22,000 packets, or 0.0046\%. Converting this packet loss ratio to a
BER illustrates the exceptional quality of these networks:

\begin{itemize}
\item Packet Loss Rate = 0.0046\% ($4.6 \times 10^{-5}$)
\item Assuming a standard 1500-byte frame: BER $\approx (4.6 \times
  10^{-5}) / (1500~\text{bytes/frame} \times 8~\text{bits/byte})
  \approx 4 \times 10^{-9}$
\end{itemize}

This result---a BER of $4 \times 10^{-9}$, meaning roughly 4 erroneous
bits in every billion---confirms that packet loss is negligible in
well-engineered backbones. Note that using jumbo frames would further
reduce the implied BER.

The operational lower bound on error rates is corroborated by
practical, large-scale experience. In September 2018, using the setup
shown in Fig.~\ref{fig:slac-testbed} of the Appendix 6.1 of
\cite{Kissel2020ESnet}, a production trial run was carried out for 1
PB transfer in 29~hours over a 5,000-mile 100~Gbps WAN loop. The setup
was designed and implemented at SLAC National Accelerator Laboratory,
using equipment that, at the time of deployment, was three years past
its initial deployment. The trial was reported in
\cite{ESnet2018Record}. Of the hardware-limited 80~Gbps bandwidth,
with full encryption and checksumming, an average utilization of
76.63~Gbps was achieved.  This result directly contradicts the notion
that packet loss is the dominant limiter in modern, well-engineered
wide-area backbones.

\subsubsection{TCP CCA}\label{sec:tcp-cca}

The empirical evidence presented previously established that packet
loss is negligible on well-engineered network backbones. Next,
we examine the related paradigm that the choice of TCP CCA is critical
for performance. The prevalence of this view is evident from the
abstract of Google Networking Research\textquotesingle s publication
\enquote{BBR: Congestion-Based Congestion Control}
\cite{Cardwell2016BBR}, which introduced the BBR algorithm (Bottleneck
Bandwidth and Round-trip Propagation Time). The abstract states:
\enquote{Physics and climate researchers need to exchange petabytes of
  data with global collaborators but find their carefully engineered
  multi-Gbps infrastructure often delivers at only \emph{a few Mbps}
  over intercontinental distances.}---a claim that implicitly
attributes poor performance to TCP behavior rather than end-system or
architectural factors.

The Google paper was published in 2016; in 2017, ESnet publicly
reported our first production trial~\cite{ESnet2017Petabyte}, 1 PB in
1.4 days. Then, in 2018, ESnet again reported our 1 PB transfer, at a
higher mean data rate (76.63~Gbps)~\cite{ESnet2018Record}, roughly a
15\% improvement over the 2017 result. Both cases had an 80 Gbps
hardware bandwidth cap~(Fig.~\ref{fig:slac-testbed}). In the 2nd trial,
96\% bandwidth utilization was achieved---practically the capped line
rate. The results were obtained not with a novel CCA, but with the
default CUBIC in the standard CentOS 7.5 distribution, which did not
yet include BBR. This demonstrates that the fundamental bottleneck was
not the CCA, but the lack of a holistically co-designed system.

Before proceeding, we establish the context regarding high-speed TCP
congestion control algorithms (CCAs) by citing relevant Internet
Engineering Task Force (IETF) Request for Comments (RFCs) and
referring to prior work \cite{ESnet2018Record}. Starting in 2018, the
suitability of CUBIC for Fast Long-Distance Networks was already
documented by RFC8312 \cite{RFC8312CUBIC}. We note that RFC9438
\cite{RFC9438CUBIC}, titled \enquote{CUBIC for Fast and Long-Distance
  Networks}, has since obsoleted RFC8312. More recently, RHEL 9.6 and
its free rebuilds now include BBR among the available CCAs, as
demonstrated by the simple command output presented below:

\begin{lstlisting}[
    basicstyle=\ttfamily\footnotesize,
    breaklines=true,
    breakatwhitespace=true,  % Break at spaces, not mid-word
    frame=none,
    tabsize=4
]
$ sysctl net.ipv4.tcp_available_congestion_control
net.ipv4.tcp_available_congestion_control = reno cubic bbr
\end{lstlisting}

\begin{figure*}[t]
  \centering
  \includegraphics[width=0.98\textwidth]{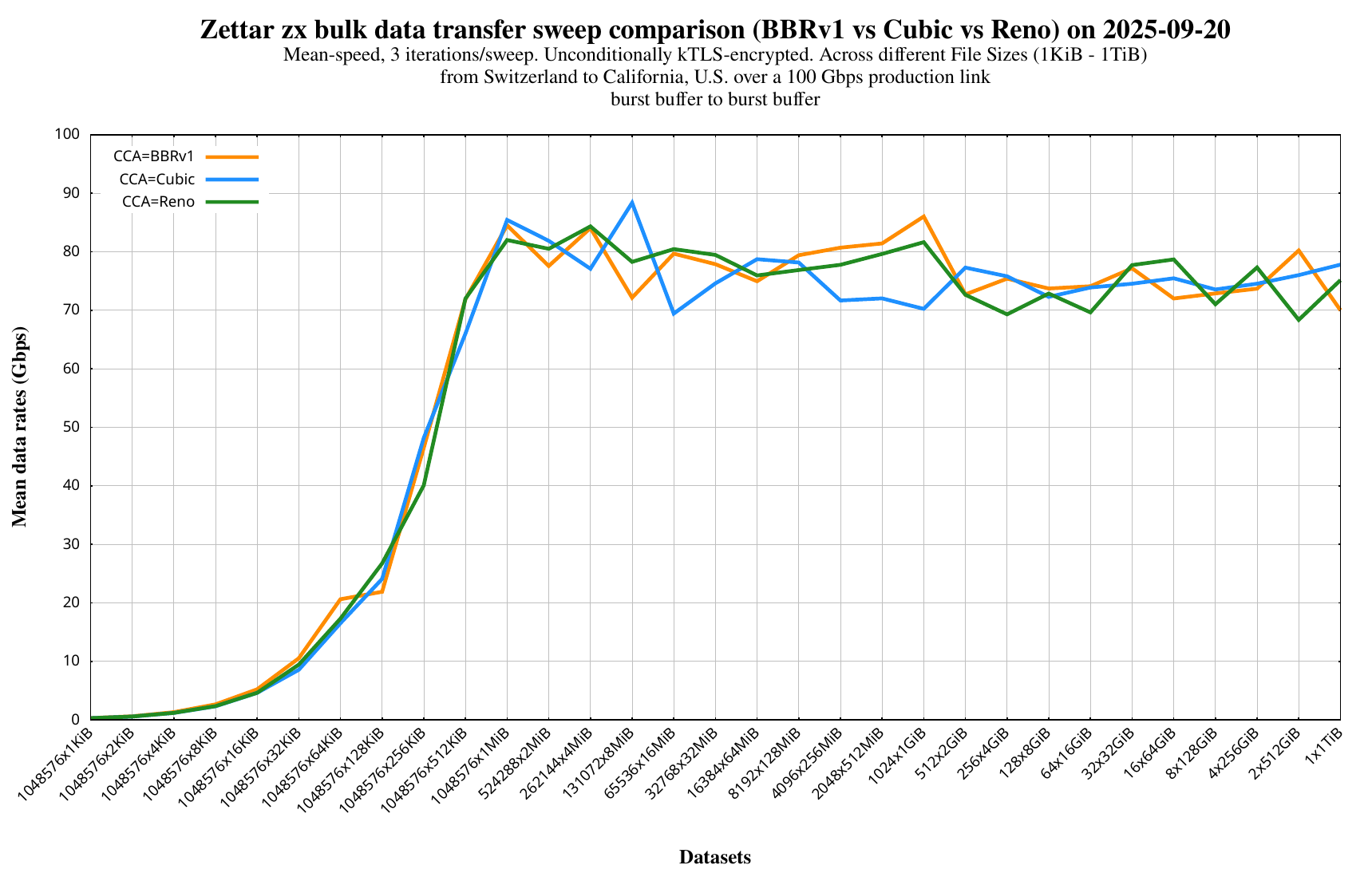}
  \caption{A bulk transfer sweep leveraging kTLS (kernel TLS) offload
    in RHEL~9.6~\cite{RedHatKTLS} to evaluate the three CCAs. BBRv1
    did not demonstrate a clear performance benefit over CUBIC or
    Reno. Incidentally, kTLS did not yield any transfer rate
    improvement in this case.}
  \label{fig:bulk-sweep-ktls}
  \Description{Performance graph showing bulk transfer rates vs. file
    sizes for BBR, CUBIC, and Reno congestion control algorithms with
    kTLS offload.}
\end{figure*}

\begin{figure*}[t!]
    \centering
    \includegraphics[width=0.98\textwidth]{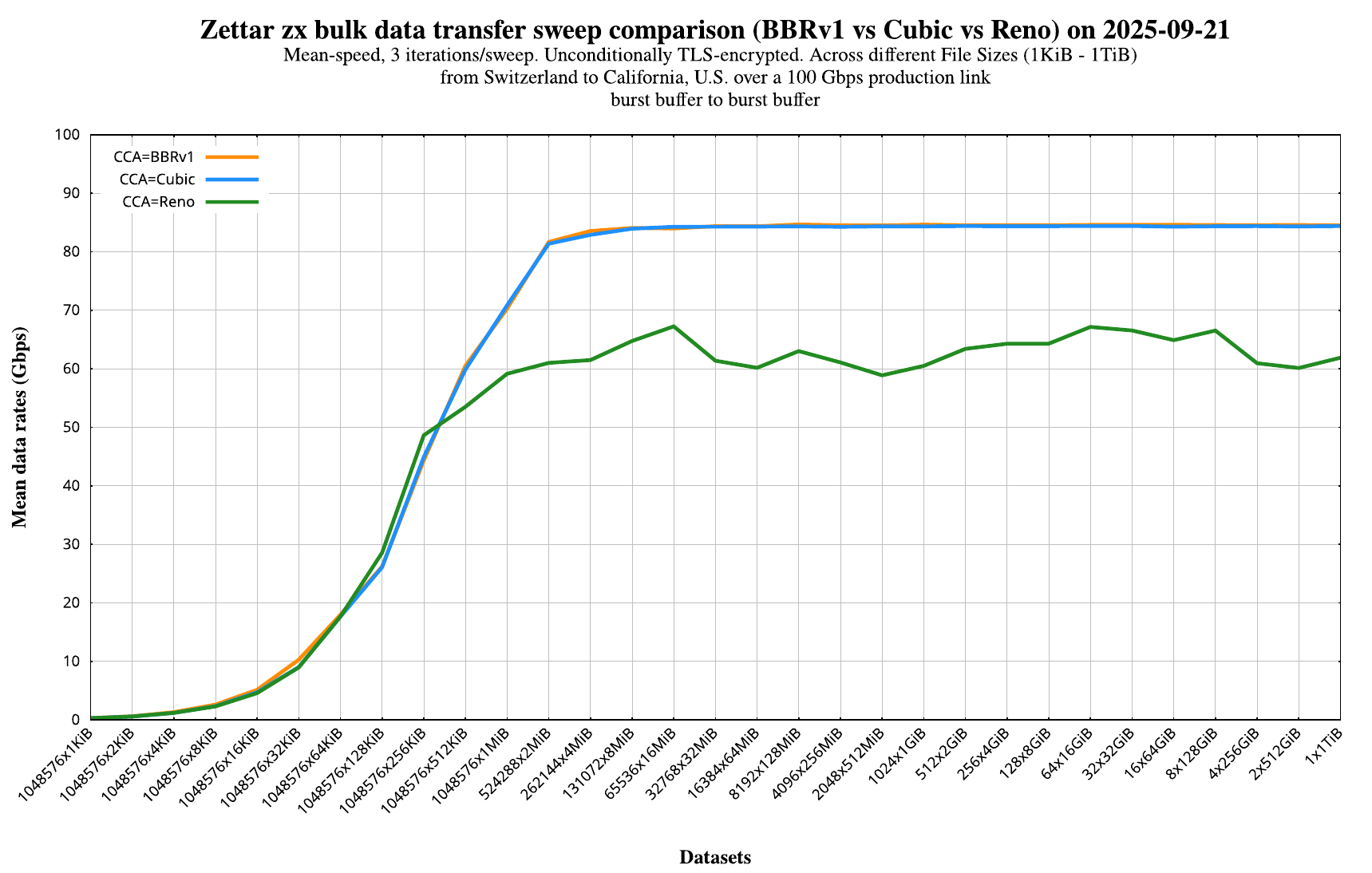}
    \caption{Subsequent bulk transfer sweeps were performed without
      kTLS due to the prior degradation. Using the RHEL 9.6
      CCAs, BBRv1 and CUBIC exhibited identical throughput across 1
      KiB--1 TiB. Reno showed degradation consistent with older
      congestion control paradigms.}
    \Description{Performance graph showing bulk transfer rates
      vs. file sizes for BBR, CUBIC, and Reno congestion control
      algorithms without kTLS.}
    \label{fig:bulk-sweep-plain}
\end{figure*}

\begin{figure*}[htbp]
    \centering
    \includegraphics[width=0.98\textwidth]{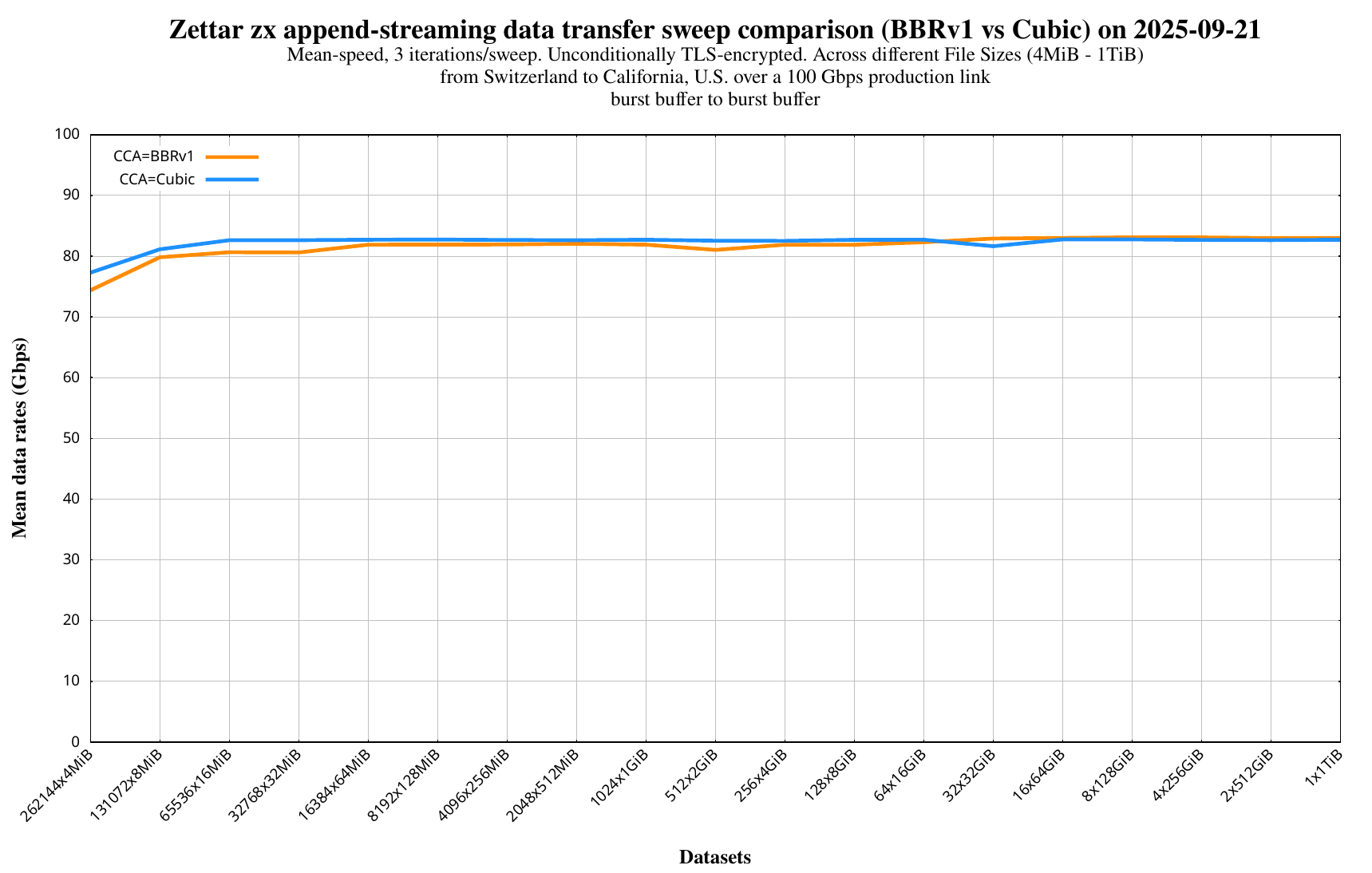}
    \caption{As a unified software data mover \cite{ZettarProducts},
      zx has built-in streaming capability, motivated by
      \cite{Thayer2019LCLS}. BBRv1 shows no advantage over CUBIC, even
      for streaming transfers. kTLS was not used.}
    \Description{Performance graph showing streaming transfer rates
      vs. file sizes for BBR, CUBIC, and Reno congestion control
      algorithms.}
    \label{fig:streaming-sweep}
\end{figure*}

It is noteworthy that while Google researchers introduced the
model-based BBR CCA in 2016---the same year CUBIC was already the
default in widely used Linux operating system distributions such as
CentOS 7---BBR has since progressed to version 3
\cite{GoogleBBR}. However, the BBR implementation that ships with RHEL
9.6 remains version 1, which is notably more aggressive than later
versions. This distinction is relevant because BBRv1's aggressiveness
should, if anything, favor higher throughput in uncongested
environments.  Here, we compared the performance of BBR, CUBIC, and
Reno (a classical TCP CCA algorithm that emerged in the 1990s) and
provided the results in
Figs.~\ref{fig:bulk-sweep-ktls}-\ref{fig:streaming-sweep}. The
following definitions and methodology are established before
presenting the three figures.

\begin{itemize}
\item
  \textbf{Hyperscale Data Set.} A synthetic
  test data set containing
  files of a uniform size, where the total number of files is
  $2^{20}$~(1,048,576) or the aggregate size is 1 TiB, or
  both.
\item
  \textbf{Data Transfer Sweep.}

  \begin{itemize}
  \item
    \textbf{Bulk Transfer Sweep:} File sizes range from 1~KiB to 1~TiB, with
    sizes incremented by powers of two, resulting in 31 distinct
    datasets.
  \item
    \textbf{Streaming Transfer Sweep:} File size ranges from 4~MiB to 1~TiB;
    sizes are incremented by powers of two, so there are 19 datasets.
  \end{itemize}
\end{itemize}

Each sweep iteration processed datasets sequentially from the smallest
to the largest file size. This process was repeated for multiple
iterations to gather statistically stable results. The statistical
analysis for each complete sweep (bulk and streaming) included the
calculation of the mean, median, and standard deviation. The mean
values are plotted automatically using a custom gnuplot command file
\cite{Gnuplot5}. See the GitHub repo \cite{FangPieee} for tables
containing raw mean, median, and standard deviation values.

Based on the demonstrated functional equivalence of high-speed
performance, we conclude that the choice among competing CCAs is not a
primary concern for high-volume, high-throughput data transfer. CUBIC
is empirically validated, using the appliances in Fig.~\ref{fig:core-bom},
as a safe and effective default
(Figs.~\ref{fig:bulk-sweep-plain}-\ref{fig:streaming-sweep}), over
high-speed networks.

It is worth noting that Reno, characterized by slower window size
adjustments than both CUBIC and BBRv1, is highly susceptible to the
non-offloaded userspace overheads of TLS. The implementation of kTLS
effectively mitigates this susceptibility, allowing Reno to achieve
throughput parity with the other CCAs by removing the userspace
processing bottleneck (Fig.~\ref{fig:bulk-sweep-ktls}).

\subsubsection{Practical Implications and Real-World Consequences}\label{sec:cca-implications}

The engineering realities discussed thus far have significant, though
often invisible, operational and economic impacts.

With AI neoclouds, a baseline 1024-GPU cluster provisions 2~PB
of usable networked storage, which must be loaded with training data
and checkpoints before the cluster can
train~\cite{SemiAnalysis2024Neocloud}. At that scale, a petabyte is
the routine floor, not the exception.

The tangible consequences of suboptimal data transfer are further
illustrated by the extreme measures reported by the Wall
Street Journal (WSJ)~\cite{WSJ2025AI}. A June 13, 2025 article
titled \enquote{Chinese AI Companies Dodge U.S. Chip
  Curbs by Flying Suitcases of Hard Drives Abroad} detailed how four
engineers flew from Beijing, China to Kuala Lumpur, Malaysia,
transporting four suitcases of hard drives containing data for
training AI models. This story highlights the severe real-world
constraints imposed by inefficient electronic data transfer.

A related article from Tom\textquotesingle s Hardware
\cite{Toms2025AI} clarified that this physical transport was a
\enquote{meticulously planned operation and took \emph{several months
  of preparation.}} In sharp contrast, the performance results
illustrated in
Figs.~\ref{fig:bulk-sweep-ktls}-\ref{fig:streaming-sweep}, combined
with the entries in Table~\ref{tab:tab5}, show that transferring
PB-scale data electronically, using Zettar co-designed data movement
appliances and a 100~Gbps link, would require \emph{at most one week
of sustained transfer}---storage-to-storage. The number assumes such a
path. Cloud-mediated transfers, subject to the architectural penalties
of Section~\ref{sec:cloud}, cannot approach it.

This dichotomy recalls the hierarchy of solutions presented in
Figs.~\ref{fig:core-bom}-\ref{fig:iperf3-latency-sweep}, which a
metaphor helps illustrate: \emph{providing water}. A balcony plant
takes a watering can; a garden, a hose. But supplying a city is
\emph{not a bigger pipe}---it is a water transport \emph{system},
e.g., the Hetch Hetchy Aqueduct \cite{BAWSCAHetch}: reservoirs,
tunnels, pump stations, and pressure regulation, co-designed to act as
one.  Replace any single piece with an off-the-shelf part, and the
city taps run dry.

Data movement follows the same hierarchy.  At $\sim$1~Gbps (the can)
or $\sim$10~Gbps (the hose), a good connection suffices. Past
100~Gbps, a high-speed link may not. An analogy: a city may have a big
transport pipe from its reservoir. But the big pipe does not ensure
good water volume at each household, because the pipe is no longer
the constraint. On the high floor of a building, if the water pressure
is low, only a trickle is delivered; a 100~Gbps link likewise delivers
a fraction of its rate when the endpoints are not co-designed to feed
it. At this scale, ad-hoc solutions do not merely underperform---they
fail.

\subsection{The Dedicated-Lines Necessity Paradigm}\label{sec:dedicated-lines}

A common paradigm holds that validating high-speed data transfer
requires a dedicated, high-bandwidth WAN link,
e.g.,~\cite{DellVcinity2021}. This belief creates a significant barrier,
as such links are costly and complex to provision, requiring
specialists to operate well. We confronted this directly in 2023 while
developing data movement appliances with Intel Corp. Neither
organization had a 100~Gbps WAN, raising a critical question: how
could we rigorously validate performance in the absence of a
production-grade link?

From 2015-2019, the zx R\&D was carried out on a testbed at SLAC
National Accelerator Laboratory, using a 5,000-mile, 100~Gbps loop
provisioned by ESnet. After 2019, collaboration with ESnet continued
\cite{Kissel2020ESnet}. Nevertheless, accessing their 100G Software
Defined Network (SDN) Testbed required a formal application process
\cite{ESnetTestbeds}, hindering agile development. This need for a
fully controlled, on-demand test environment led us to evaluate
commercial WAN simulators, which we found to be prohibitively
expensive and functionally limited.

\begin{figure*}[!t]
  \centering
  \includegraphics[width=\linewidth]{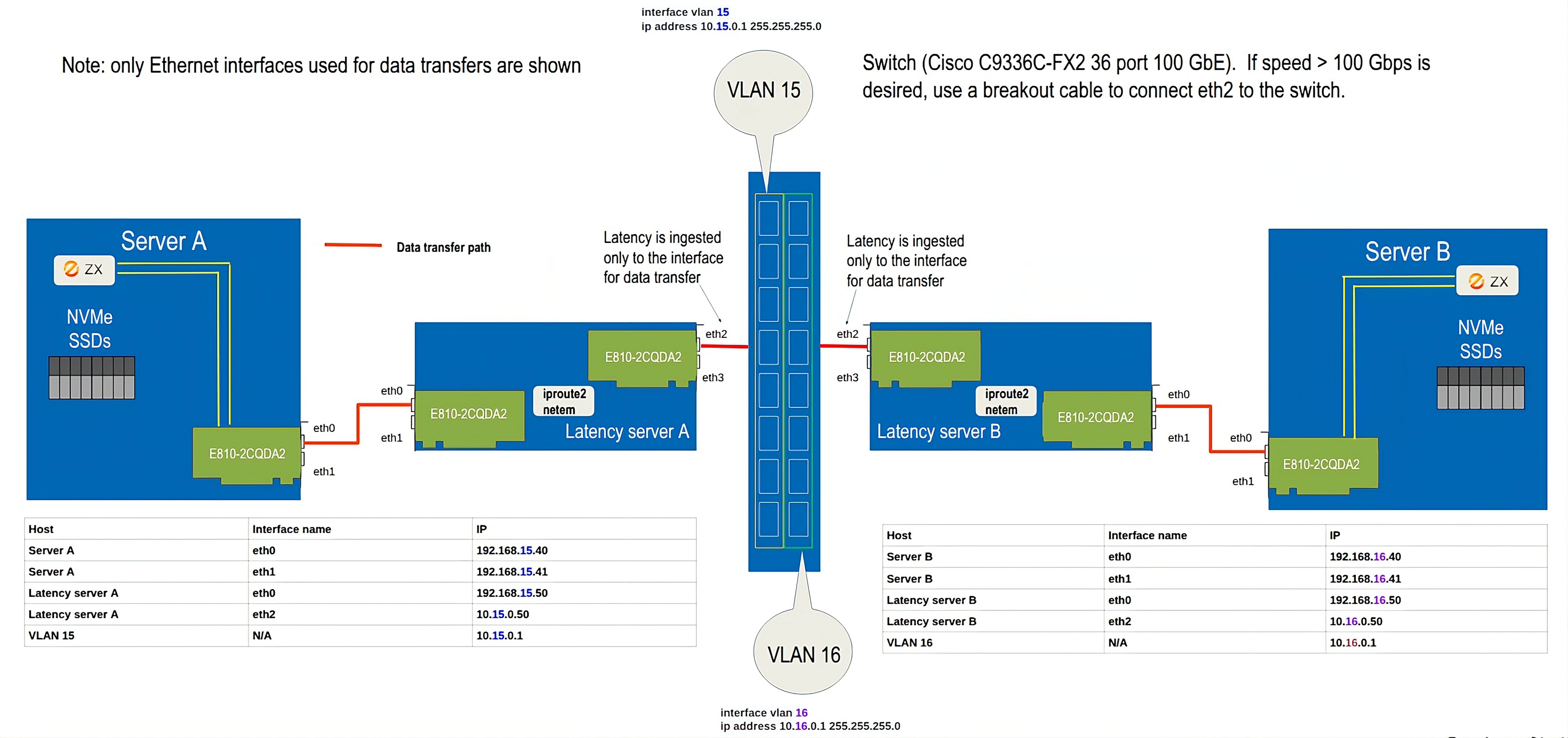}
  \caption{Intel Corp. arranged its former Swindon Lab in Swindon,
    U.K., to collaborate with Zettar using an enhanced version of the
    approach created by Dr.~Ezra Kissel. The essential components are
    labeled explicitly.}
  \label{fig:swindon-testbed}
  \Description{The topology of the testbed setup at Intel Swindon Lab
    showing two HPE servers connected via latency simulation servers.}
\end{figure*}

A turning point was the discovery of a 2012 technical report,
\enquote{Validating Linux Network Emulation} \cite{Kissel2012Netem},
which outlined a method using Linux\textquotesingle s built-in traffic
control (\texttt{tc} \cite{LinuxManTC}) and network emulation
(\texttt{tc-netem} \cite{LinuxManNetem}) tools. Recognizing the
potential of this software-based approach to create a high-fidelity,
100~Gbps~capable testbed at a fraction of the cost, we implemented an
enhanced version. This version extended functionality and automation
beyond the original method. The resulting testbed
(Fig.~\ref{fig:swindon-testbed})---built at the former Intel Swindon
Lab in collaboration with Intel and HPE---consists of two HPE DL380
Gen 11 servers~(Fig.~\ref{fig:core-bom}) configured as the appliances
under test connected via a switch to two Intel \enquote{latency
  servers}. These servers use \texttt{tc} and \texttt{tc-netem} to
impose precise delay, accurately simulating transcontinental links in
an automated, remotely manageable setup.

\begin{figure*}[htbp]
    \centering
    \includegraphics[width=0.98\textwidth]{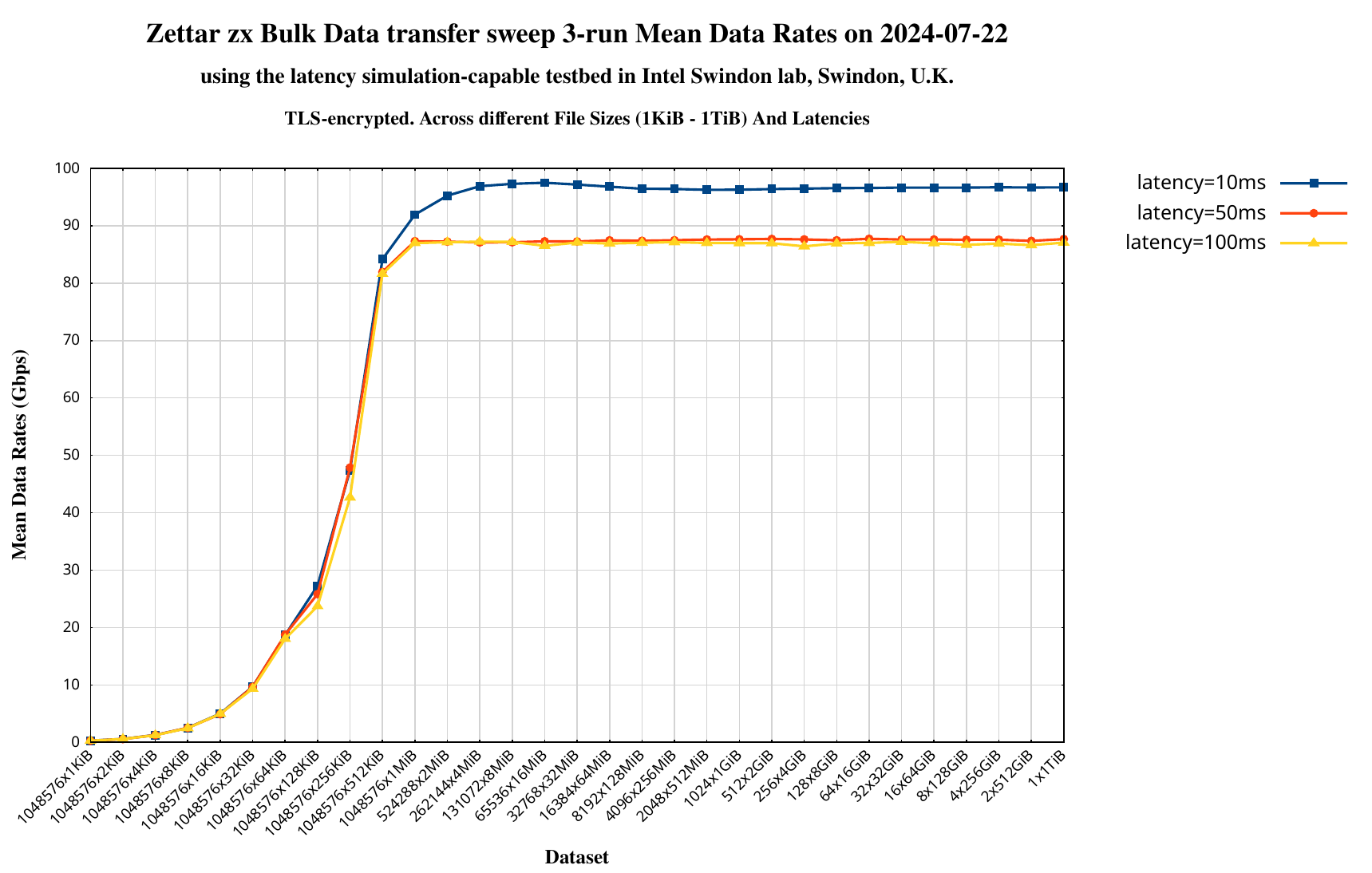}
    \caption{Bulk transfer sweeps vs three simulated latency values:
      10~ms, 50~ms, and 100~ms, corresponding respectively to Los
      Angeles to San Francisco, California, U.S.; Singapore to Alaska,
      U.S.; and Singapore to Atlanta, Georgia, U.S.}
    \Description{Graph showing bulk transfer performance
      vs. simulated latency values (10~ms, 50~ms, 100~ms) on the
      100~Gbps-capable testbed.}
    \label{fig:bulk-vs-latency}
\end{figure*}

The results, shown in Fig.~\ref{fig:bulk-vs-latency} (bulk transfers)
and Fig.~\ref{fig:streaming-vs-latency} (streaming transfers)
demonstrate the testbed\textquotesingle s capability to sustain high
data rates under simulated latencies of 10~ms, 50~ms, and 100~ms. The
critical validation comes from comparing the testbed\textquotesingle s
performance across this simulated latency range with the results from
a production 100~Gbps link, which operates at an approximate latency
of 74~ms
(Figs.~\ref{fig:bulk-sweep-plain}-\ref{fig:streaming-sweep}). The
achieved data rates and performance profiles are strikingly similar,
demonstrating the testbed\textquotesingle s predictive reliability
across the critical WAN latency range.

The true significance of this Linux-based emulation approach: it
transforms the high-speed development environment from a logistical
burden into a strategic, compact, cost-effective, and versatile
engineering platform. It is capable of reliably predicting performance
profiles, making it a useful engineering tool for development and
optimization across numerous critical functions:

\begin{figure*}[htbp]
    \centering
    \includegraphics[width=0.98\textwidth]{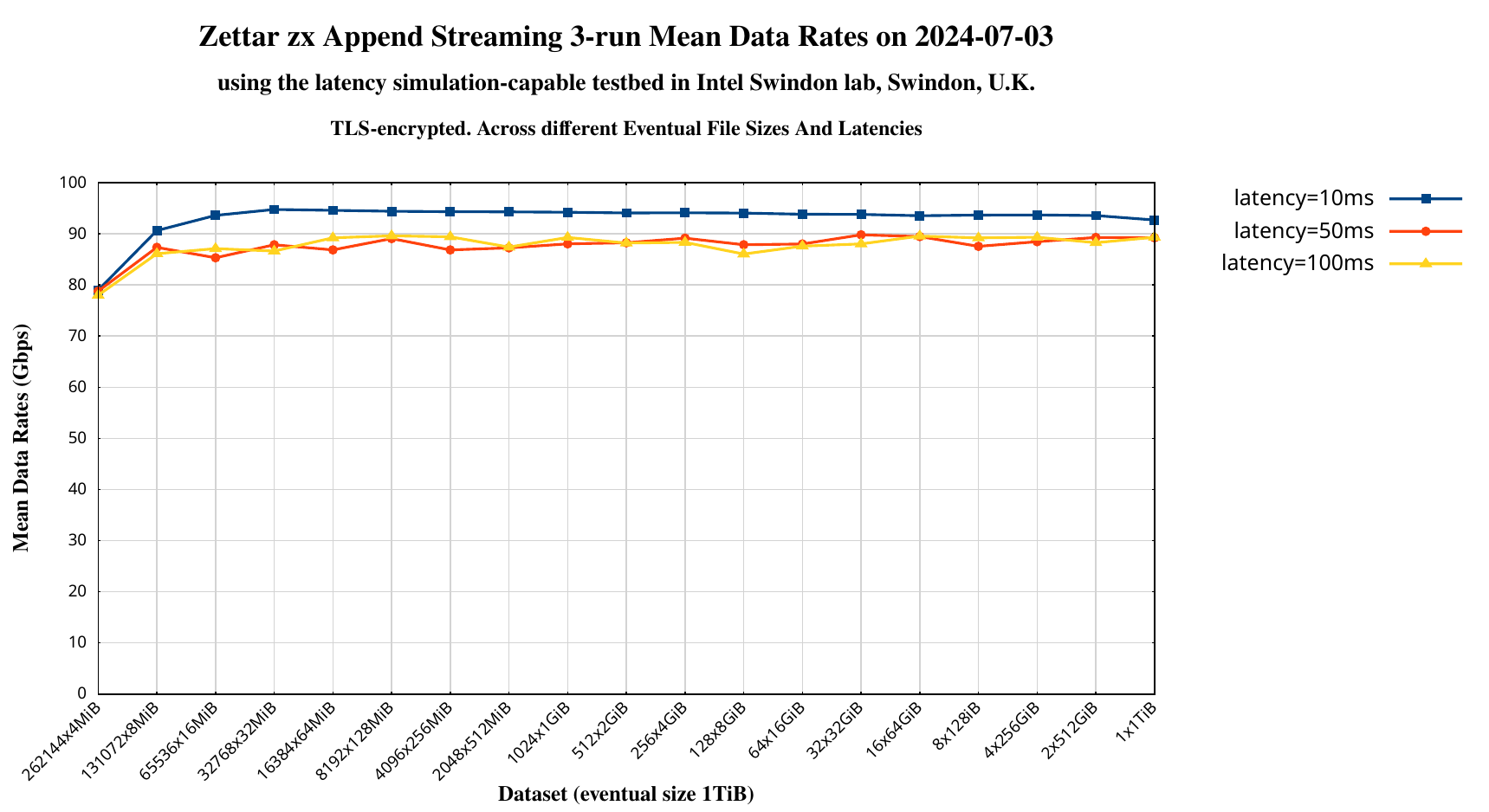}
    \caption{Streaming transfer sweeps vs three simulated latency
      values, 10~ms, 50~ms, and 100~ms. Note that the data rate levels
      are quite close to their counterparts from the bulk transfer
      sweeps shown in Fig.~\ref{fig:bulk-vs-latency}.}
    \Description{Graph showing streaming transfer performance
      vs. simulated latency values (10~ms, 50~ms, 100~ms) on the
      aforementioned 100~Gbps-capable testbed.}
    \label{fig:streaming-vs-latency}
\end{figure*}

\begin{itemize}
\item
  \textbf{Systems Focus:} Unlike commercial WAN simulators or network
  testbeds that isolate network performance (often using Random Access
  Memory (RAM)-to-RAM methods), this platform is uniquely suited to
  measure true BB-to-BB data transfer rates. This is essential
  because, as established in Section~\ref{sec:bandwidth}, persistent
  I/O is the dominant constraint for petascale transfers.
\item
  \textbf{Agile Development:} The platform supports rapid software
  iteration through automated regression testing for Continuous
  Integration/Continuous Deployment (CI/CD) and accelerated
  development of new components, such as AI Model Context Protocol
  (MCP) with guardrails for connecting zx to a natural language input
  for improved usability, without risking a live production network.
\item
  \textbf{Operational Utility:} Its reproducibility enables critical
  regulation validation for compliance-driven sectors, supports WAN
  planning through latency simulation, facilitates training/education
  for new systems engineers, and provides an efficient demo showcase
  environment.
\end{itemize}

While this method validates core performance metrics under controlled
conditions, it is important to note that it does not replace the
necessity of an active SDN testbed for validating complex
control-plane, security, or multi-flow operational scenarios typical
of a fully instrumented network like ESnet. However, for core
high-throughput systems performance engineering, the software-defined
WAN emulation offers a cost-effective, high-fidelity alternative.

The full testbed benefits and implementation details, together with
expanding it to speed $\ge 100$~Gbps, warrant a separate, in-depth
tutorial. See also Section~\ref{sec:conclusions}.

\subsection{The Network-Bandwidth Panacea Paradigm}\label{sec:bandwidth}

Before any further discussion, the premise is already weak. Research
and education networks typically operate at only
20--30\% of capacity~\cite{Nickolay2018Bridging}, the low utilization
condition predicted more than two decades ago and still holding
today~\cite{Odlyzko2003Utilization}. Bandwidth is seldom the scarce
resource.

In essence, this paradigm fails because the rest of the
infrastructure---primarily storage and compute, combined with the
parallelism and concurrency of the software data mover---must be in
balance with the available network bandwidth. A chain is only as
strong as its weakest link; once the network ceases to be that link,
introducing more bandwidth yields no benefit.

Foremost among these limiting factors, and often underestimated, is
storage I/O. Once network bandwidth exceeds the system\textquotesingle
s ability to read from or write to persistent storage, further
increases are futile. The bottleneck most frequently occurs during
write operations, as virtually all storage media deliver lower write
than read performance.

\begin{figure*}[htbp]
    \centering
    \includegraphics[width=0.8\textwidth]{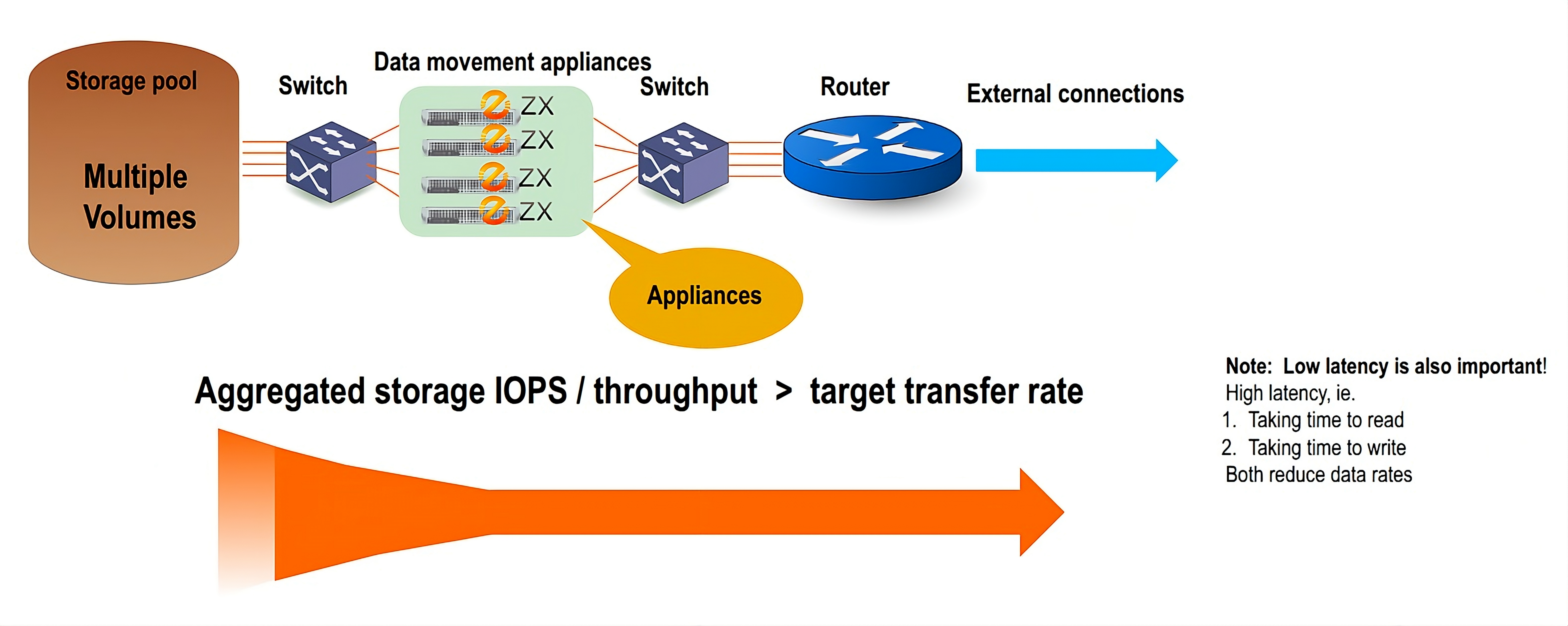}
    \caption{For production storage to support demanding data
      transport (bulk and streaming) well, it must have high enough
      throughput and low latency.}
    \label{fig:storage-headroom}
    \Description{A diagram illustrating storage throughput and latency
      requirements needed to support data transfer at speed and
      scale.}
\end{figure*}

Other system-level factors also contribute, e.g., the processing of
TCP/IP stacks, filesystem metadata operations, and
encryption/checksumming, all of which consume finite CPU cycles that
could otherwise be allocated to sustaining storage I/O. As indicated in Fig.~\ref{fig:storage-headroom}, to
attain high and sustainable data transfer performance, the storage
service must meet two criteria: 1) high enough storage Input/Output
Operations Per Second (IOPS)/throughput and 2) low latency. Both are
elaborated below.

In Fig.~\ref{fig:storage-headroom}, the term \enquote{high enough}
means the overall aggregated storage IOPS/throughput $>$ the target
transfer rate. \enquote{Low latency} is particularly crucial because,
for highly concurrent data movers, storage latency effectively
sabotages concurrency---a scenario often associated with random
I/O. What appears as a software bottleneck is frequently a storage
latency problem in disguise.

Closely related and often overlooked is the dataset's file size
distribution. Two opposite shapes defeat raw bandwidth, each for a
different reason:

\noindent
  \begin{minipage}{\linewidth}
    \centering
    \small
    \begin{tabularx}{\linewidth}{|p{3.2cm}|X|X|}
      \hline
      \textbf{Dataset shape} & \textbf{Root cause} &
      \textbf{Why more bandwidth does not help} \\
      \hline
      Many small files
      & Per-file overhead: metadata operations, open/close costs, and
      short-lived transfers
      & Pipelining never reaches steady state, so available bandwidth
      sits idle \\
      \hline
      A few very large files
      & Concurrency starvation: the file is the fundamental I/O
      scheduling unit, so achievable concurrency is bounded by the file
      count
      & Too few independent I/O operations to saturate a high-bandwidth
      link; even aggressive tuning cannot compensate \\
      \hline
    \end{tabularx}
  \end{minipage}

\begin{figure*}[htbp]
   \centering
   \includegraphics[width=0.8\textwidth]{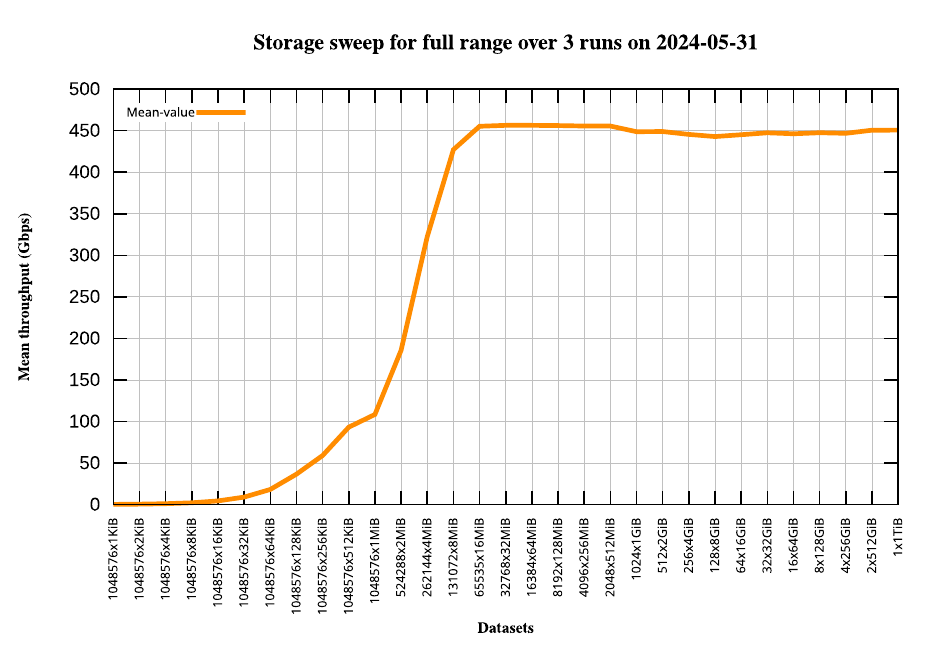}
   \caption{Storage sweep across the full 1\,KiB--1\,TiB range under a
     single configuration: throughput stays near zero in the
     small-file regime, then climbs and plateaus near 455~Gbps once
     files are large enough to keep the pipeline full. Neither the
     small-file floor nor the storage-bound plateau moves with more
     network bandwidth.}
   \label{fig:storage-sweep}
   \Description{A performance curve of mean throughput versus dataset
     file size, near zero for small files and plateauing around
     455~Gbps for large files.}
\end{figure*}

The first effect is evident in
Figs.~\ref{fig:bulk-sweep-plain}--\ref{fig:streaming-sweep} and
\ref{fig:bulk-vs-latency}--\ref{fig:streaming-vs-latency}, where added
parallelism or bandwidth yields no proportional gain, and most
directly in the storage sweep of Fig.~\ref{fig:storage-sweep}, which
isolates the storage ceiling on its own; the second has been observed
in operational DTN environments~\cite{Kissel2020ESnet}, where the
baseline data mover's rate declines as files grow larger and
fewer. Note that zx's concurrency architecture eliminates the second
effect.

This limit is not merely our own observation; it surfaces
independently wherever high-bandwidth links meet real endpoints. As
examined in Section~\ref{sec:aggregation}, a 400~Gbps benchmarking
effort found a single server capped near 260~Gbps even with the
network idle and the data served from memory, i.e., the endpoint, not
the link, set the ceiling~\cite{XRootD2025}. Provisioning more
bandwidth cannot move a wall that sits inside the server.

The same lesson appears in a piece of guidance that routes around
storage rather than confronting it. A cloud provider's migration
tutorial stages transfers through a RAM disk and candidly concedes it
is \enquote{a cool hack \ldots more than \ldots a recommended
  production solution}, with \enquote{limitations if there are
  extremely large files or datasets with millions of small
  files}~\cite{NebiusSkyPilot2025}; an industry post proposes a cache
to \enquote{absorb repeated reads}~\cite{Watts2026Cache}, a strategy
that is inapplicable to bulk movement, which reads each object exactly
once. In each case the fast path is measured and the durable-storage
path is avoided.

Such arrangements can \emph{mitigate}, but never \emph{mask}, the
throughput impact of slow or high-latency production storage. By
Eq.~(\ref{eq:stage-capacity}), the reprieve a buffer affords is
finite: a benchmark whose dataset fits within it measures the
buffer, whereas a petascale migration measures the storage.
  
\subsection{The CPU Primacy Paradigm}\label{sec:cpus}

A common consensus in high-performance computing is that powerful CPUs
are essential for high transfer rates. During 2023, while discussing
with our industry collaborators regarding CPU selection for data
movement appliances, the prevailing feedback was to select high-end,
high-core-count models. We held reservations about this
approach. Furthermore, we had skepticism about suggestions to
utilize CPUs with embedded hardware acceleration such as Quick Assist
Technology (QAT) \cite{IntelQAT}, based on the observation that such
acceleration can tie software to specific vendor hardware and drivers,
increasing complexity and potential \enquote{software bloat}.

This skepticism was grounded in years of deployment experience, which
revealed that zx consistently achieves high performance without
requiring a high core count; typically, 12-24 cores are
sufficient. While resources like ESnet\textquotesingle s fasterdata
recommend high clock rates for good encryption performance
\cite{ESnetHardware}, our observations suggested an alternative
path. We hypothesized that with zx's asynchronous I/O design, fewer
cores with CPU-built-in encryption acceleration and moderate clock
rates would reduce context switches (improving software efficiency)
and lower energy consumption, thereby reducing the TCO. This led to
the selection of the Intel Xeon 5418N, a mid-range model
\cite{IntelXeonGold}.

In 2024, more extensive testing with the two HPE DL380 Gen11
server-based appliances confirmed that even with full encryption, QAT
was unnecessary. The CPU\textquotesingle s native instruction set
extensions were sufficient. The results shown in
Figs.~\ref{fig:bulk-sweep-ktls}-\ref{fig:streaming-sweep},
\ref{fig:bulk-vs-latency}-\ref{fig:streaming-vs-latency} were all achieved
without specialized hardware acceleration and using this modest CPU.

As such, we can conclude that CPU raw computing power matters, but
software efficiency and storage architecture matter more. The right
software makes adequate CPUs shine; the wrong software dims the most
powerful CPUs.

% The 6th paradigm starts here.
\subsection{The Cloud-Universality Paradigm}\label{sec:cloud}

We received an inquiry about evaluating zx for a site in South Africa
needing to transfer data to Europe at higher rates. The
site\textquotesingle s proposal to use a virtual machine (VM)
immediately raised concerns, as it revealed a common misconception:
that virtualized environments can deliver the performance required for
high-speed data movement.

Virtualization technologies have been a mainstay for over two decades
\cite{RedHatVirt}. Nevertheless, a critical determinant of their
performance for intensive data movement workloads is the level of
system control accessible to the user. Table~\ref{tab:tab6} summarizes
the scope of Linux kernel parameters across virtual machines and
containers.

\begin{table}[!b]
  \centering
  \caption{Linux Kernel Parameter Scope: Virtual Machine vs. Container}
  \label{tab:tab6}
  \small
  % Distributing width: 20% for first two, 60% for the last
  \begin{tabularx}{\textwidth}{%
    |>{\hsize=0.7\hsize}X 
    |>{\hsize=0.7\hsize}X 
    |>{\hsize=1.6\hsize}X |}
    \hline
    \textbf{Location} & \textbf{Platform} & \textbf{Scope} \\
    \hline
    \textbf{Inside VM} & VM (e.g. KVM) & Affects \emph{only} the 
    guest OS kernel. \\
    \hline
    \textbf{Container} & Docker/Podman & Affects namespaced subset; 
    no host change. \\
    \hline
    \textbf{Privileged} & Docker/Podman & Can change host settings, 
    affecting all. \\
    \hline
    \textbf{On Host} & Physical host & Affects physical server and all 
    VMs and containers. \\
    \hline
  \end{tabularx}
\end{table}

This hierarchy of control directly leads to three critical performance
implications for high-speed data transfer:

\begin{enumerate}
\def\labelenumi{\arabic{enumi}.}
\item
  \textbf{Interrupt Overhead:} Even with Single Root I/O
  Virtualization (SR-IOV) \cite{NvidiaSRIOV}, interrupt handling in VMs incurs
  measurable overhead due to interception and remapping by the
  hypervisor.
\item
  \textbf{I/O Performance Degradation:} High-speed I/O operations,
  which generate a high volume of interrupts, consequently achieve
  lower performance in virtualized environments.
\item
  \textbf{Host-Level Dependency:} If the host kernel is not tuned to
  mitigate latency (as established in Section~\ref{sec:latency}), adjustments within
  the guest OS are largely ineffective.
\end{enumerate}

Achieving consistent, latency-insensitive data transfer performance
within a VM is intrinsically constrained by the architecture.
Virtualized cloud offerings inherit these limitations. Coordinating
the kernel parameter
adjustments necessary for maximum VM performance is
application-dependent (see Table~\ref{tab:tab3}) and
practically infeasible for a heterogeneous customer base.

\subsubsection{The Architectural Cost of Cloud Abstraction}\label{sec:cloud-abstraction}

While virtualization and cloud abstraction deliver significant
benefits in manageability and elasticity, they impose substantial
performance penalties on data movement. The canonical cloud data
path---encumbered by layers of overheads such as hypervisor,
virtualized storage, and HTTP/REST APIs---incurs a substantial
performance inefficiency compared to a tuned bare-metal environment.
This inefficiency is empirically observed in our tests to routinely
reach 30--50\%. For instance, on 10~Gbps Elastic Compute Cloud (EC2)
instances in 2019, we achieved a maximum of $\sim$6~Gbps per
VM, a 40\% loss from line rate. In 2025, as detailed in
Fig.~\ref{fig:kek-cryoem-cloud}, the native cloud tooling
(\texttt{aws-cli}~\cite{AWSCLIGithub}) performed at a small fraction
of the available capacity, while our co-designed appliance, even when
partially firewalled, consistently attained a significantly higher
percentage of the physical link\textquotesingle s potential.

Standard cloud workarounds, such as Amazon Web Services (AWS)
multipart uploads \cite{AWSS3MPU}, fragment a simple file transfer into a
multi-step coordination process---effectively a highly fragmented
workflow \cite{Acharya2010Rube} for data movement.

Even bare metal instance offerings~\cite{AWSBareMetal, GCPBareMetal,
  AzureBareMetal} may remove the abstraction overhead, but note that
such provisioned hardware is not a co-designed data movement
appliance.

\subsubsection{A Critical Examination of Cloud Provider Metrics}\label{sec:cloud-metrics}

The prevailing paradigm of general-purpose cloud data paths,
regardless of provider, introduces inherent architectural conflicts
for high-performance data movement workflows. This challenge becomes
acute at the petascale level. Analysis of major U.S. cloud providers
reveals a focus on theoretical network bandwidth, often decoupled from
storage I/O and end-to-end performance. The info in
Tables~\ref{tab:tab7}-\ref{tab:tab9}, extracted from official provider
documentation around mid-2025, illustrates this focus.

\paragraph{The Disconnect Between Theoretical Bandwidth and Practical Throughput}\label{sec:cloud-bw-gap}

Maximum network bandwidth figures---such as 1,000~Gbps or
400~Gbps---are often theoretically impressive but unrealizable in
practice, as storage subsystems cannot always sustain the throughput.
The published info from three leading hyperscalers, shown in
Tables~\ref{tab:tab7}-\ref{tab:tab9} should make this aspect clearer.

\begin{table}[!b]
\centering
\caption{Google Cloud Platform (GCP) virtual instance parameters.}
\label{tab:tab7}
\small
\setlength{\tabcolsep}{2pt}
\begin{tabularx}{\textwidth}{|>{\raggedright\arraybackslash}X|
                             >{\raggedright\arraybackslash}X|
                             >{\raggedright\arraybackslash}X|}
\hline
\textbf{Service} & \textbf{Maximum Bandwidth} & 
\textbf{Key Features \& Optimizations} \\
\hline
Compute Engine VMs (e.g., A3, G4, C4, H4D) & Up to 
\mbox{1,000~Gbps} (A3-highgpu-8g) and \mbox{400~Gbps} 
(G4-standard-384) & Tier1 networking is required on selected machine 
types to reach 100/200~Gbps per-VM egress. Google Virtual NIC 
(gVNIC) and Fast Socket improve performance for distributed 
workloads. \\
\hline
Dedicated \mbox{Interconnect} & \mbox{10~Gbps} and \mbox{100~Gbps port} 
speeds & Direct, private physical connection from on-premises network to 
Google for high-throughput hybrid connectivity. \\
\hline
\end{tabularx}
\end{table}

\begin{table}[tbp]
\centering
\caption{AWS virtual instance parameters.}
\label{tab:tab8}
\small
\setlength{\tabcolsep}{2pt}
\begin{tabularx}{\textwidth}{|>{\raggedright\arraybackslash}X|
                             >{\raggedright\arraybackslash}X|
                             >{\raggedright\arraybackslash}X|}
\hline
\textbf{Service} & \textbf{Maximum Bandwidth} & 
\textbf{Key Features \& Optimizations} \\
\hline
EC2 Instances (\mbox{e.g., HPC,} GPU, Network-optimized) & Up to 3,200~Gbps (P5, via EFA) &
Requires specific high-end instances. Enhanced Networking (ENA) for
high packet-per-second, low-latency performance. \\
\hline
\mbox{AWS Direct} Connect & 1--\mbox{100 Gbps} (Dedicated) &
Dedicated network from premises to AWS. Supports Link Aggregation
Groups (LAG) and single 100 Gbps connections. \\
\hline
\end{tabularx}
\end{table}

\begin{table}[tbp]
\centering
\caption{Azure virtual instance parameters.}
\label{tab:tab9}
\small
\setlength{\tabcolsep}{2pt}
\begin{tabularx}{\textwidth}{|>{\raggedright\arraybackslash}X|
                             >{\raggedright\arraybackslash}X|
                             >{\raggedright\arraybackslash}X|}
\hline
\textbf{Service} & \textbf{Maximum Bandwidth} & 
\textbf{Key Features \& Optimizations} \\
\hline
\mbox{Azure VMs} (HPC/GPU series) & Varies by instance, up to 100 Gbps+ &
Accelerated Networking reduces latency, maximizes throughput;
HPC often uses InfiniBand. \\
\hline
Azure ExpressRoute & \mbox{Up to 100 Gbps} (Dedicated) &
Private connection from infrastructure to Azure; ExpressRoute Premium
supports higher route/virtual network limits. \\
\hline
Azure Firewall Premium & \mbox{Up to 100 Gbps} throughput &
Handles high-volume network traffic with deep packet inspection enabled. \\
\hline
\end{tabularx}
\end{table}

\paragraph{Focusing only on computing, ignoring data availability, and I/O}\label{sec:cloud-io}

In the context of using the cloud for HPC, the role of data I/O is
defined by the following premises:

\begin{itemize}
\item Without data, compute is useless.
\item Data is produced predominantly at the edge (e.g., LCLS-II \cite{Thayer2019LCLS}).
\item Efficiently moving this data to the cloud remains unsolved due
  to the intrinsic overheads.
\item Regardless of computational speed, data transport imposes a fundamental delay.
\item The current \enquote{sneakernet} workarounds described in
  \cite{WSJ2025AI} and \cite{Toms2025AI} deserve scrutiny.
\end{itemize}

\begin{table}[tbp]
  \centering
  \caption{Typical performance information published by 
    hyperscalers.}
  \label{tab:tab10}
  \small
  \begin{tabularx}{\textwidth}{|>{\hsize=0.6\hsize\raggedright\arraybackslash}X 
                               |>{\hsize=0.7\hsize\raggedright\arraybackslash}X 
                               |>{\hsize=0.8\hsize\raggedright\arraybackslash}X 
                               |>{\hsize=1.9\hsize\raggedright\arraybackslash}X
                               |}
    \hline
    \textbf{Cloud Provider} & \textbf{Workload} & \textbf{Metric} & 
    \textbf{Throughput Implication} \\
    \hline
    AWS & Western Digital HPC Simulation & 2.5M tasks on 1M vCPU 
    cluster in 8 hours & Massive-scale, low-latency, high-bandwidth 
    communication; reduces job time from 20 days on-prem to 8 hours 
    in cloud. \\
    \hline
    GCP & PGS Seismic Processing & 202k on-prem cores $\rightarrow$ 
    1.2M vCPUs in cloud & $\sim$4$\times$ increase in compute/data-transfer; 
    among world's largest supercomputers if continuous. \\
    \hline
    Azure & Univ. of Bath HPC & 250 on-prem nodes $\rightarrow$ 
    thousands in Azure & Provides on-demand, high-speed burst 
    capacity; ultra-high data access elasticity. \\
    \hline
  \end{tabularx}
\end{table}

\paragraph{Lack of reproducibility}\label{sec:cloud-repro}

Results are presented primarily in marketing form~(Table~\ref{tab:tab10}) and omit key
engineering details required for reproducibility, including storage,
network, system configuration.

\subsubsection{The Antidote: A Co-Designed Data Path for the Cloud}\label{sec:cloud-antidote}

Intrinsic overhead cannot be optimized away. We propose to architect a
parallel data path that coexists with the current cloud
model. Although zx can be embedded in NVIDIA
DPUs~\cite{NvidiaDPUAbout,Zettar2024HPCwire,NvidiaHPCEdge} and coupled
with NVMe over Fabrics (NVMe-oF) as a workaround, there is a far
better way: almost all major cloud providers now offer bare-metal
instances~\cite{GCPBareMetal,AWSBareMetal,AzureBareMetal} and file
storage~\cite{GCPManagedLustre,AWSManagedLustre,AzureManagedLustre}. Because
data movement appliances are built from COTS hardware, a provider can
deploy them on that bare-metal fleet directly. From there the
appliances can be offered as a metered service and leased to end
users, paired with the provider's existing direct-connection
services~\cite{AWSDirectConnect,GCPInterconnect,AzureExpressRoute},
already provisioned at 100~Gbps and, for AWS Direct Connect, 400~Gbps.
The commercial model is thus as complete as the technical one: a
high-throughput offering that rides infrastructure the provider
already owns. These are rates a co-designed appliance cluster
saturates rather than strains against; a 400~Gbps direct connection is
headroom, not a ceiling. Furthermore, the flexible hardware
BOM~(Fig.~\ref{fig:core-bom}) accommodates different customer needs.

This path provides:

\begin{itemize}
\item \textbf{File Transfer Semantics:} Direct file-to-file
  transfer, avoiding complex HTTP chunking.
\item \textbf{New Transfer Modes:} Unbounded bulk plus streaming and
  incremental transfers~(Figs.~\ref{fig:streaming-sweep}
  and~\ref{fig:streaming-vs-latency}), none of which the object-based
  S3 REST API can express.
\item \textbf{Predictable Performance:} Latency-insensitive, encrypted
  movement at full line rate.
\item \textbf{Co-existence:} Zero changes to REST APIs, improving
  performance without disrupting services.
\end{itemize}

This architecture mitigates the cloud's data-movement penalties and
gives providers a high-throughput data path they can run alongside
their existing services.

\begin{figure}[h!]
    \centering
    \includegraphics[width=0.92\textwidth]{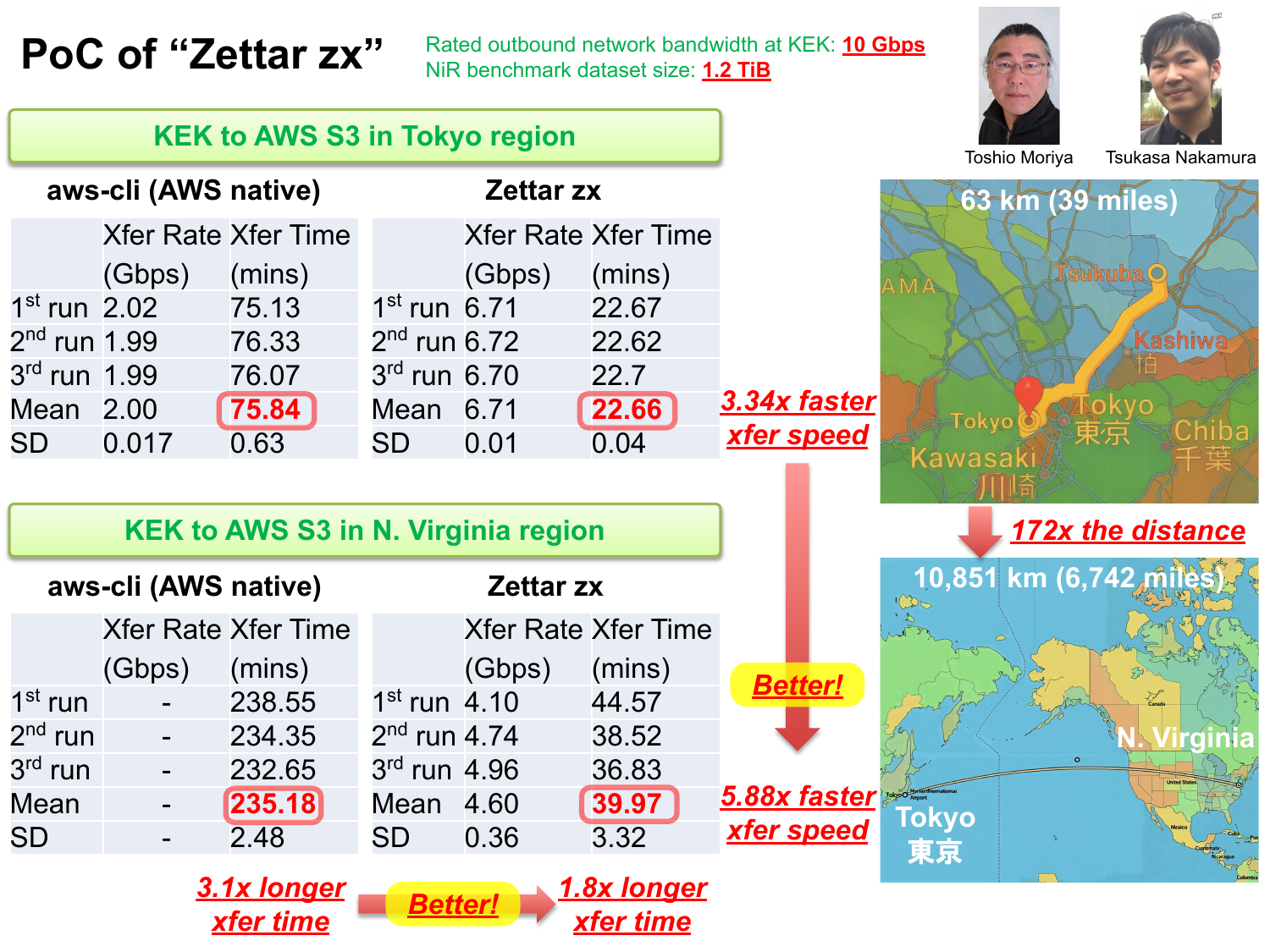}
    \caption{Transfer comparison of a 1.2 TiB Cryo-EM dataset from KEK
      to AWS Regions using \texttt{aws-cli} and zx.  Over the same
      10,851~km trans-Pacific path to N.~Virginia, zx finished in
      39.97~min against \texttt{aws-cli}'s 235.18~min, nearly
      6$\times$ faster.}
    \Description{A bar chart comparing transfer performance of zx
      vs. aws-cli for Cryo-EM dataset transfers from KEK to AWS Tokyo
      and Virginia.}
    \label{fig:kek-cryoem-cloud}
\end{figure}

\subsubsection{Empirical Evidence of the Cloud Data Movement Limitations}\label{sec:cloud-evidence}

The performance penalties imposed by standard cloud data paths are
measurable. Using the GoToCloud platform on AWS
\cite{Moriya2024GoToCloud}, a comparative analysis conducted by High
Energy Accelerator Research Organization (KEK) \cite{KEKAbout} vividly
illustrates this point. As summarized in
Fig.~\ref{fig:kek-cryoem-cloud}, the transfer of a 1.2 TiB
Cryo-electron microscopy (Cryo-EM) dataset was measured with zx vs
\texttt{aws-cli}.

Results:

\begin{itemize}
\item Over 63 km (KEK $\rightarrow$ AWS Tokyo), zx completed in 22.66
  minutes.
\item Over 10,851 km trans-Pacific (KEK $\rightarrow$ AWS N. Virginia),
  zx completed in 39.97 minutes, a 1.76$\times$ increase in transfer
  time.
\end{itemize}

This performance differential is notable given the 10~Gbps network at
KEK. While partially constrained by a firewall, zx utilized
the bandwidth effectively. In comparison, \texttt{aws-cli} transfer to
N. Virginia required 235.18 minutes.

Two technical observations:

\begin{enumerate}
\item \textbf{Cloud's Self-Imposed Bottleneck:} The
  underperformance of \texttt{aws-cli} indicates that the bottleneck
  is not the physical network, but the inefficient software stack and
  limited end-to-end control.
\item \textbf{Architectural Solution:} zx demonstrates that
  long-distance transfers can be latency-insensitive. Data movement
  appliance adoption by cloud providers ought to bring more consistent
  and predictable data movement to the cloud.
\end{enumerate}
% The 6th paradigm ends above.
% \FloatBarrier 
\section{Reproducibility}\label{sec:reproducibility}

For reproducibility, the exact kernel parameters, latency simulation
values, and visualization commands are publicly
available in a GitHub repository \cite{FangPieee}. The \texttt{iperf3}
benchmark tool used in comparative analysis is also publicly available
from its official repositories \cite{iperf3Github}.

The endpoints, not the wire, set the true performance limit, so
characterizing the storage service deserves the same care as
characterizing the network. Where \texttt{iperf3} above measures the
network path, the storage path can be measured with
\texttt{elbencho}~\cite{Elbencho, Breuner2021Elbencho}, an open-source
benchmark that reports throughput, IOPS, and latency for both read and
write operations, and which ESnet's own DTN I/O-benchmarking guidance
likewise recommends~\cite{ESnetIOBench}. Because storage throughput
depends strongly on file (or object) size---the small-file cliff and
the large-file plateau---we contributed
\texttt{storage\_sweep}~\cite{StorageSweep} in 2020, a set of
\texttt{bash} wrappers around \texttt{elbencho} in its
\texttt{contrib} directory that automate a power-of-two sweep across
sizes and plot the resulting curve. Together they let a reader locate
the knee of their own storage before attributing a shortfall to the
network, and give the Open Invitation of Section~\ref{sec:aggregation}
a concrete, storage-side referent.

A caveat on \emph{what} to measure. The storage industry rates devices
by small-block (e.g., 4~KiB, 8~KiB) figures, but most applications do
not address raw devices; they read and write \emph{files} through a
filesystem, and bulk movers do so in large blocks, e.g., we often use
16~MiB. A device's 4~KiB random-I/O rating therefore says little about
the throughput an application will actually see.
\texttt{fio}~\cite{fioGithub} can drive filesystem I/O and handles
large-block sequential work, but it is unreliable in the single-KiB
small-file regime---precisely where the small-file cliff
lives. \texttt{elbencho} is dependable there, which is why we rate
storage with it, at application-relevant sizes. Nor is the filesystem
incidental: an ill-chosen or misconfigured filesystem will hobble the
application even with good underlying device benchmarks.

The zx data mover and associated deployment tooling represent
proprietary commercial technology reserved for customers and
partners. The provided materials allow independent verification of the
benchmarking methodology without requiring
access to the commercial zx software. The principle: \textbf{open
  methodology, not open product}.

\begin{figure*}[!t]
  \centering
  \includegraphics[width=0.98\textwidth]{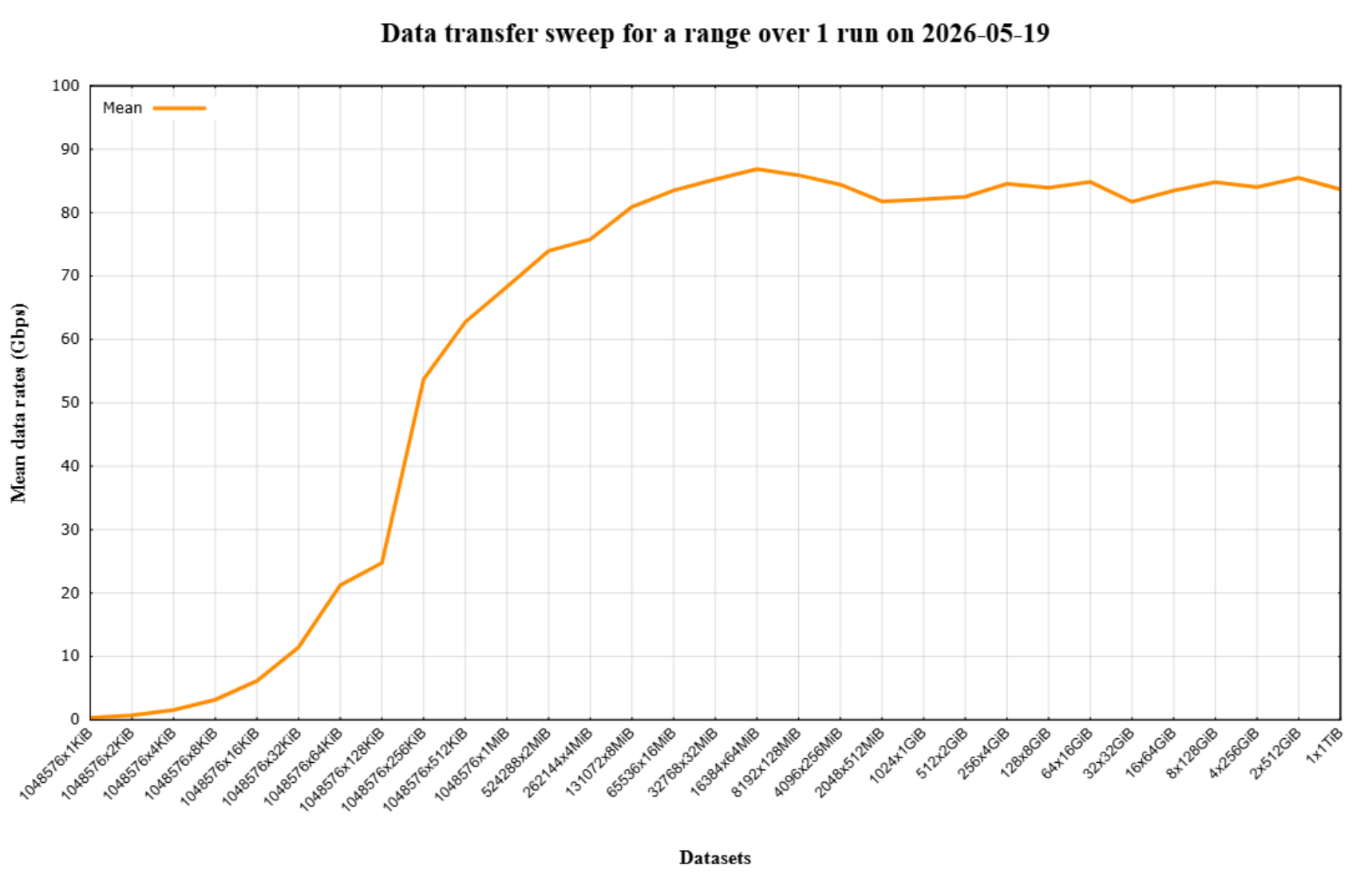}
  \caption{Independent reproduction by an HPE Technical
    Consultant on a transatlantic 100~Gbps link (U.S. Midwest
    $\leftrightarrow$ Central Europe). Operating remotely through a
    challenging enterprise control path (VPN, Web-based remote
    desktop, virtual machine), the operator achieved $>$80~Gbps on the
    first independent attempt using only a simple automation
    \texttt{bash} script by Zettar. The native timestamps (2026-05-19)
    and distinct visual style attest to its independent provenance.}
\label{fig:hpe-reproduction}
\end{figure*}

The two differ in kind, not degree. A methodology is a finite,
statable set of principles that can be described and published.  The
product's value lies elsewhere: in the co-design interactions among
endpoints, network, and storage, which grow as $\binom{n}{2}$
(Fig.~\ref{fig:cn2}, Eq.~\ref{eq:cn2}). The elements of these
interactions are tacit, in the phrase of Zhuangzi (\zhname{莊子}),
\zhname{只可意會，不可言傳} \cite{ZhuangziTiandaoC4BCE}: apprehended
only in the doing, never carried across in words.

The standard challenge runs: open-source the software and we will
figure out the rest. It mistakes the artifact for the capability. Two
examples: After the Second World War the United States seized the V-2
blueprints and shipped captured hardware home by the trainload, yet
still judged the documents insufficient and brought some 1,600 German
scientists and engineers across the Atlantic, von Braun's rocket team
among them; in the Smithsonian's own account, the people who designed
the weapons ``were needed to help transfer the
technology''~\cite{SmithsonianPaperclip}. Extending the 1944 rocket
design in the blueprints, von Braun went on to build the Saturn~V that
carried Apollo to the Moon. Qian Xuesen (\zhname{錢學森}) is another
substantial case: the U.S. government detained him for five years,
then deported him in 1955, after which he became the father of China's
missile and space program~\cite{CASQianXuesen}. The method is the
tell. As his Caltech colleague Frank Marble recorded, the five-year
detention was designed to cut him off from the ongoing classified work
so that what he knew would go stale. It was a deliberate effort to age
out the knowledge in his head~\cite{MarbleTsienOralHistory}. A
government that spent five years trying to make a man's knowledge
obsolete had already conceded where the value lies: it is not in the
paperwork; the effort failed. The two cases bracket the point. Having
the source is not enough (Paperclip), and the real asset walks out on
two legs (Qian). If superpowers (the U.S. and USSR) with the hardware,
the blueprints, and generous budgets could not depend on the
artifacts alone, ``We will figure it out'' underestimates the gap
badly. Qian's journey is the Zhuangzi principle lived, not told.

The same combinatorial law also explains why so few open-source
projects endure at all. Projects that last, for example, Ansible,
elbencho, Linux, the MIT X11R5 release, are each stewarded by a single
leader or a few, or have sponsorship from established organizations,
along a clear chain of decision. This is true for any large
organization, including the military. Well-known generals such as Bai~Qi
(\zhname{白起})~\cite{BaiduBaiqi} or a powerful ancient Roman
legion~\cite{RomanBritainLegion} adhere to this principle, never a
flat committee. That structure is the source of the strength, not
incidental to it: when the order was lost, even Rome's legions could
be annihilated, as three were in the Teutoburg
Forest~\cite{BritannicaTeutoburg}. Without bounded command,
coordination collapses under its own $O(n^2)$ weight and the effort
joins the graveyard of abandonware. Open methodology is safe to
publish precisely because bounded stewardship is what keeps it coherent.

Furthermore, many prominent independent parties have verified the
results, e.g.,
ESnet~\cite{Kissel2020ESnet}, KEK's GoToCloud
results~\cite{Moriya2024GoToCloud}, and a separate HPE reproduction
(Fig.~\ref{fig:hpe-reproduction}). The last one is notable for several
reasons beyond the raw performance figure:

\begin{itemize}
  \item \textbf{Independent operator:} The operator had no prior experience 
    with the system---he succeeded on his first attempt using only a simple 
    \texttt{bash} automation script and its documentation.
  
  \item \textbf{Hostile control path:} To reach the appliances, he had
    to tunnel through a multi-layered enterprise remote access
    facility (VPN, web-based remote desktop, and a virtual machine)---%
    elements that introduce serious automation productivity degradation.
  
  \item \textbf{Configuration drift:} The local IT team had, despite explicit 
    warnings, applied their perceived ``standard adjustments'' for 
    general-purpose servers to the appliances, altering the carefully 
    designed configuration.
\end{itemize}
  
Yet the system still delivered $>$80~Gbps on the transatlantic
link. This combination of circumstances---independent operator,
hostile control path, configuration drift---transforms
Fig.~\ref{fig:hpe-reproduction} from a mere performance plot into a
stress test of operational robustness.

\section{Deployments}\label{sec:deployments}

\subsection{Validation Platforms}\label{sec:platforms}

This testing is part of a decade-long, sustained effort to validate
the co-design principle across diverse production environments, as
documented in prior studies
\cite{Thayer2019LCLS,ESnet2017Petabyte,ESnet2018Record,ICM2019Poland,NSCC2019DMC,Fang2016SLAC,Fang2019Samsung}.

Testing for this study was performed on multiple platforms,
demonstrating the adaptability of the co-design principle. The
validation platforms included:

\begin{itemize}
   \item {A testbed built using two HPE DL380 Gen 11 servers and two
	 Intel M50CYP2UR208 servers at the former Intel Swindon Lab (U.K.).}
   \item {Two HPE DL380 Gen 11 server-based appliances located in
     Switzerland and California, U.S. data centers, connected via a
     shared production 100~Gbps link carrying other network traffic.}
   \item {Two mini appliances deployed at two distinct sites in
     Taiwan, connected via the Taiwan Advanced Research and Education
     Network (TWAREN) \cite{TWARENAbout}.}
   \item {Independent validation conducted at KEK in Japan using the
     GoToCloud platform on AWS, as reported in Section~\ref{sec:cloud-evidence}.}
\end{itemize}
    
\subsection{Data Availability and Licensing}\label{sec:data-availability}

In the spirit of reproducible research, all elements required to
replicate the testing and benchmarking methodology presented in this
report are provided in a public GitHub repository \cite{FangPieee}. The
zx software data mover is a commercially licensed product and is not
available for public distribution. \textit{Note that the tacit
  knowledge earned from first-hand intense work is unavailable
  either.} A report and its code are the explicit artifacts of a
solution; its essence resides in the people who created it, i.e., the
``soul'' as elucidated by the 4th century BCE Chinese philosopher
Zhuangzi~\cite{ZhuangziTiandaoC4BCE} (\zhname{莊子}), who held \emph{that
the deepest knowledge can only be sensed, not conveyed through words}
(\zhname{只可意會，不可言傳}). This view is also portrayed much later
(1981) by Tracy Kidder's \emph{The Soul of a New
Machine}~\cite{Kidder1981Soul}.

\section{Transfer Task Forwarding Across Security Perimeters}
\label{sec:forwarding}

\begin{figure*}[tp]
\centering
\includegraphics[trim=0 64 0 60,clip,width=0.98\textwidth]{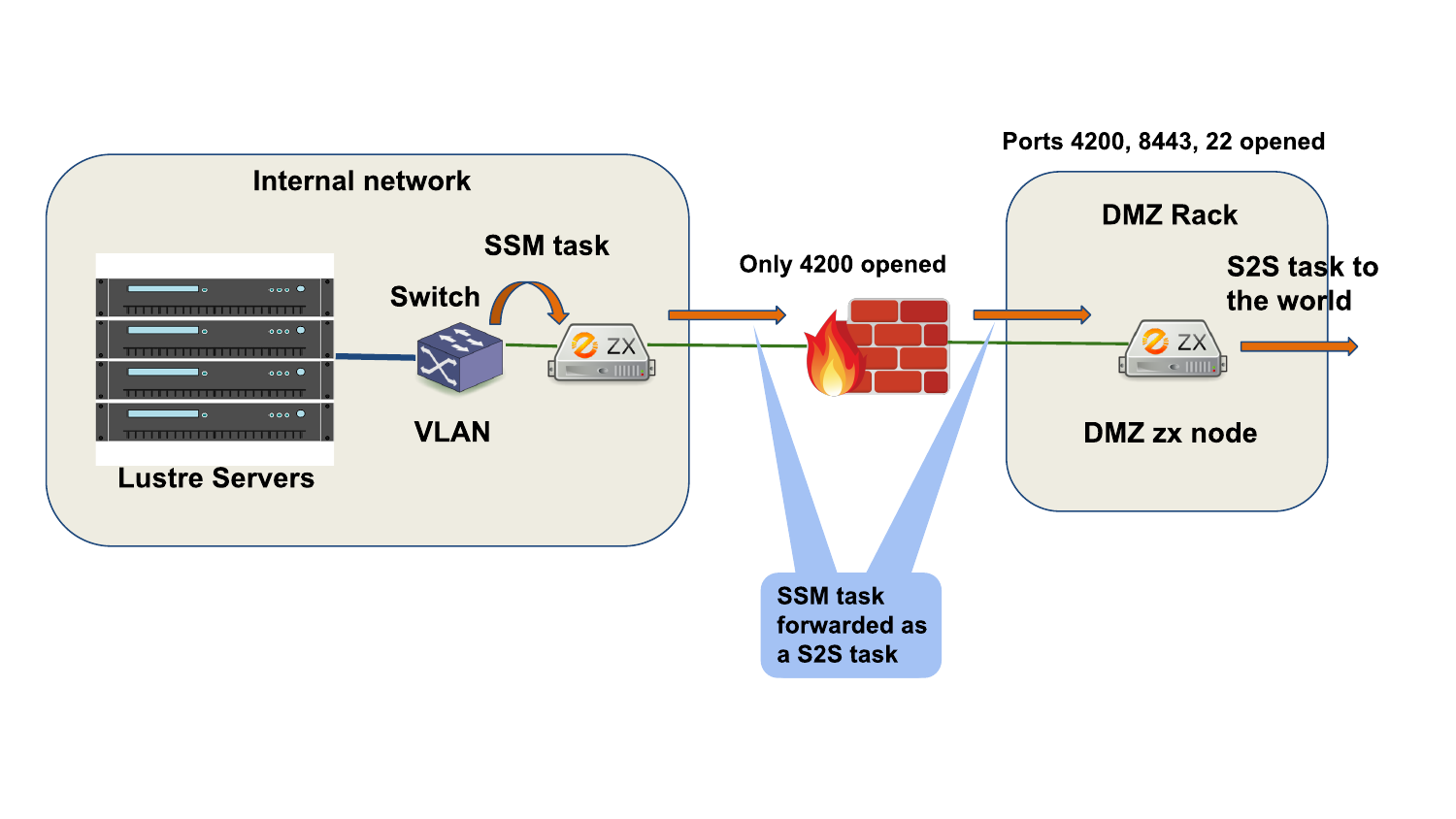}
\caption{Transfer task forwarding preserves the enterprise security
    posture: production storage stays inside the perimeter and never
    enters the DMZ. Being a unified software data mover, zx first uses
    its single-site mode (SSM) to stage data into the BB of
    the left appliance, which resides inside the perimeter. zx then
    forwards the staged data to the right appliance over its
    site-to-site (S2S) mode; only that appliance---holding transient
    staged data---faces the external world, before delivering the data
    to the external destination. Because the flow crosses a single
    opened S2S port (default 4200) rather than bypassing the firewall,
    the existing security controls remain in force.}
\label{fig:transfer-forwarding}
\end{figure*}

\begin{table*}[tp]
  \centering
  \small
  \caption{Crossing a security perimeter at high throughput: the
    typical approach versus zx transfer task forwarding.}
  \label{tab:forwarding}
  \begin{tabularx}{\linewidth}%
    {|>{\raggedright\arraybackslash}p{3.0cm}|L|L|}
    \hline
    & \textbf{Typical Approach}
    & \textbf{Transfer Forwarding (zx)} \\
    \hline
    \textbf{Appliances}
    & One transfer node plus dedicated perimeter gear
    & Two data movement appliances: one inside the perimeter, one in
    the DMZ \\
    \hline
    \textbf{Production storage}
    & Exposed in the DMZ, or the transfer is blocked by security
    policy
    & Stays inside the perimeter; never enters the DMZ \\
    \hline
    \textbf{Firewall}
    & Bypassed via ACLs on a dedicated switch, or kept in-path and
    throttled
    & Kept in place; data crosses a single opened S2S port
    (default 4200) \\
    \hline
    \textbf{Extra infrastructure}
    & Dedicated high-performance switch/router, transfer node, and
    measurement service; network re-architecture
    & None; configured through the standard zx task workflow \\
    \hline
    \textbf{Throughput}
    & Throttled and data-mover dependent
    & Below native fast/high-speed rate---bounded by the firewall and
    the file-size histogram---but generally tolerable, see e.g.,
    Fig.~\ref{fig:kek-cryoem-cloud}\\
    \hline
    \textbf{Security acceptance}
    & Resisted by many enterprises: storage exposure or firewall
    bypass
    & Preserves the existing security posture \\
    \hline
    \textbf{Indicative cost}
    & Enclave switch/router + transfer node + measurement +
    network-engineering effort
    & $2 \times$ mini-appliance ($\sim$\$5{,}500 class) + zx
    licenses \\
    \hline
  \end{tabularx}
\end{table*}

Although different approaches for security exist, the most common one
is a straightforward use of firewalls. Nevertheless, even though data
movement appliances can be placed in a DMZ, placing resources such as
production storage in such a zone tends to conflict with existing
security policies. Moreover, a firewall placed in front of an
appliance typically throttles the attainable data transfer rates. To
address this, zx has a feature: transfer task
forwarding~(Fig.~\ref{fig:transfer-forwarding}). Please note that crossing an opened
firewall port, forwarding does not reach the rates of a native fast or
high-speed link---the firewall's architecture and the data's file-size
histogram both bound it---but the resulting throughput is generally
tolerable. Because this approach works with the enterprise firewall
already in place, forwarding requires no dedicated data path or
perimeter re-architecture: an operator configures it through the same
zx task workflow used for any transfer.

For human organizations, a firewall is analogous to the watershed of
the Drainage Basin Pattern.  Forwarding uses the guiding principle
in Section~\ref{sec:foundations} to flow data through a single
permitted port and leaves the boundary standing, like the Grand Canal.

\section{Conclusions}\label{sec:conclusions}

The arguments presented in this report against over-reliance on
isolated, formulaic optimizations are rooted in a system
perspective. In the design of composite laminates
\cite{Fang1993Monte}, the macroscopic properties of a composite
plate result from the arrangement of discrete plies. Likewise, the
performance of a data movement system emerges from the interplay of
discrete, non-linear components: CPU cores, memory hierarchies,
storage devices, network interfaces, among others.

The holistic co-design principle central to this work represents the
evolution of a decade-plus sustained and continuing engineering
effort. This lineage traces back to foundational architectural
blueprints developed at SLAC National Accelerator Laboratory
\cite{Fang2016SLAC} and subsequently validated through formal
DOE/ESnet evaluation \cite{Kissel2020ESnet}. The architecture's
efficacy has been demonstrated through high-stakes production trials
at LCLS~\cite{Thayer2019LCLS} and
through extensive public discourse, e.g., \cite{Fang2019Samsung},
\cite{Kissel2021Rice}, \cite{Fang2021HPCKP}. Recently, the same
co-design principle has proven portable to specialized form factors
such as Data Processing Units (DPUs)~\cite{Fang2022Nvidia,
  Zettar2024HPCwire}---reaffirming that end-to-end performance is set
by holistic system design rather than by any single component.

\begin{figure}[b]
    \centering
    \includegraphics[width=0.80\textwidth]{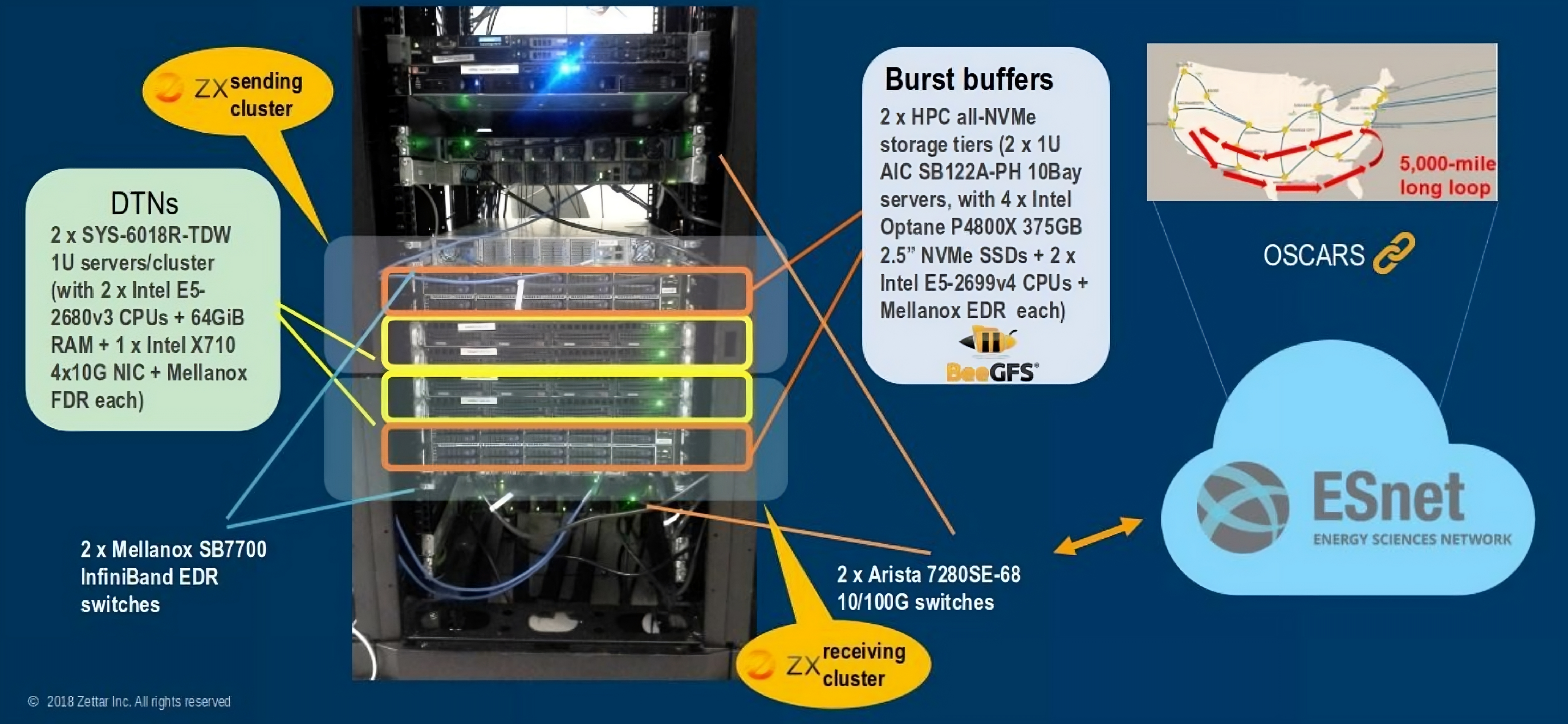}
    \caption{Zettar's testbed at SLAC from 2015 to 2019. See page 19
      of \cite{Kissel2020ESnet} for more details of the
      configuration. (Copyright Zettar Inc. Reproduced with permission
      for this publication.)}
    \label{fig:slac-testbed}
    \Description{Photograph of Zettar's testbed at SLAC National
      Accelerator Laboratory from 2015-2019 showing server rack
      setup.}
\end{figure}

\begin{figure}[!tb]
    \centering
    \includegraphics[width=0.80\textwidth]{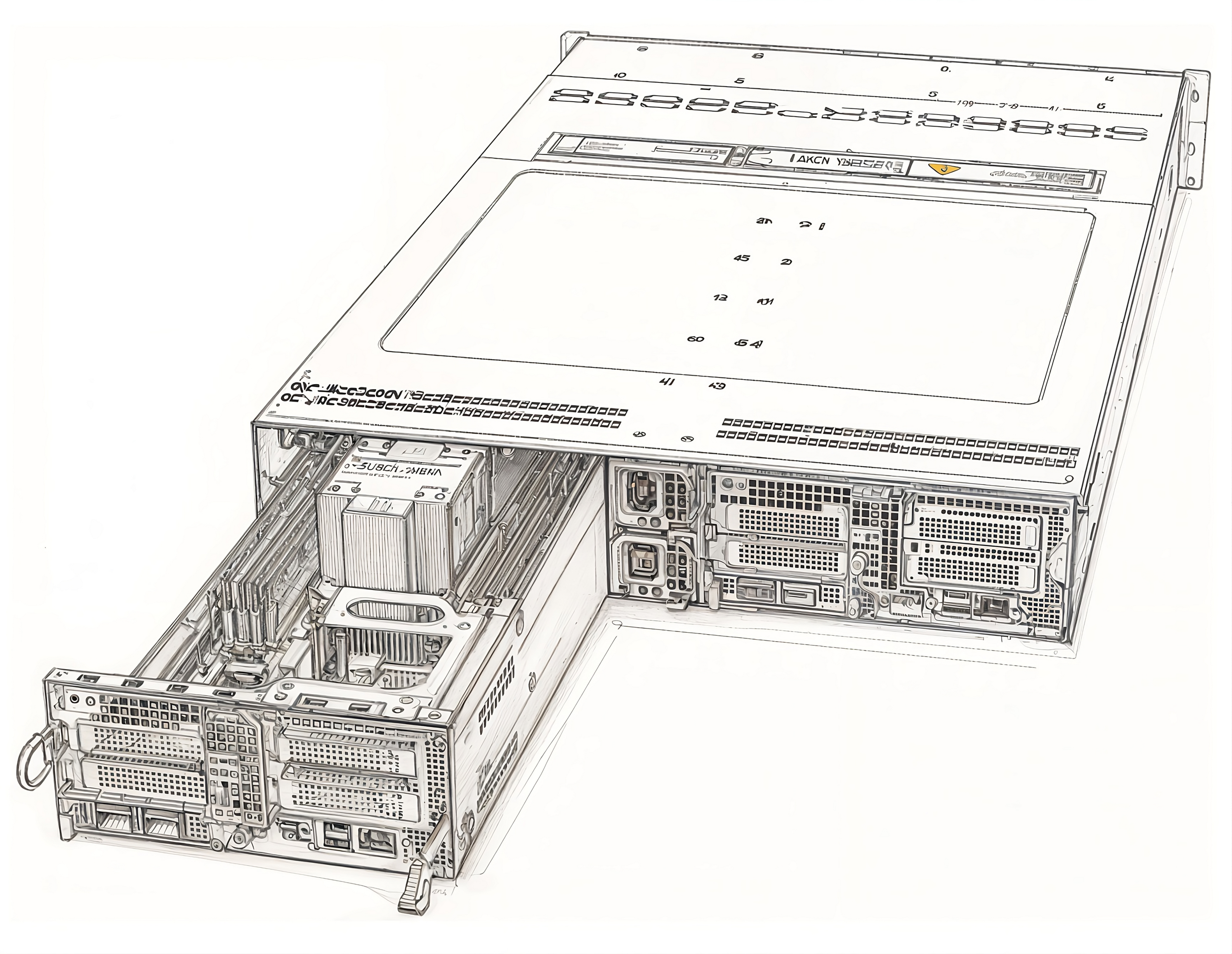}
    \caption{The Supermicro 2U 2-Node BigTwin with 12 hot-swap 2.5"
      NVMe drives per node (Diagram by Chin Fang).}
    \Description{Photograph of Supermicro 2U 2-Node BigTwin server
      with 12 hot-swap 2.5" NVMe drives per node.}
    \label{fig:bigtwin-hardware}
\end{figure}

This work also introduces a validated path toward democratizing
high-performance data movement. The principle of co-design integrates
the data mover software, host OS, and hardware stack. It
eliminates the reliance on complex, manual, and often proprietary
system tuning. The result: a straightforward and cost-effective
deployment model that provides predictable outcomes from the
resource-constrained edge sites to the high-throughput core. This
consistency is evidenced by the use of commonly available hardware
valued at approximately \$5,500 per device for the 1--10~Gbps
mini-appliances. This capital efficiency shows that rich data
transfer capabilities are no longer confined to environments with
expensive hardware and dedicated network engineering staff. The
resulting architecture is not merely faster, but fundamentally more
accessible to the Research and Education communities at the perimeter.

The enduring lesson is that seeking a simple, elegant formula to
govern such systems is a mirage. The true path to performance lies not
in mathematical purity applied to a simplified model, but in embracing
the inherent complexity through empirical, holistic co-design. It
consists of two fundamental aspects: optimizing the internal host
architecture and ensuring efficient, scale-out coordination among peer
appliances. This report demonstrates that the same systems-thinking
principles that govern the design of physical materials are equally
critical for architecting the high-performance data systems of the
digital age.

The scalability of this architectural approach was demonstrated as
early as 2017 in the petabyte-transfer setup \cite{ESnet2017Petabyte}
and \cite{ESnet2018Record} shown in Fig.~\ref{fig:slac-testbed}
(reproduced from \cite{Kissel2020ESnet} Appendix 6.1). This scale-out
design used multiple older servers with zx-aggregated 10~Gbps
interfaces, making it evident that the software architecture could
distribute workload across nodes without cluster management
overhead. The implication is that the design---whether implemented as
a 10~Gbps mini-appliance or a 400~Gbps chassis (2 $\times$ 200~Gbps
nodes)---remains uniform across scales. An example is the
SYS-222BT-DNR \cite{SupermicroBigTwin}
(Fig.~\ref{fig:bigtwin-hardware}). It also helps extend our
testbed-based methodology to \textbf{400~Gbps} and beyond, using the
same architecture that successfully moved petabytes in 2018.

The practical impact of this approach extends beyond technical
metrics: predictable, high-throughput data movement enables research,
education, and industrial workflows at scales previously reserved for
multi-million-dollar infrastructures, now reachable with commodity
appliances starting at \$5,500.

\section{Acknowledgment}\label{sec:ack}

The authors extend their sincere gratitude to the following colleagues
for their invaluable insights and for graciously reviewing early drafts
of this manuscript:

\begin{itemize}
\item \textbf{\href{https://www.es.net/about/esnet-staff/esnet-leadership/Chin-Guok/}%
  {Chin Guok}} (CTO, ESnet) for reviewing the manuscript and
  providing network expertise.
\item \textbf{\href{https://www.es.net/about/esnet-staff/testbeds-and-prototypes/ezra/}%
  {Dr.~Ezra Kissel}} (Network Research Engineer, ESnet) for
  his thoughtful review of our testbed methodology and for his
  foundational 2012 work on Linux network emulation at the University
  of Delaware.
\item \textbf{\href{https://www.frontiersin.org/journals/high-performance-computing/articles/10.3389/fhpcp.2024.1414569/full}%
  {Dr.~Amedeo Perazzo}} (formerly LCLS-II Controls and Data
  Systems Director, SLAC) for offering insights from large-scale
  scientific facilities.
\item \textbf{\href{https://profiles.stanford.edu/jana-thayer}%
  {Dr.~Jana Thayer}} (LCLS Experimental Data Systems Division
  Director, SLAC) for her review and for insights into the LCLS-II data
  system architecture and real-time processing requirements that
  motivated our focus on streaming data movement and holistic system
  design.
\item \textbf{\href{https://inspirehep.net/authors/1001765}%
  {Dr.~Wilko Kroeger}} (LCLS-II Information Specialist,
  SLAC) for his insights and review of the data management challenges
  associated with the LCLS-II, which informed the architectural
  perspectives in this work.
\item \textbf{\href{https://pawsey.org.au/about-us/staff-list/}%
  {Mark Gray}} (Head of Strategic Partnership, and formerly
  Head of Scientific Platforms, Pawsey Supercomputing Centre) for
  contributing HPC and storage perspectives, sharpened by our
  long-standing discussions on SKA data movement challenges since 2018.
\item \textbf{\href{https://github.com/breuner}%
  {Sven Breuner}} (creator of BeeGFS and the
  \texttt{elbencho} tool) for his review and for insights from a
  storage architecture and performance perspective.
\item \textbf{\href{https://pmc.ncbi.nlm.nih.gov/articles/PMC10139958/}%
  {Chih~Chuan Shih}} (Platform Lead, Genome Institute of
  Singapore) for providing genomics and data lifecycle insights.
\item \textbf{\href{https://nrid.nii.ac.jp/ja/nrid/1000040852471}%
  {Dr.~Tsukasa Nakamura (\zhname{中村 司})}} (KEK, Institute of Materials
  Structure Science) for his assistance in establishing the KEK IT and
  internal network environment for the zx implementation, leveraging
  his expertise in large-scale computational research and
  bioinformatics.
\end{itemize}

Their feedback, drawn from deep expertise across networking,
large-scale scientific facilities, high-performance computing,
high-performance storage, and genomics, significantly strengthened this
work. The views and conclusions presented herein are, of course, solely
those of the authors.

The authors also gratefully acknowledge the generous in-kind support
that enabled critical empirical validation of the architecture: SLAC
National Accelerator Laboratory for providing testbed space and
facilities (2015--2019); ESnet for provisioning a dedicated
5{,}000-mile 100~Gbps On-demand Secure Circuits and Advance
Reservation System (OSCARS) loop (2015--2019); Mellanox Technologies
(now part of NVIDIA) and Intel Corporation for providing essential
hardware components; and AIC for providing the storage servers.

Chin Fang wishes to thank his colleague Riccardo Veraldi for valuable
assistance with TCP congestion control and kernel optimization in
2024. While Chin was building the data movement appliances
(Section~\ref{sec:dedicated-lines}), Riccardo's guidance contributed
to the success of this work.

Special thanks to Dr.~Roger Leslie (Les)
Anderton Cottrell, whose mentorship prior to his retirement from SLAC
in 2017 played a pivotal role in enabling further collaboration with
the DOE national laboratory community.

Acknowledgments are due to two Zettar colleagues, Igor Solovyov and
Alexander Nazarenko, for their critical engineering contributions to
the zx software, which proved instrumental in the work presented
herein. Mr.~Nazarenko, paired with Chin Fang, earned the Overall
Winner title at the SCA19 Data Mover Challenge
\cite{NSCC2019DMC}.

Finally, Chin Fang wishes to express gratitude to his wife for her
unwavering personal support throughout the duration of this sustained
effort.

\section{Author Biographies}\label{sec:bios}
\begingroup\footnotesize
\setlength{\intextsep}{2pt}
\vspace{1\baselineskip}
\phantomsection\label{bio:fangchin}
\noindent\par
\begin{wrapfigure}[7]{l}{3.5cm}
  \vspace{-4pt}
  \includegraphics[height=2.0cm,keepaspectratio]{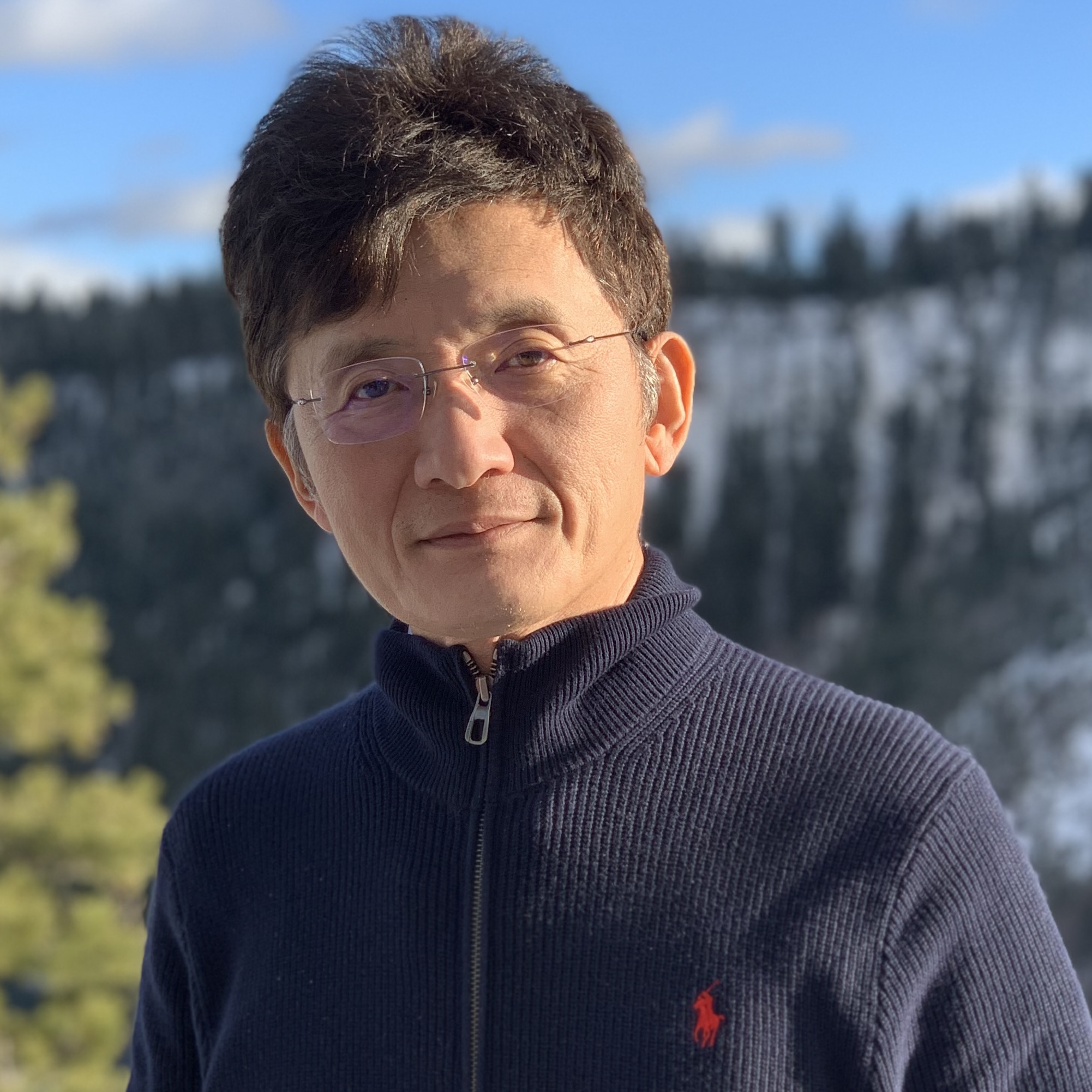}
\end{wrapfigure}
\noindent\textbf{Chin Fang (\zhname{方智})} is the Founder and CEO of
  Zettar Inc. He originated the co-design principle for demanding data
  transport (bulk and streaming), together with the ``Drainage Basin
  Pattern'' conceptual model (Fig.~\ref{fig:dbp}).  The two form the
  conceptual foundation of this work. His expertise spans
  system architecture, systems engineering, and performance
  validation. This technical background underpins many landmark
  production-scale results, for example
  \cite{ESnet2017Petabyte,ESnet2018Record,ICM2019Poland,Kissel2020ESnet}.
  During a collaboration with Intel Corp., he gained extensive
  experience in building and testing high-throughput data movement
  appliances and latency-simulation testbeds. Zettar's technologies,
  including the zx data mover, are commercially integrated into
  appliances sold by partners such as Hewlett Packard Enterprise
  (HPE).  Intel Corporation has also collaborated with Zettar on
  related development projects. He holds M.S. and Ph.D. degrees in
  Mechanical Engineering from Stanford University.\par

\needspace{7\baselineskip}
\phantomsection\label{bio:stitt}
\begin{wrapfigure}[7]{l}{3.5cm}
  \vspace{-4pt}
  \includegraphics[height=2.0cm,keepaspectratio]{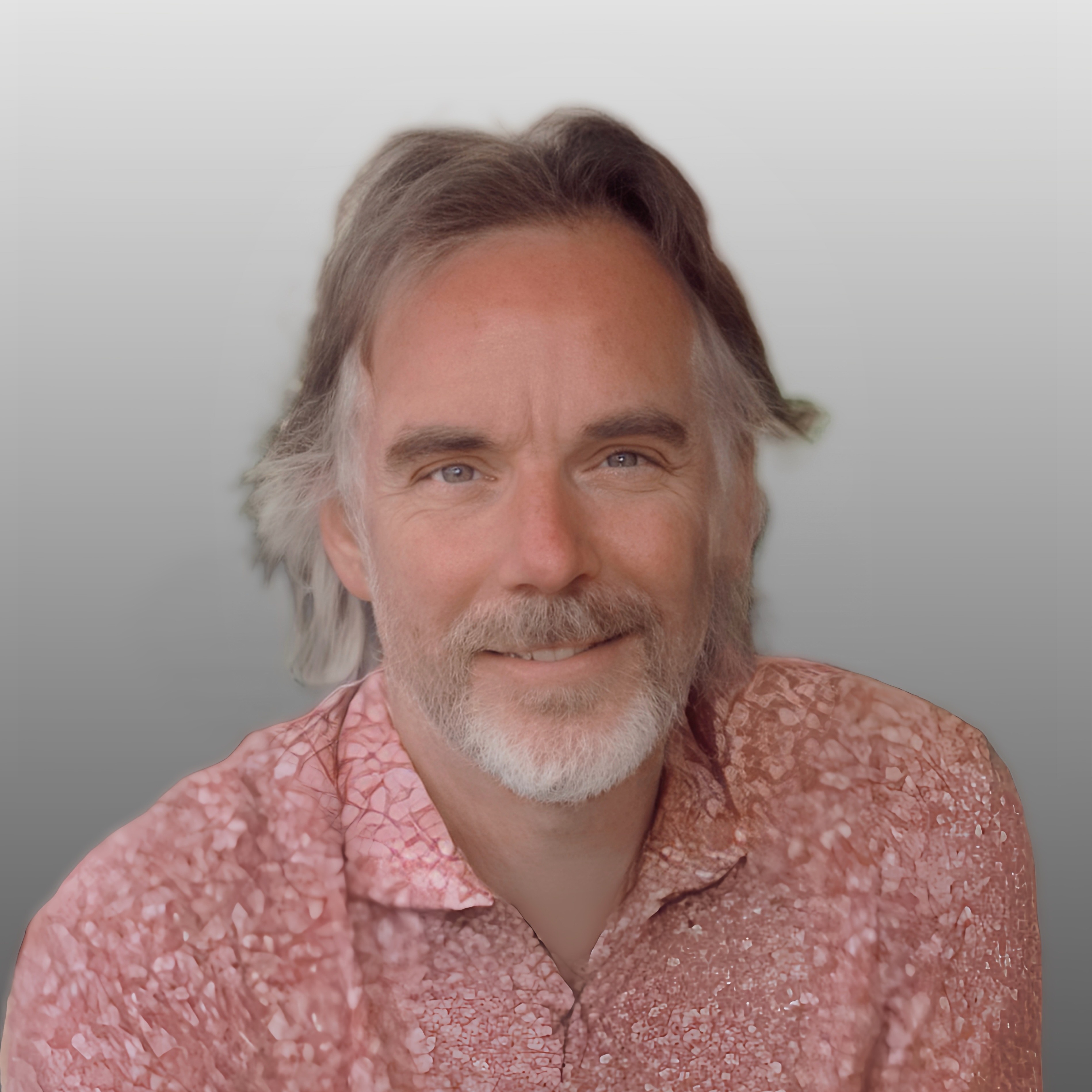}
  \end{wrapfigure}
  \noindent\textbf{Timothy Stitt} brings over two decades of cross-disciplinary
  expertise in strategizing, procuring, and optimizing HPC and AI
  infrastructure for data-intensive scientific workflows. His career
  spans premier academic and corporate research institutions, where he
  has led the architectural planning and integration of large-scale,
  composable compute and storage services. This extensive experience
  with the full stack of HPC technologies---from application-level
  tuning to global service design---provides a critical real-world
  perspective on the infrastructural barriers and requirements for
  end-to-end data movement. His contributions have been recognized with
  awards, including \enquote{Best Use of HPC in Life Sciences} (SC16) and
  \enquote{Best Practice in IT Infrastructure/HPC} (Bio-IT World 2017). Other
  than his current position at a tier-1 biopharma business, Dr.~Stitt's
  background includes roles as a Research Assistant Professor at the
  University of Notre Dame, a Lecturer in Computer Science, and an HPC
  Application Scientist at the Swiss National Supercomputing Centre
  (CSCS), underpinned by a Ph.D.~in Computational Science.\par

\needspace{7\baselineskip}
\phantomsection\label{bio:mcmanus}
\begin{wrapfigure}[7]{l}{3.5cm}
  \vspace{-4pt}
  \includegraphics[height=2.0cm,keepaspectratio]{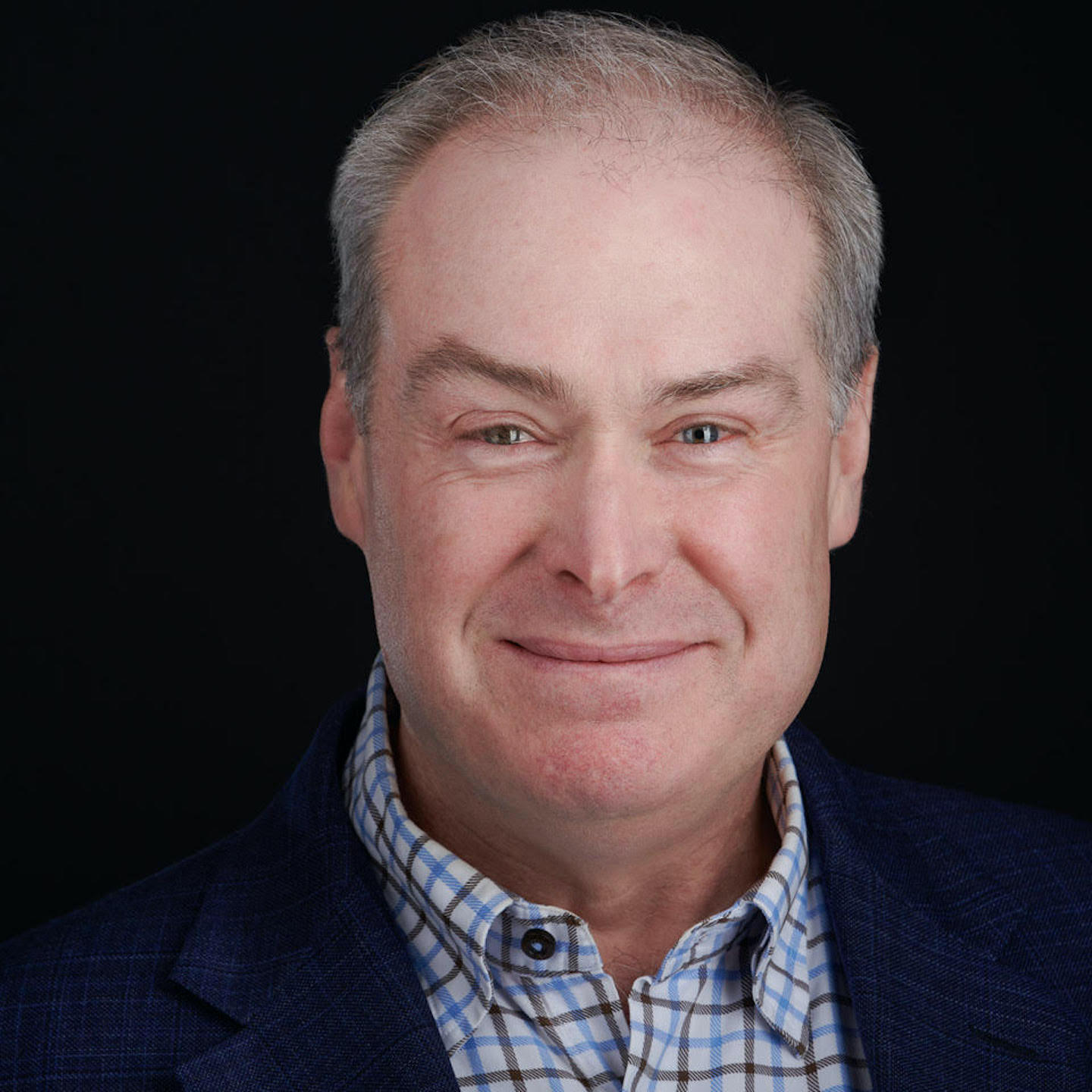}
  \end{wrapfigure}
  \noindent\textbf{Michael J.~McManus} brings a cross-disciplinary perspective
  from a career spanning deep science and enterprise IT. With a Ph.D.~in
  synthetic organic chemistry (MIT) and a B.S.~in polymer chemistry
  (UMass Amherst), his experience---from the U.S. Army and Intel to
  Fujitsu, Kodak, and six scientific software startups---embodies the
  cross-domain integration this report advocates as an antidote to siloed
  optimization. His role as Principal Engineer and Director of Precision
  Medicine at Intel, coupled with his tenure on the NIH AI Working
  Group, provides critical context for the discussion on data movement
  in computational science and AI/ML workflows.\par

\needspace{7\baselineskip}
\phantomsection\label{bio:moriya}   
\begin{wrapfigure}[7]{l}{3.5cm}
  \vspace{-4pt}
  \includegraphics[height=2.0cm,keepaspectratio]{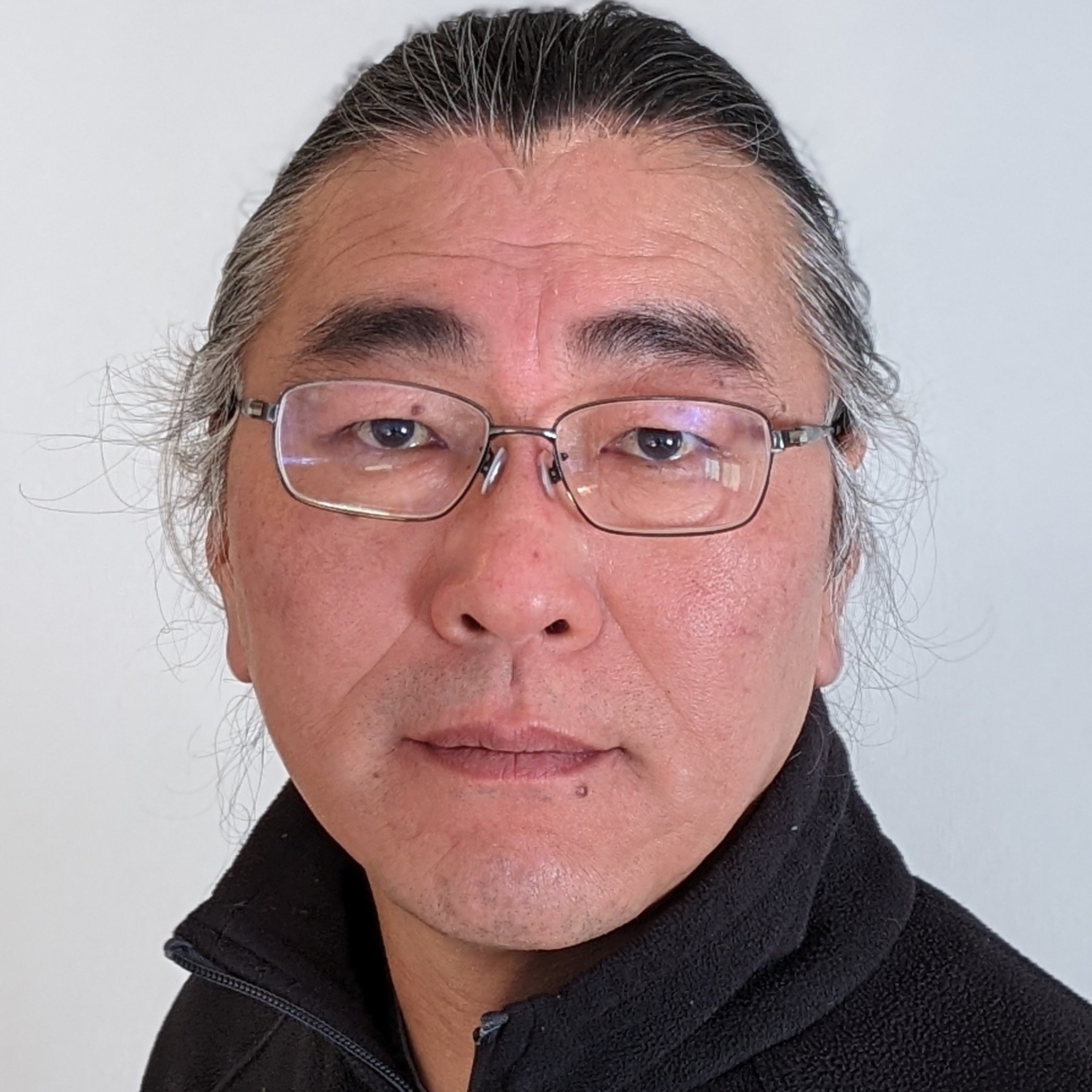}
  \end{wrapfigure}
  \noindent\textbf{Toshio Moriya (\zhname{守屋 俊夫})} is a Project Associate
  Professor at the High Energy Accelerator Research Organization (KEK)
  in Japan. He provided an independent scientific validation
  environment for the current work, specifically through the Cryo-EM
  data use case, enabling independent scientific replication and
  domain-specific benchmarking. His research focuses on automating
  Cryo-EM single-particle analysis to remove human involvement from
  the workflow and to overcome practical barriers in applying Cryo-EM
  to compound screening for structure-based drug design (SBDD) and
  other industrial applications. In line with this objective, he is
  establishing an IoT-based Cryo-EM network across Japan. Using AWS
  cloud services as the hub and zx as the main data mover
  (Fig.~\ref{fig:kek-cryoem-cloud}), this system will fully automate data
  processing between nationwide Cryo-EM facilities and the cloud. This
  effort aims to rapidly build a large-scale database of
  compound-bound protein structures and enable future big-data-driven
  discovery.\par
\endgroup
% ===================================================================
% Appendix: Data Movement and AI Infrastructure Efficiency
%
% INSERT: append this whole block at the very END of
% converted_content.tex, just after the final \endgroup that closes
% the Author Biographies section. Because all four main*.tex variants
% \input{converted_content}, putting it there gives every build the
% appendix from one source.
% ===================================================================
\appendix
% Number appendix floats as A.1, A.2, ... instead of continuing the
% body's sequence, so they read clearly as appendix material.
\setcounter{table}{0}
\renewcommand{\thetable}{\thesection.\arabic{table}}
\setcounter{figure}{0}
\renewcommand{\thefigure}{\thesection.\arabic{figure}}

\section{Data Movement and AI Infrastructure Efficiency}
\label{sec:ai-infra}

Efficient data movement fundamentally changes AI infrastructure
economics. Given a fixed budget (e.g., \$1M), two approaches emerge
(Table~\ref{tab:ai-infra}).

\begin{table}[!ht]
  \centering
  \caption{Two approaches to AI infrastructure under a fixed \$1M
    budget. All figures are illustrative; see the notes below the
    table.}
  \label{tab:ai-infra}
  \small
  \begin{tabularx}{\textwidth}{@{}l X X@{}}
    \hline
    \tblheader{Feature} & \tblheader{Traditional Approach}
      & \tblheader{Optimised Approach} \\
    \hline
    GPU Model & 32 $\times$ NVIDIA B200 & 8 $\times$ NVIDIA B200 \\
    Data Infrastructure & Standard / bottlenecked\textsuperscript{*}
      & Co-designed high-speed \\
    Target Utilisation\textsuperscript{\S} & 20\% (data starvation)
      & 85\% (continuous) \\
    Effective Compute\textsuperscript{**} & 6.4 GPU-equiv.\
      & 6.8 GPU-equiv.\ \\
    Idle Power (total)\textsuperscript{\textdagger} & $\sim$6,400~W
      & $\sim$1,600~W \\
    Operating Power (total)\textsuperscript{\textdagger}
      & $\sim$12,800~W & $\sim$7,040~W \\
    Annual Energy\textsuperscript{\textdagger} & $\sim$112,128~kWh
      & $\sim$61,670~kWh \\
    Est.\ Hardware Cost & $\sim$\$1,600,000+
      & $\sim$\$750,000\textsuperscript{\textdaggerdbl} \\
    vs.\ \$1M Budget & Over budget & Under budget \\
    Throughput per Dollar\textsuperscript{**} & 1.0$\times$ (baseline)
      & $\sim$2.3$\times$ \\
    \hline
  \end{tabularx}

  \par\smallskip
  {\footnotesize\raggedright
   \textsuperscript{*}\,Without co-designed data-movement appliances
   and burst buffers, production storage I/O becomes the bottleneck
   (Sections~\ref{sec:burst-buffers} and~\ref{sec:bandwidth}).\par
   \textsuperscript{\S}\,Data starvation is the dominant
   \textbf{addressable} cause of the utilisation gap. Kernel
   efficiency, batch size, and interconnect also bear on GPU
   utilisation; the figures here isolate the data-movement
   contribution rather than claiming it as the sole lever.\par
   \textsuperscript{**}\,Effective compute $=$ GPU count $\times$
   utilisation. The optimised approach delivers $\sim$6\% higher
   effective throughput at roughly 47\% of the capital cost and 55\%
   of the energy, about a \textbf{2.3$\times$ improvement in
   throughput-per-dollar}.\par
   \textsuperscript{\textdagger}\,Illustrative, from publicly
   available TDP estimates. The power and energy rows count the GPU
   nodes only; the co-designed appliances draw additional power that
   is not included in these figures.\par
   \textsuperscript{\textdaggerdbl}\,Includes GPU nodes plus
   co-designed data-movement appliances.\par}
\end{table}

The optimised model removes most data-starvation idle time, and the
same result serves two kinds of operator. Where budget is the binding
constraint, it delivers comparable effective throughput at roughly
half the capital and operational cost: the same outcome for less.
Where scale is the goal, that same spend converts idle GPU-hours into
useful work, so the accelerators already deployed deliver more. The
underlying gain is one and the same; whether it is taken as savings
or as added capacity is the operator's to decide.

% Keep Table A.1 anchored within the appendix; without this the
% float drifts forward into the References list.
\FloatBarrier

% References section
\section{References}
\begingroup
\footnotesize
\bibliographystyle{unsrturl}
\setlength{\bibsep}{0pt plus 0.5ex}
% pdflatex uses \bibliography; LaTeXML (arXiv HTML) ignores the .bbl
% and rebuilds from references.bib in file order, which renumbers the
% list. \input the citation-ordered .bbl instead so the HTML numbering
% matches the PDF. \jobname resolves to main_arxiv locally and main on
% arXiv, so the same line finds the right .bbl in both builds.
\iflatexml
\else\bibliography{references}\fi
\endgroup
\end{document}